\documentclass[fleqn,usenatbib]{mnras}
\usepackage{newtxtext,newtxmath}

\usepackage[T1]{fontenc}

\DeclareRobustCommand{\VAN}[3]{#2}
\let\VANthebibliography\thebibliography
\def\thebibliography{\DeclareRobustCommand{\VAN}[3]{##3}\VANthebibliography}

\usepackage{graphicx}
\usepackage{caption}
\usepackage{subcaption}
\usepackage{booktabs}
\usepackage{lscape}
\usepackage{footnote}
\usepackage{ulem}
\usepackage{color}  
\usepackage{amsmath}
\usepackage{rotating}

\newcommand{\alii}{Al{\sc ii}$\lambda$1670\/}
\newcommand{\aliiMg}{Al{\sc ii}$\lambda$2670\/}
\newcommand{\aliii}{Al{\sc iii}$\lambda$1860\/}
\newcommand{\al}{Al{\sc iii}}

\newcommand{\chisq}{$\chi^{2}$}
\newcommand{\cii}{{\sc{Cii}}$\lambda$1335\/}
\newcommand{\ciii}{{\sc{Ciii]}}$\lambda$1909\/}
\newcommand{\cnl}{{\sc{Ciii]}}}
\newcommand{\civonly}{C{\sc iv}\/}
\newcommand{\aliiionly}{Al{\sc iii}\/}
\newcommand{\mgiionly}{Mg{\sc ii}\/}

\newcommand{\civ}{{\sc{Civ}}$\lambda$1549\/}

\newcommand{\ergss}{ergs s$^{-1}$\/}

\newcommand{\feiiopt}{{Fe \sc{ii}}$_{\rm opt}$\/}
\newcommand{\feii}{{Fe\sc{ii}}\/}
\newcommand{\feI}{{Fe\sc{i}}\/}
\newcommand{\feUV}{{Fe\sc ii~}UV191}

\newcommand{\feiiq}{\rm Fe{\sc ii}$\lambda$4570\/}

\newcommand{\feiii}{{Fe\sc{iii}}\/}
\def\coneq{{c($\frac{1}{4}$)}\/}
\def\ctwoq{{c($\frac{1}{2}$)}\/}
\def\cthreeq{{c($\frac{3}{4}$)}\/}

\newcommand{\gtsima}{$\; \buildrel > \over \sim \;$}
\newcommand{\gtsim}{\lower.5ex\hbox{\gtsima}}

\newcommand{\hb}{{\sc{H}}$\beta$\/}

\newcommand{\heii}{He{\sc{ii}}$\lambda$1640}
\newcommand{\heiionly}{He{\sc{ii}}}

\newcommand{\heiiuv}{He{\sc{ii}}$\lambda$1640}

\newcommand{\kms}{km s$^{-1}$}
\newcommand{\lbol}{$L_{\rm bol}$\/}

\newcommand{\ledd}{$L_{\rm Edd}$\/}

\newcommand{\lledd}{$L/L_{\rm Edd}$}
\newcommand{\ltsima}{$\; \buildrel < \over \sim \;$}
\newcommand{\ltsim}{\lower.5dex\hbox{\ltsima}}  
\newcommand{\lya}{{ Ly}$\alpha$}

\newcommand{\mbh}{$M_{\rm BH}$\/}
\newcommand{\mgii}{{Mg\sc{ii}}$\lambda$2800\/}

\newcommand{\msun}{M$_\odot$\/}

\newcommand{\oi}{O{\sc ii}$\lambda$1304\/}

\newcommand{\oiii}{{\sc [Oiii]}}
\newcommand{\oiiiMg}{O{\sc iii}$\lambda$2672\/}
\newcommand{\oiiiopt}{{\sc{[Oiii]}}$\lambda\lambda$4959,5007\/}

\newcommand{\oiiiuv}{{\sc{Oiii]}}$\lambda$1663\/}
\newcommand{\oiv}{O{\sc iv]}$\lambda$1402\/}
\newcommand{\oivonly}{O{\sc iv]}\/}

\newcommand{\rfe}{$R_{\rm FeII}$}

\newcommand{\siiii}{Si{\sc iii]}$\lambda$1892\/}
\newcommand{\si}{Si{\sc iii]}}
\newcommand{\siiv}{Si{\sc iv}$\lambda$1397\/}
\newcommand{\siivonly}{Si{\sc iv}}
\newcommand{\siiiuuv}{Si{\sc ii}$\lambda$1306\/}

\defcitealias{MS14}{MS14}
\defcitealias{vp06}{VP06}
\defcitealias{marziani22}{M22}
\defcitealias{BG92}{BG92}
\defcitealias{marzianietal17}{M17}

\title[LBT super-Eddington quasars]{LBT IR observations of candidate super-Eddington quasars}

\author[Buendia-Rios et al. ]{T. M. Buendia-Rios$^{1}$, P. Marziani$^{2}$, C. A. Negrete$^{3}$ and D. Dultzin$^{1}$
\\
$^{1}$ Instituto de Astronomía, UNAM, México CDMX. 04510, Mexico\\
$^{2}$INAF, Osservatorio astronomico di Padova, 35122, Padova, Italy\\
$^{3}$ CONACyT Research Fellow - Instituto de Astronomía, UNAM, México CDMX. 04510, Mexico
}

\date{Accepted XXX. Received YYY; in original form ZZZ}

\pubyear{2024}

\begin{document}
\label{firstpage}
\pagerange{\pageref{firstpage}--\pageref{lastpage}}
\maketitle

\begin{abstract}
Quasars accreting at very high rates are believed to be prime movers of galactic evolution because of their high radiative and mechanical output. The study presented in this paper investigates a sample of six highly accreting quasars at redshifts \( z = 2-3 \) using near-infrared observations from the LUCI spectrograph at the Large Binocular Telescope (LBT). The aim is  obtain a precise measure of the quasar systemic redshift and accretion parameters (black hole mass and Eddington ratio) primarily from the  \hb\ line, \ and  on second stance from other intermediate and low ionization lines.  Outflow dynamical parameters (mass rate of outflowing gas, its kinetic power and momentum rate) were estimated from the \civ\ emission line that is perhaps the most easily accessible tracer of high-ionization  winds from the accretion disk, obtained from the Sloan Digital Sky Survey. In addition, the joint analysis of the rest-frame optical and UV spectra allowed us to estimate the chemical composition of the broad line emitting gas. The high metal content of the outflowing gas ($Z \gtrsim 10 Z_\odot$) and the high values of thrust and kinetic power may induce a chemical feedback effect in the quasar host, in addition to mechanical feedback. 


\end{abstract}

\begin{keywords}
quasars: supermassive black holes --quasars: emission lines -- line: profiles -- line: formation
\end{keywords}



\section{Introduction}

Highly accreting quasars are the most luminous quasars for a given black hole mass  \citep{mineshigeetal00,ohsugaetal05,ohsugamineshige07,sadowski11}. They may be among the prime agents of galactic evolution \citep[e.g.,][]{hopkinsetal06,fabian12} because of their extreme radiative and mechanical output: nuclear outflows, and radiative feedback effects on their host galaxies,  are maximized and reach impressive values of kinetic power and momentum rate \citep[e.g.,][]{marzianietal16a,vietrietal18}.  Very luminous quasars radiating close to the Eddington limit do not exist anymore in the local Universe \citep[e.g.,][]{shankaretal04,shankaretal09}, but were relatively frequent at $z \approx$ 2, an epoch where also the volume integrated star formation rate reaches its peak \citep[][and references therein]{conselice14}.  Therefore,  quasar feedback effects that may be more relevant for galactic evolution should be studied  at relatively high $z$.


A signature of quasar outflows is the Doppler shifting toward the blue of absorptions and emission lines \citep[e.g.][]{MS12}.  The \civ\ line in emission is observed in almost every quasar spectrum, is prominent (W\ $\sim 5 - 200$\AA),  relatively clear from contaminant lines, and is often -- but not always --  significantly blueshifted with respect to rest frame  \citep[e.g.,][]{marziani96,leighly04,leighlymoore04,sulentic07,richards11,marzianietal16a,vietrietal20}. The \civ\ line profile is perhaps the most accessible diagnostics of nuclear outflows (i.e.,  traces the outflow at the origin,  as an accretion disk wind) in quasars at $z >$ 1.4, where \civ\ shifts to the optical range. For this reason, the quasar rest-frame (RF) has to be measured with high precision ($< 100$ \kms\ at 1 $\sigma$\ confidence level) in order to estimate the dynamical parameters of the wind (such as mass flow rate, thrust and kinetic power) that may be able to escape the black hole sphere of influence.   To reliably set the quasar RF, low-ionization narrow lines such as the narrow component of \hb\ are best suited \citep{eracleous04,bon20}, as recent work confirms that it is not possible to obtain accurate redshift estimates from broad UV lines, especially from the ones whose parent species ionization potential is high  \citep{sulentic07,hewettwild10,shen16}. 

  

We remark that the FWHM of H$\beta$\ is an appropriate virial broadening estimator for the computation of the black hole mass \citep[][and references therein]{mejiarestrepo16}, while  \civ\ line is unsuitable especially for sources radiating at high Eddington ratio \citep{sulentic07,netzer07,richards11,marzianietal19}, as the broadening is most likely due to a velocity shear and not to virialized gas motion. Even if a correction to the line width can be applied \citep{brothertonetal15,coatman16,marzianietal19}, the process is often cumbersome and may still require an independent estimate of the quasar rest frame to work at best.   Other low-and intermediate ionization lines such as \mgii\ \citep{trakhtenbrot12,marzianietal13a} and \aliii\ \citep{marziani22} need further testing, because an outflow component has been revealed in both lines for highly-accreting sources \citep{marziani13}.

One can now count on the improved ability to distinguish sources radiating close to the Eddington limit, thanks to the exploitation of E1 of quasars \citep[E1, e.g.][]{BG92,sulentic00a,shenho14}.  However, there are still relatively few   observations covering   \hb, \oiii, and \feii\  redshifted into the IR for   highly accreting quasars {(hereafter xA, for extreme Population A, \citealt{MS14})}. An empirical selection criterion is based on the value of the ratio between the singly-ionized blend centered at $\lambda$4570 and the Balmer line \hb. A source is highly accreting if \rfe$\gtrsim 1$\ \citep{MS14,duetal16}. 
Sources at low $z$\ meeting this selection criterion  are mostly radiating close or above Eddington ratio \lledd $\approx$ 1 \citep{MS14,duetal16,negrete18}. 

Criteria have been suggested to select sources radiating close to their Eddington limit from diagnostic ratios computed from UV lines \citep{MS14}, as outlined in Section \ref{sec:sam_sel}.  However, the selection criterion based on UV ratios  is as yet untested on large samples. This circumstance is unfortunate, as it makes the identification of highly accreting sources more prone to spurious classification. In addition, other properties needs confirmations. Among all, the high value of the metallicity inferred by several previous studies \citep{sniegowskaetal20} can be more safely verified if ratios involving \mgii, \feiiopt\  and \hb\ are available \citep{florisetal24}. As mentioned, highly accreting quasars should produce the highest radiative output per unit black hole mass. Even if radiation pressure drives a wind, estimates of the wind dynamical parameters as traced by the optical and UV high-ionization lines differ wildly \citep[e.g.,][]{bisognietal17,fioreetal17,marzianietal17,vietrietal17,vietrietal18,kakkadetal20}, the wind dynamical parameters of xA sources at the highest luminosity should exceed the threshold for significant feedback effect \citep{dimatteoetal05}, if AGN  are to have a feedback effect at all. This is said leaving aside the most complex part of the problem, namely the dispersion of thrust and kinetic power along the interstellar medium \citep{kingpounds15}.   

 {The coverage of the \hb\ spectral range is essential for verifying the consistency of the UV/optical selection criteria, conducting additional tests at high $z$ values, and analyzing the dynamical parameters of the winds. However, there is still a lack of data covering the optical/UV range at high redshifts, especially for super-Eddington candidates. The LBT sample makes a significant contribution to the study of xA quasars by expanding the observational coverage available to date (available for a few tens of sources at most), and providing useful information of the physical properties, emission mechanisms, and evolutionary trends regarding the highly accreting sources.}
After the description of the sample selection in Section \ref{sec:sam_sel}, Section \ref{obsred} presents the new LUCI \citep[LBT Utility Camera in the Infrared,][]{seifertetal03} NIR observations at LBT that, as summarized before,  are fundamental to measure the dynamical properties of the wind from the accretion disk as they relate them to the accretion parameters. Section \ref{sec:data_an} describes the analysis of the redshift estimation, the multi-component fitting of the \hb\ and \feii\ spectral ranges, and the UV-RF from \civ\ to \mgii. In Section \ref{sec:results} we present the empirical results, including the inter-comparison between \civonly\ and \hb\ and other lines. Section \ref{sec:discussionb} is focused on the analysis of the derived physical parameters, and discusses the virial \mbh\ obtained from \hb, \mgii\ and \aliii. Wind parameters including mass outflow rate, thrust, kinetics power, estimates of metal content, and  the inclusion of these objects in an Hubble diagram are also discussed. 

\section{Sample description and observations}\label{sec:sam_sel}

\subsection{Selection of  xA quasars}

As the sample for the NIR (using LUCI observations), we selected six sources among the ones of the \citet{MS14} SDSS-based sample in the redshift range $2.35 < z < 2.44$\ in order to ensure maximum S/N and full coverage in the \hb\ spectral range  4600 \AA\ -- 5100 \AA.  \civ, \aliii\  and \siiii\ are already covered by high S/N optical spectra from the Sloan survey. 

A previous analysis (\citetalias{MS14}) suggested working criteria to identify highly accreting quasars {(i. e. extreme quasars or 'xA' quasars)}: (a) intensity ratio between \feii\ emission at $\lambda$4570 (or equivalently at $\lambda$5130) and \hb\ \rfe\ $>$1.0, in the optical;  (b) intensity ratios \aliii / \siiii\ $\geq$ 0.5 and \siiii / \ciii\ $\leq$1.0, in the UV. The sources of \citetalias{MS14} shown Eddington ratio distribution of low-$z$ and high-$z$\ highly accreting objects is sharply peaked around 1, with a dispersion of just 0.13 dex, and were selected by applying criteria (a) and (b). The sources of the present sample were selected for observations by applying criteria (b) only due to the lack of pre-existing IR observations covering the \hb\ spectral range. 

Table \ref{tab:sample} gives source identifications and basic properties including: the SDSS name (Col. 1), apparent V magnitude (Col. 2), apparent magnitudes in the H-band obtained from 2MASS and LUCI-LBT (Col. 5), redshift and uncertainty (Cols. 6 and 7), airmass (Col. 8) and comments (Col. 9). Three of the sources of our sample are radio-detected in the Faint Images of the Radio Sky at Twenty-cm Survey (VLA-FIRST Survey\footnote{\url{https://sundog.stsci.edu/index.html}}), {and were found to be formally radio-quiet, although close to the minimum Kellermann's ratio for radio-intermediate $R_\mathrm{K} \sim 10$\ \citep{zamfiretal08}. The $R_\mathrm{K}$ reported in Table \ref{tab:sample}  was computed according to the original defintion by \citet{kellermannetal89},  by taking the FIRST flux and dividing it by the $H$\ band flux which correspond to the de-redshifted 5 GHz and $B$ \ band fluxes.  }

 \subsection{Preliminary analysis of Sloan data}

The optical data were acquired through the Baryon Oscillation Spectroscopic Survey (BOSS), the extended Baryon Oscillation Spectroscopic Survey (eBOSS), and the Legacy Survey programs of the Sloan Digital Sky Survey (SDSS). The spectra, which had been calibrated for both wavelength and flux, serve as the optical counterpart to the NIR sample.  For redshift and flux conversions to the rest frame, we divided the wavelength scale by $(1+z)$ and increased the flux by a factor of $(1+z)^3$, using the redshift values provided by SDSS as initial estimates. 
 


\begin{table*}
\centering
\caption{Source identification and basic properties of the LUCI-LBT sample.}
\label{tab:sample}
\resizebox{\textwidth}{!}{%
\begin{tabular}{lllllllll}
\hline
\multicolumn{1}{c}{SDSS ID} & \multicolumn{1}{c}{$m_V$} & \multicolumn{1}{c}{$i_\mathrm{SDSS}$} & \multicolumn{1}{c}{$z_\mathrm{SDSS}$} & \multicolumn{1}{c}{$H_\mathrm{2MASS}$} & \multicolumn{1}{c}{$z_\mathrm{sys}$} & \multicolumn{1}{c}{$\delta z_\mathrm{sys}$} & \multicolumn{1}{c}{AM} & \multicolumn{1}{c}{Comments} \\
\multicolumn{1}{c}{(1)} & \multicolumn{1}{c}{(2)} & \multicolumn{1}{c}{(3)} & \multicolumn{1}{c}{(4)} & \multicolumn{1}{c}{(5)} & \multicolumn{1}{c}{(6)} & \multicolumn{1}{c}{(7)} & \multicolumn{1}{c}{(8)} & \multicolumn{1}{c}{(9)} \\ \hline
J084502.73+081214.3 & 18.357 & 18.280 & 18.048 & 16.862 & 2.37104 & 0.00077 & 1.67 & \begin{tabular}[c]{@{}l@{}}RE: 1.49$\pm$0.13 mJy, $R_\mathrm{K} \approx 8$\\ BAL QSO\end{tabular} \\
J093403.96+315331.3 & 17.166 & 17.111 & 16.869 & 15.01*--14.68* & 2.42368 & 0.00026 & 1.47 & \begin{tabular}[c]{@{}l@{}}RE: 4.68$\pm$0.13 mJy, $R_\mathrm{K} \approx 3.5$\\ FBQS BAL\end{tabular} \\
J105427.17+253600.8 & 17.361 & 17.644 & 17.361 & 16.126 & 2.41429 & 0.00057 & 1.09 & \begin{tabular}[c]{@{}l@{}}RE: 2.99$\pm$0.13 mJy, $R_\mathrm{K} \approx 8$\\ FBQS BAL\end{tabular} \\
J125914.83+672011.8 & 17.983 & 17.947 & 17.776 & 16.257 & 2.44480 & 0.00051 & 1.45 &  \\
J144218.09+484101.8 & 18.800 & 18.798 & 18.629 & \ldots & 2.43337 & 0.00013 & 1.42 & \begin{tabular}[c]{@{}l@{}}$ < z-H > \approx 1.28 \pm 0.14$\\ $\rightarrow H \approx 17.35$\end{tabular} \\
J210831.56-063022.5 & 17.234 & 17.176 & 17.064 & 15.775 & 2.37180 & 0.00031 & 1.67 &  \\ \hline
\end{tabular}%
}
\vspace{1ex}

\textbf{Notes:} Columns are as follows: (1) SDSS coordinate name. (2) Apparent V magnitude. (3) SDSS magnitude in the $i$ band. (4) SDSS magnitude in the $z$ band. (5) Apparent IR magnitude in the H-band (2MASS), *Apparent magnitude in the H-band using our LBT estimates. (6) Systemic redshift. (7) Redshift uncertainty. (8) AM = airmass. (9) Quasar classification: first bright quasar survey (FBQS); broad absorption lines (BAL); radio emission (RE) detected in FIRST; $R_\mathrm{K}$ is Kellermann's ratio \citep{kellermannetal89}.
\end{table*}

\section{LUCI/LBT Observations and data reduction}
\label{obsred}

Table \ref{tab:log_obs} contains a summary of the observations, including the SDSS identification, observation date, grism employed, total exposure time, and the camera employed for each LUCI configuration. The observations required the LUCI1 N1.8 camera with a plate scale of 0.25 pixels and the 210 lines/mm high-resolution grating, covering the H band (central wavelength 1.65$\mu$m) for each run. For the grism, it was employed G210HiRes and G200LoRes with resolutions of 5900 and 1900/2600 [$\lambda_\mathrm{cen}/\delta\lambda$], respectively. The $\lambda_\mathrm{cen}$ were 1.65 and 1.93$\mu$m for the high-resolution grism and low, respectively.

The calibration strategy and requirements include standard stars for spectrophotometric calibration and telluric band subtraction, which were deemed necessary for each target. These calibrations were to be conducted at a similar airmass as the science exposure ($\sim$1.5), either immediately before or after the science exposure (seeing requested 1.5[arcsec]). The exposure time for each calibration was set at 3-5 minutes, suitable for a typical star with a magnitude of r $\sim$ 10. Consequently, the total exposure time on each target comprised approximately 1-3 hours (3-5 × 2 for telluric and spectrophotometric calibration × number of observations).


Data reduction was carried out in a standard way using the Image Reduction and Analysis Facility ({\fontfamily{lmtt}\selectfont IRAF}, \citealt{tody86}). Bias subtraction correction was performed nightly. Sequence of frames with a given detector integration time were obtained with the source at different positions (e.g. A,B) along the slit. Spectra were taken as a sequence of exposures recorded at  A and B position.  Wavelength calibration was obtained using the night-sky spectral emission lines from \cite{rousselotetal00}, and the spectra were corrected for dispersion direction/CCD frame axis and geometric distortion using the IRAF tasks {\tt rotate} and {\tt transform}.  The A and B  calibrated 2D spectra  were then subtracted one to the other. 1D spectra were traced on the difference frame, and then re-subtracted. 

Instrumental response and specific flux calibration were obtained nightly with observations of the spectrophotometric standard stars (see Table \ref{tab:log_obs}). 
Telluric absorptions, that affect our spectra, were also corrected using the standard stars and IR stellar spectra from \cite{pickles98}, to obtain a normalized wavelength-calibrated template of the absorption features. Each target spectrum was then divided by its corresponding standard star spectrum in order to correct for the atmospheric absorption features. This was achieved with the {\tt telluric} task, which allows to optimize the correction with slight adjustments in shift and scaling of template of the absorption features.
Finally, the correct flux calibration of each spectrum was achieved by scaling it according to the magnitude of the standard star and the source divided by the integration time.  For this, we used the {\tt sbands} task. The standard stars were observed with a wider slit that the sources, so significant light loss occurred. We compared our estimates with the Two Micron All Sky Survey \citep[2MASS,][]{Skrutskie06} magnitudes for our spectra. If we encountered  differences $>$1.5mag,  we employed the values from 2MASS. In  Table \ref{tab:sample}, the only source that was calibrated with our LBT estimates is J093403.96, and we present two values corresponding to each camera data, as this objects was observed in fraternal mode.

\begin{table*}
\centering
\caption{Log of observations}
\label{tab:log_obs}
\resizebox{\textwidth}{!}{%
\begin{tabular}{cccccccc}
\hline
ID    & \multicolumn{2}{c}{Obs. Date} & Grism     & ET   & Camera & Standard Star    & Comments         \\
(1)        & (2)           & (3)           & (4)       & (5)  & (6)    & (7)              & (8)              \\ \hline
J084502 & 2017-04-18    & 2011-11-28    & G210HiRes & 1560 & LUCI1  & HIP41751 (G4V)   & seeing=0.7 -0.9  \\
           & 2017-04-18    &               & G210HiRes & 1560 & LUCI2  &                  & fraternal mode   \\
J093403 & 2019-01-27    & 2018-12-11    & G200LoRes & 1440 & LUCI1  & HIP 44027 (G2V)  &                  \\
           & 2019-01-27    &               & G200LoRes & 1440 & LUCI2  &                  & fraternal mode   \\
J105427 & 2016-12-04    & 2013-02-17    & G210HiRes & 1800 & LUCI1  & HIP 52192 (G2V)  & seeing=1.7 - 3.2 \\
J125914 & 2017-04-17    & 2014-04-01    & G210HiRes & 1800 & LUCI1  & HIP64451 (F7V)   &                  \\
           & 2017-04-17    &               & G210HiRes & 2320 & LUCI2  &                  & fraternal mode   \\
J144218 & 2017-06-13    & 2013-04-13    & G210HiRes & 2900 & LUCI2  & HIP71172 (A0V)   &                  \\
J210831 & 2016-09-17    & 2001-06-21    & G210HiRes & 1440 & LUCI1  & HIP 106382 (A0V) & seeing=0.49-0.65 \\ \hline
\end{tabular} %
}
\\ 
\textbf{Notes:} Columns are as follows: (1) SDSS identification. (2) LBT Observation Date; 
(3) SDSS Observation Date. (4) Grism. (5) Exposure time (ET) in seconds. (6) Camera used, LUCI configuration. (7) Standard star used, obtained from our LBT observations and extracted from \cite{pickles98}; spectral type is specified in parenthesis.
\end{table*}

\section{Data analysis}\label{sec:data_an}

\subsection{Redshift estimations}\label{ssec:z_est}
We found that the narrow component of \hb\ and the \oiii\ emission are very weak in our spectra, in accordance with the results that broad \hb\ is mainly affected close to the line base in highly accreting sources \citep{negrete18}. Redshifts were measured using the wavelength of the \hb\ peak intensity, with {\tt splot} task from {\fontfamily{lmtt}\selectfont IRAF}. A posteriori cross-check towards the redshift derived from \hb\ with the ones of \mgii\ and \aliii\ was done, and a good agreement was found. The redshift estimates are reported for each object in Table \ref{tab:sample}.

\subsection{Multi-component fitting}\label{sec:multifitting}

\begin{figure*}
    \centering
    \includegraphics[width=\textwidth]{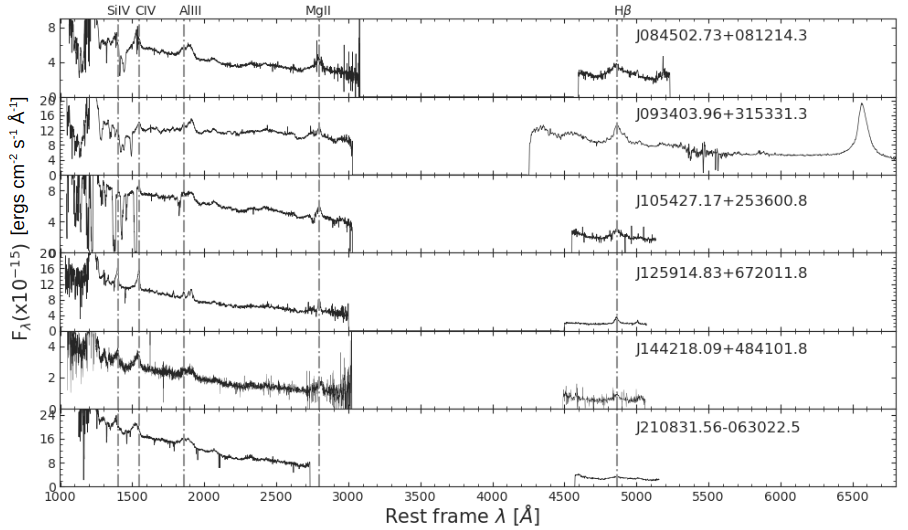}
    \caption{Rest-frame spectra of our six type 1 AGN, with UV and optical regions after joining the observed spectra from LBT and the SDSS. Abscissas are rest-frame wavelength in \AA\ and ordinates are specific flux in units of 10$^{-15}$\ergss\ cm$^{-2}$ \AA$^{-1}$. Dot dashed vertical lines trace the rest-frame wavelength, from left to right, of \siiv, \civ, \aliii, \mgii, and \hb, respectively.} 
    \label{fig:RF_specs}
\end{figure*}

To analyze the emission lines of the spectra, multi-component fits were carried out using the task {\tt specfit} \citep{kriss94}. This routine allowed us to simultaneously fit all the components present in the spectrum: continuum, \feii\ features, and emission lines, computing the \chisq\ parameter that measures the difference between the original spectra and the fitted one. The task {\tt specfit} minimizes the \chisq\ to find the best fit.

The primary continuum source in the optical-UV region is well known to be originated from the accretion disk {\citep[{e.g.,}][]{malkan82,wandel88,capellupo16}}. In the absence of extinction, the most widely-used  model for the continuum is a single power-law over a limited spectral range \citep[see e.g.,][]{sniegowskaetal20}. In Fig. \ref{fig:RF_specs} are shown the quasar spectra with the coverage from the LBT and SDSS observations. We fitted a local continuum for the spectral ranges centered on the most important emission lines shown in Fig. \ref{fig:RF_specs}: 1400\AA\ blend, \civ, 1900\AA\ blend, \mgii, and \hb, using a single power-law for each range centered in 1350, 1700, 3050, and 5100\AA, respectively.

\subsubsection{Region 1: 1250-1450\AA.} \label{sssec:siiv_fit}  

The dominant emission in this region involves \siiv+\oiv\ (the 1400 \AA\ blend), and is accompanied by weaker Si{\sc ii}$\lambda$1263, \siiiuuv, \oi, and \cii.

\paragraph*{\siiv+\oiv\ -- } The underlying assumption of the blend modeling is that the broad component (BC) emission is dominated by \siiv\ due to collisional deexcitation of the inter combination \oiv\ multiplet \citep{wills&netzer79}, while the blue shifted component is due to an inextricable contribution of both \oiv + \siiv. The continuum is fitted with a single power-law in the region 1450-1700\AA, using a continuum window at 1700 \AA\ as defined by \cite{francis91}.
The \siiv+\oiv\  feature is a high-ionization blend, and shows a blueshifted, asymmetric profile not unlike \civonly. The broad component was modeled with the same emission components of \civonly, whenever possible. If the blue side of the 1400\AA\ blend was strongly contaminated by absorption features (J0934 and J1054)  only the line flux was free to vary but the blueshifted \siiv+\oiv\ emission was modeled still taking  the parameters of the asymmetry, width, and profile of \civonly\ Blue component. {The  S/N between 30 and 40  always allows for the deblending  of the broad and blue components when the blend is covered by the SDSS spectra}.  

\subsubsection{Region 2: 1450-1700\AA.} \label{sssec:civ_fit}

The \civ\ emission line dominates this region and is accompanied by \heii, \oiiiuv and \alii. We followed the method of \cite{MAetal18}.

\paragraph*{\civ\ --}  The BC of \civonly\ is modeled by a Lorentzian profile fixed at the rest-frame. The flux of the \civonly\ BC is free to vary and FWHM is assumed to be the same or larger as \aliii\ and \siiii. All the \civonly\ profiles in our sample show a blueshift and or blueward asymmetry. In order to model it with {\tt specfit}, we used one or two blueshifted skewed Gaussian profiles. The flux, FWHM, asymmetry, and shift were unconstrained. 

\paragraph*{\heiiuv --} We assumed a profile similar to the one of \civ: Lorentzian and skewed Gaussian profiles for the BC and blueshifted components, respectively. The FWHM, shift and asymmetry were assumed equal to those of \civ, but the flux varies freely. 

\paragraph*{\oiiiuv,\alii\ --}  Components modeled with unshifted Lorentzian profiles with their fluxes and FWHM varying freely.

\subsubsection{Region 3: 1650-2150\AA.} \label{sssec:1900_fit}

In the 1900\AA\ blend, the most prominent emission lines are: \aliii, \siiii\ and \ciii. We followed the considerations made by \cite{buendia2023}. 

\paragraph*{\feiii\ and \feii\ --} Emission of the \feiii\ multiplets can be strong in the vicinity of \ciii, as seen in the average quasar spectrum from \citet{vanden01}. They appear to be strong when \al\ is also strong \citep{HB86,MAetal18,mediavillaetal19,templeetal20}. We adopted the \feiii\ template model obtained by \cite{vestergaard01}.
The {\tt specfit} task scaled and broadened the template to reproduce the observed emission \citep{BG92}. We fitted the multiplet \feUV\ (seen in the blueward of the 1900\AA\ blend, \citealt{moore45}) as an isolated Gaussian in the rest-frame. An extra component is added: \feiii $\lambda1914$ to fit an excess seen near the red wing of \ciii\ associated with unresolved \feiii\ template emission \citep{negrete12}. This \feiii$\lambda$1914 emission is a fitted using a Lorentzian profile to be consistent with the profile of the BCs \citep{negrete12,negrete13}. This criterion rests on the assumption that \feiii $\lambda$1914 and the \feii\ UV multiplet \#191 are enhanced by \lya\ fluorescence \citep{sigutpradhan98,johansson00}.

\paragraph*{\ciii\ --}  The BC is modeled by a Lorentzian profile fixed at the rest-frame, same as \siiii\ and \aliii. Strengths and FWHM were left free to vary in the \texttt{specfit} model with one restriction: FWHM(\ciii) $\leq$ FWHM(\aliii) or FWHM(\siiii).

\paragraph*{ \siiii, \aliii\ --}  Strengths and FWHM were free to vary, with one restriction: FWHM(\siiii) $\geq$ FWHM(\ciii). The \aliii\ doublet was resolved and the blue component shifts, FWHM, and intensity were allowed to vary, with the red one tied to the blue by identical FWHM and fixed wavelength ratio. The ratio between the intensity of the red and blue component of the doublet was kept fixed 0.8 \citep{laor97}. Doublet total strengths were left free to vary. We also added a blueshifted skewed Gaussian profile. The flux, FWHM, asymmetry and shift were unconstrained.

\subsubsection{Region 4: 2600-3050\AA.} \label{sssec:mgii_fit}  
    
The \mgii\ emission line domains in this region and is accompanied by \aliiMg, \oiiiMg, and \feii\ emission. We followed the method of \cite{marzianietal13a}.

\paragraph*{\feii\ and \feI --} \cite{bruhweiler&verner08} provides \feii\ UV emission templates computed from cloudy simulations and using an 830 level model of the  Fe$^+$ ion. The \cite{bruhweiler&verner08} template results in a systematic flux excess near 2950\AA. The excess of flux observed was fitted with \feI\ UV emission at $\lambda\lambda$2937.5,2970.5. \feI\ emission has been predicted by photoionization models \citep{sigutetal04} and was suggested by previous observations \citep[e.g.][]{kwanetal95,marzianietal13a}.
\paragraph*{\mgii\ --} The doublet was resolved and the blue component shifts, FWHM, and intensity were allowed to vary, with the shorter wavelength (blue) one tied to the longer (red) one by identical FWHM and fixed wavelength ratio. Lorentzian profiles at rest-frame were employed for  both components. The ratio between the intensity of the red and blue component of the doublet was kept fixed at 0.8 \citep{marzianietal13a}, and in some cases it was assumed to be equal \citep{laor97,vestergaard01}. The doublet total strengths were left free to vary. We also added a blueshifted skewed Gaussian profile. The flux, FWHM, asymmetry, and shift were unconstrained.
\paragraph*{\aliiMg, \oiiiMg} Components modeled with unshifted Lorentzian profiles with their fluxes and FWHM varying freely.

\subsubsection{Region 5: 4500-5200\AA.} \label{sssec:Hb_fit}   

The dominant emission in this region is \hb\ accompanied by \oiiiopt\ and \feii. The method we used was the one of \cite{negrete18}.
\paragraph*{\hb}: The BC is modeled by a Lorentzian profile fixed at rest-frame. We also added a blueshifted skewed Gaussian profile. The flux, FWHM, asymmetry and shift were unconstrained. 

\paragraph*{\feii}: We used the semi-empirical template by \cite{marzianietal09}, obtained from a high-resolution spectrum of I Zw 1, with a model of the \feiiopt\ emission computed by a photoionization code in the range of \hb.

\paragraph*{\oiii$\lambda\lambda$4959,5007}: As for the narrow component (NC), we fitted this doublet with two Gaussians, considering the ratio of theoretical intensities of 1:3 \citep{dimitrijevicetal07}, the same FWHM, and the same line shift. We added a second blueshifted semi broad component if necessary to the fit.  We have cases with no detectable \oiii$\lambda\lambda$4959,5007 NC or semi-broad emission. 

\subsubsection{Error estimates}

Data used in this analysis come from two different  instruments yielding spectra with widely different S/N. In addition, the comparison is between emission lines, which are relatively strong (W(\hb)$>$50\AA) and ones of which are fainter in the UV (W(\al), W(\siivonly), W(\heiionly)$\leq$10\AA\ in most cases). So, we used the quality parameter $\mathcal{Q}$ \citep{marziani22}. {The uncertainties of parameter $\mathcal{Q}$ were estimated following a Bayesian approach with a likelihood function (their Eq. A1) dependent on the specific flux (as a function of wavelength), the S/N over the spectra and the expectation values of the multi-component model using \texttt{specfit}. For a detailed description of the prior and posteriori model parameters in the $\mathcal{Q}$ parameter see Appendix A of \citep{marziani22}}.
This parameter is defined as the ratio between the line equivalent width and its FWHM multiplied by the S/N measured on the local continuum fitted for each region. By multiplying the S/N by the ratio W/FWHM, a computation of the S/N is more appropriate for a line depending on its strength and width. The parameter $\mathcal{Q}$ is larger for sharp lines in spectra with high S/N in the continuum. Especially for large $\mathcal{Q}$ values, the scatter is relatively modest,
and the relation between the parameter FWHM, ﬂux, and shift
and log $\mathcal{Q}$ can be written in a linear form, save for the fractional uncertainty of FWHM(\al) that is best ﬁt by $\delta$FWHM/FWHM $\approx$ 1 / (\textit{a} + \textit{b} log $\mathcal{Q}$). Table 3 of \citet[][Appendix A]{marziani22}  provides the coefficients $a$ and $b$ of the best fits along with the $\mathcal{Q}$ domain. 

\section{Results}\label{sec:results}

The results are presented starting from multi-component fits for each  spectral region (Section \ref{res_inmed}),  and  measurements reported for each object, accompanied by a detailed description of the most significant emissions from each region, including the blue component detected in several lines. Section \ref{profiles} provides the measurements of the line profile parameters for the most prominent lines considered in the analysis In Section \ref{consis_optUV}, we analyze the consistency of selection criteria in the optical and UV for the LBT sample. 



\subsection{Immediate Results}\label{res_inmed}

\begin{figure*}
  \sbox0{\begin{tabular}{@{}cc@{}}
    \includegraphics[scale=0.2]{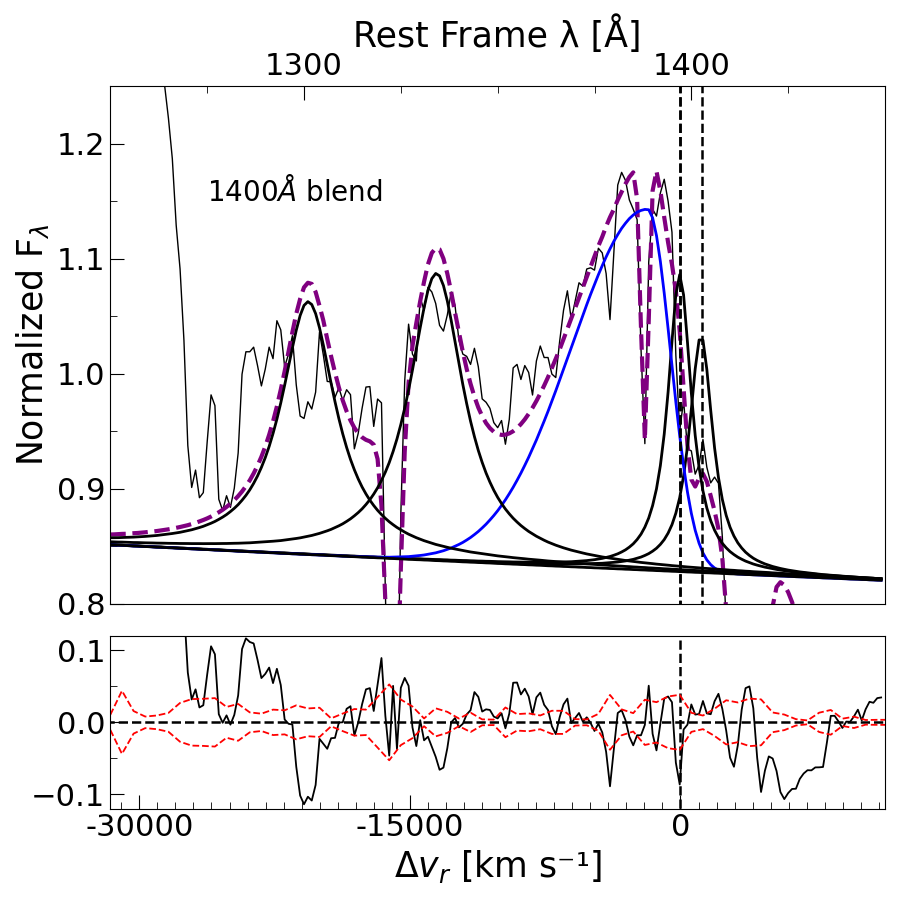}
    \includegraphics[scale=0.2]{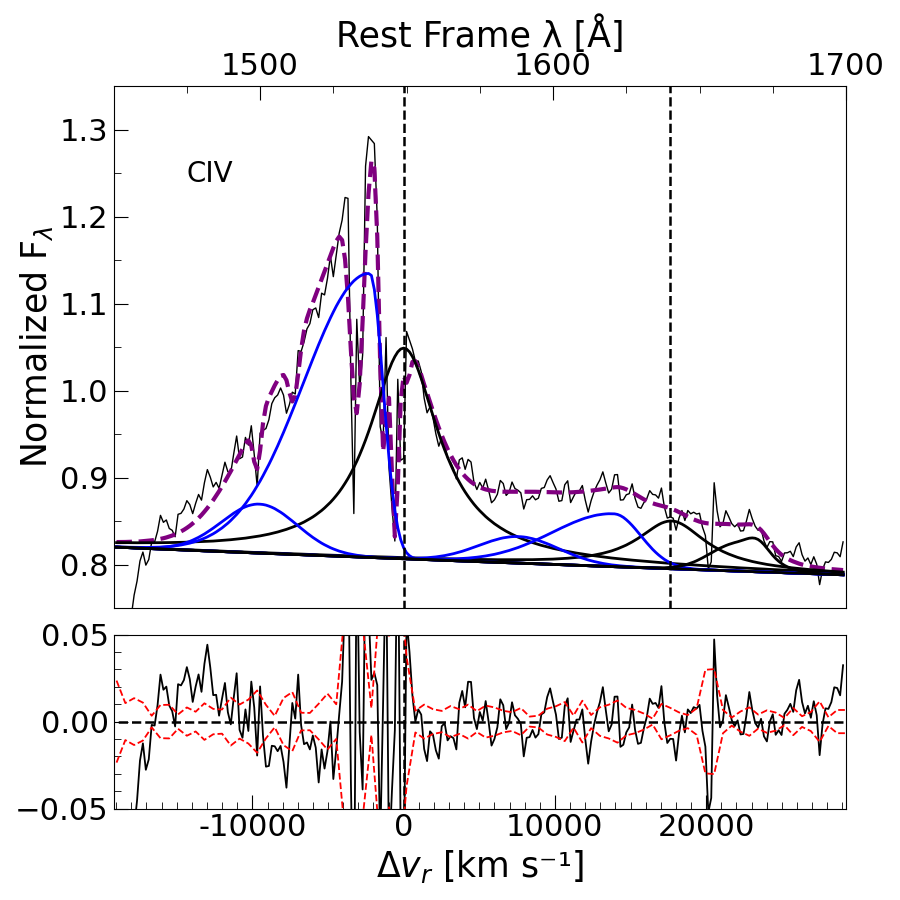}
    \includegraphics[scale=0.2]{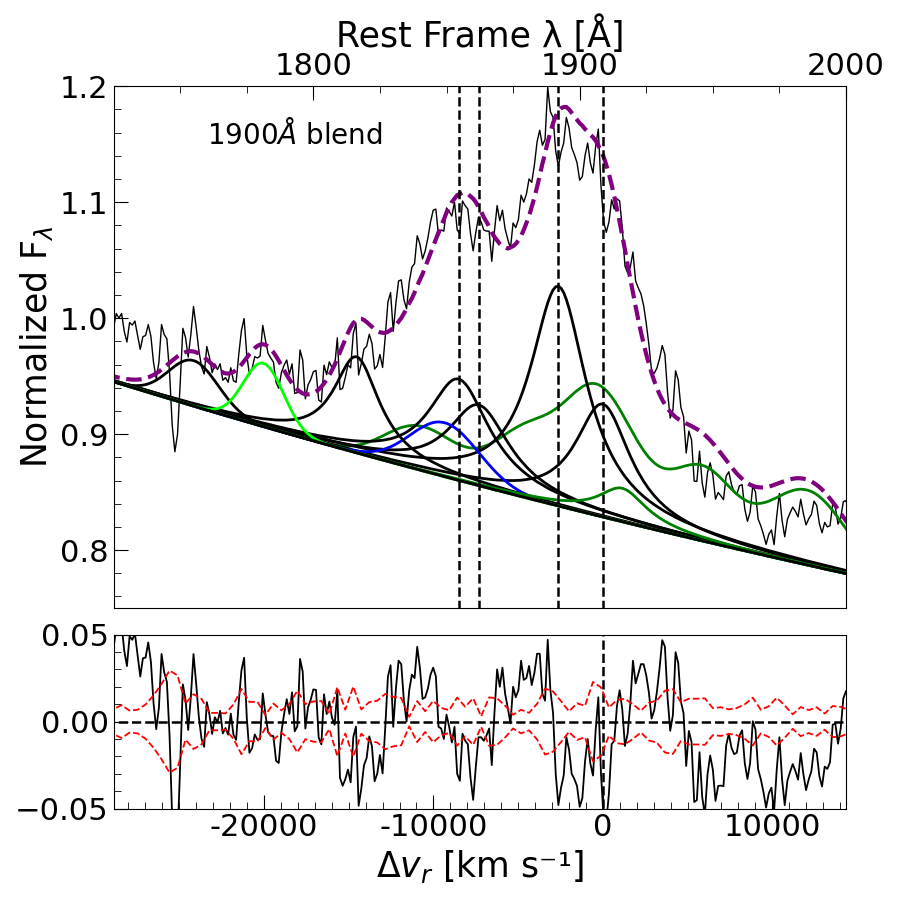}
    \includegraphics[scale=0.2]{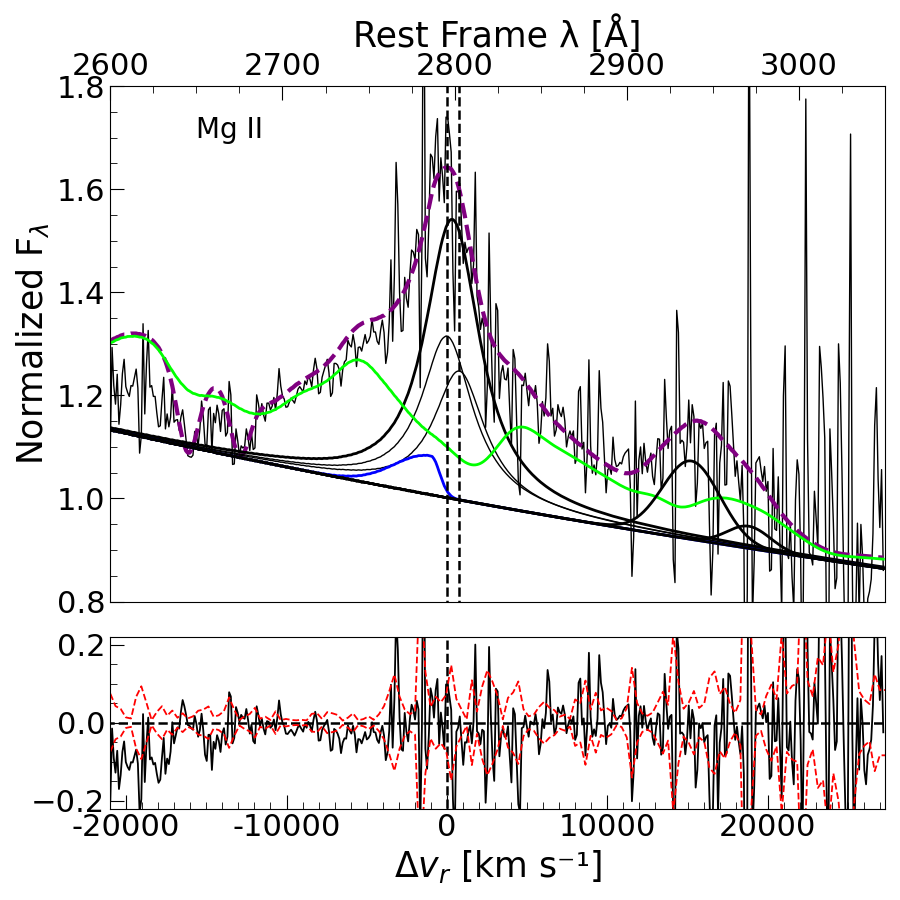}
    \includegraphics[scale=0.2]{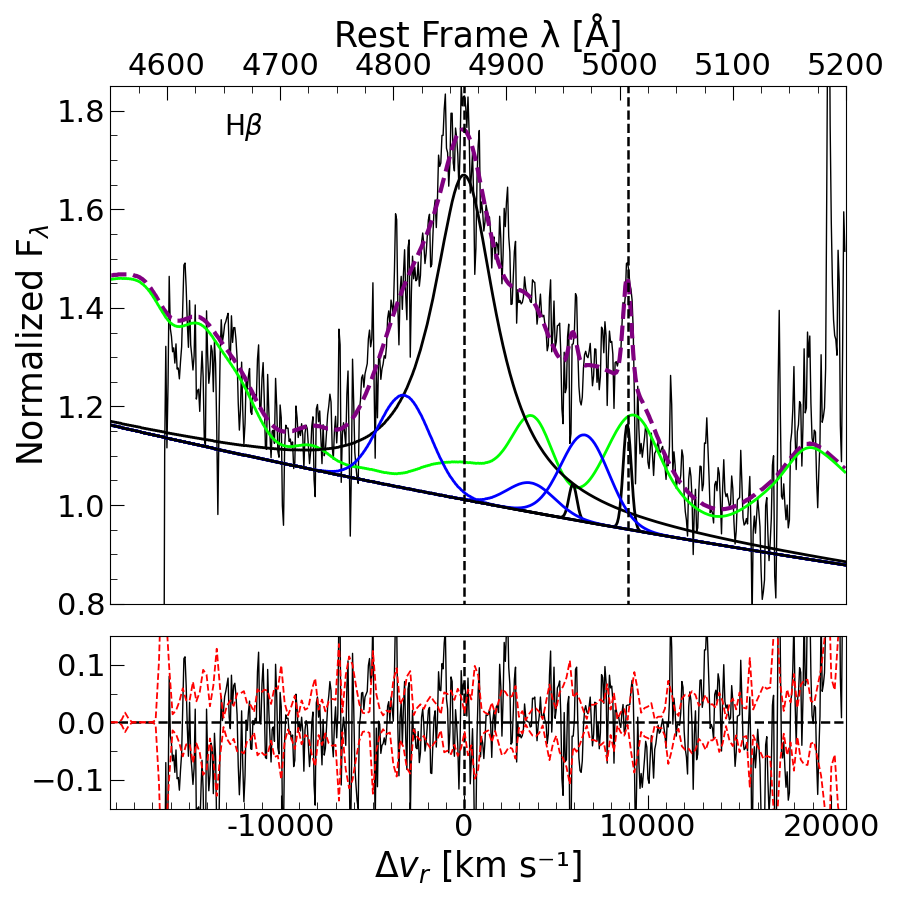} \\
    \includegraphics[scale=0.145]{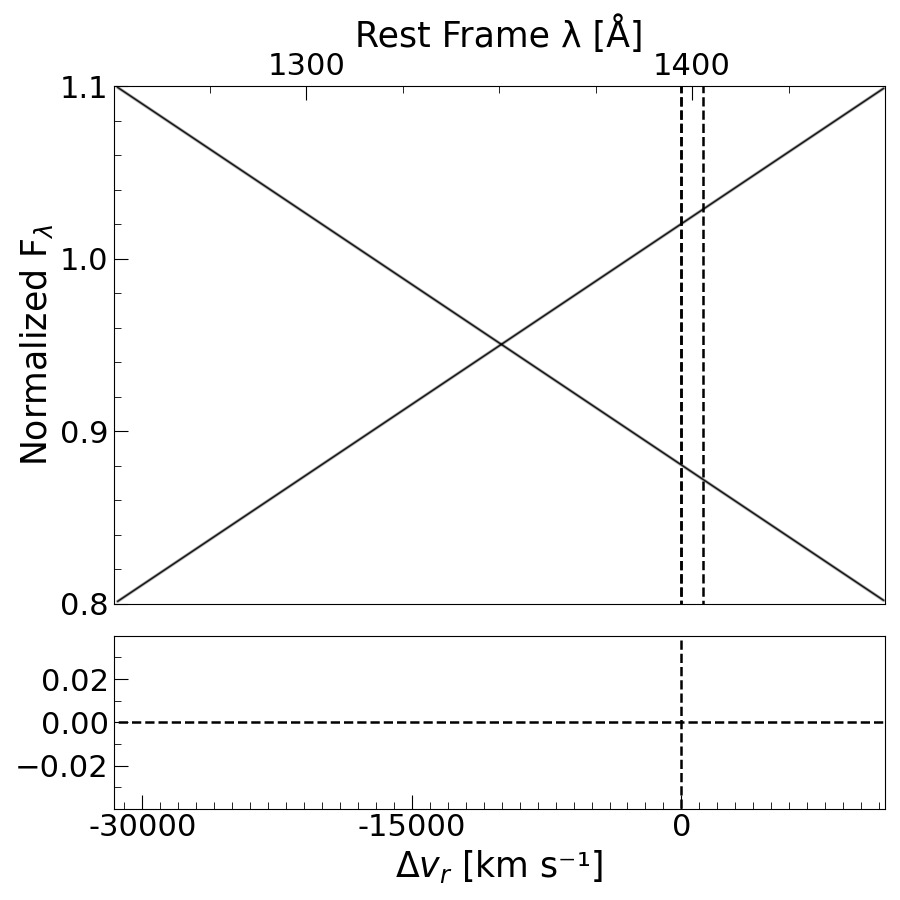} 
    \includegraphics[scale=0.2]{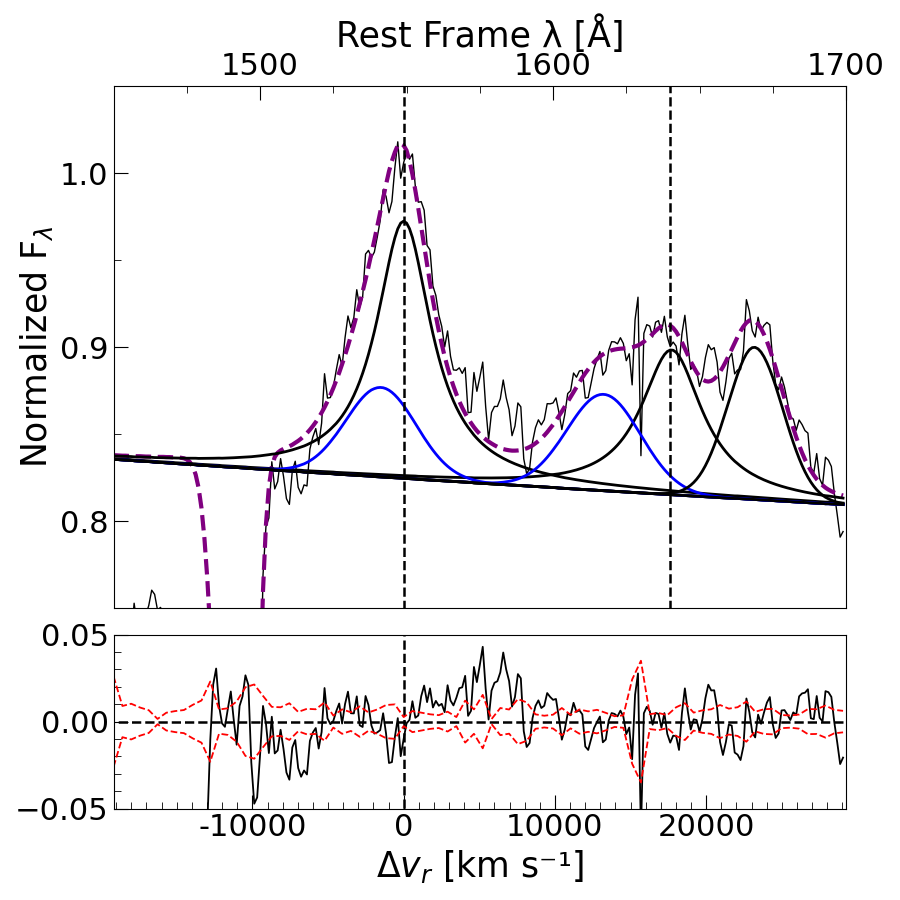}
    \includegraphics[scale=0.2]{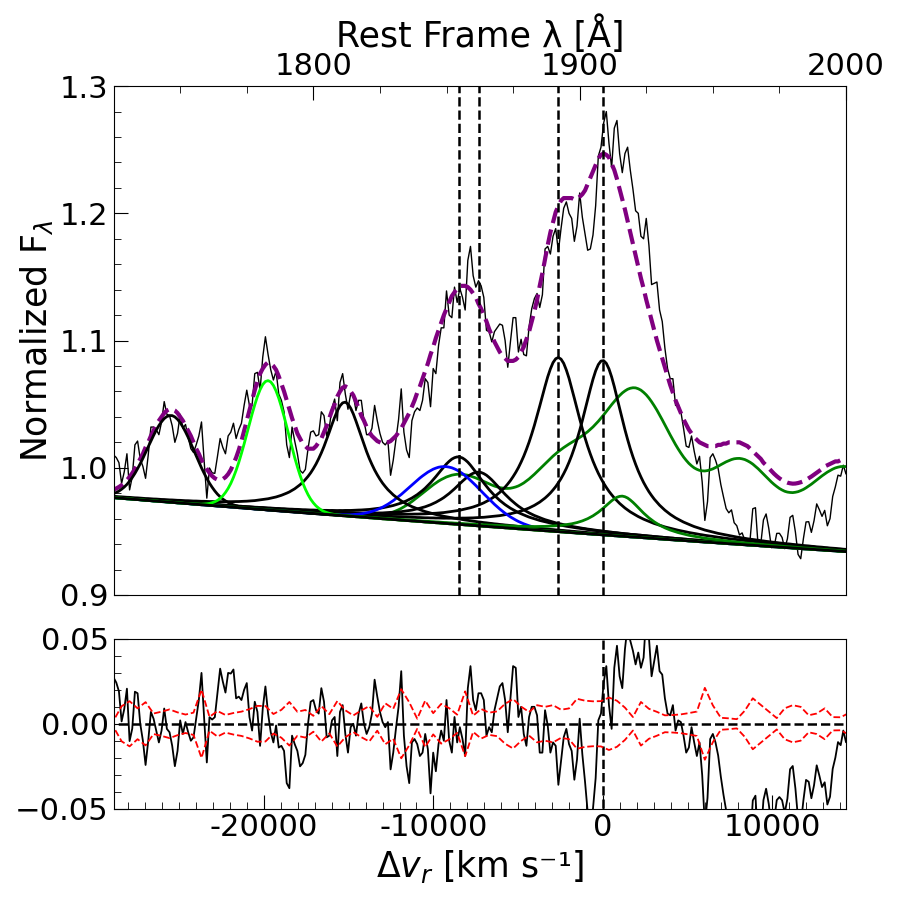}
    \includegraphics[scale=0.2]{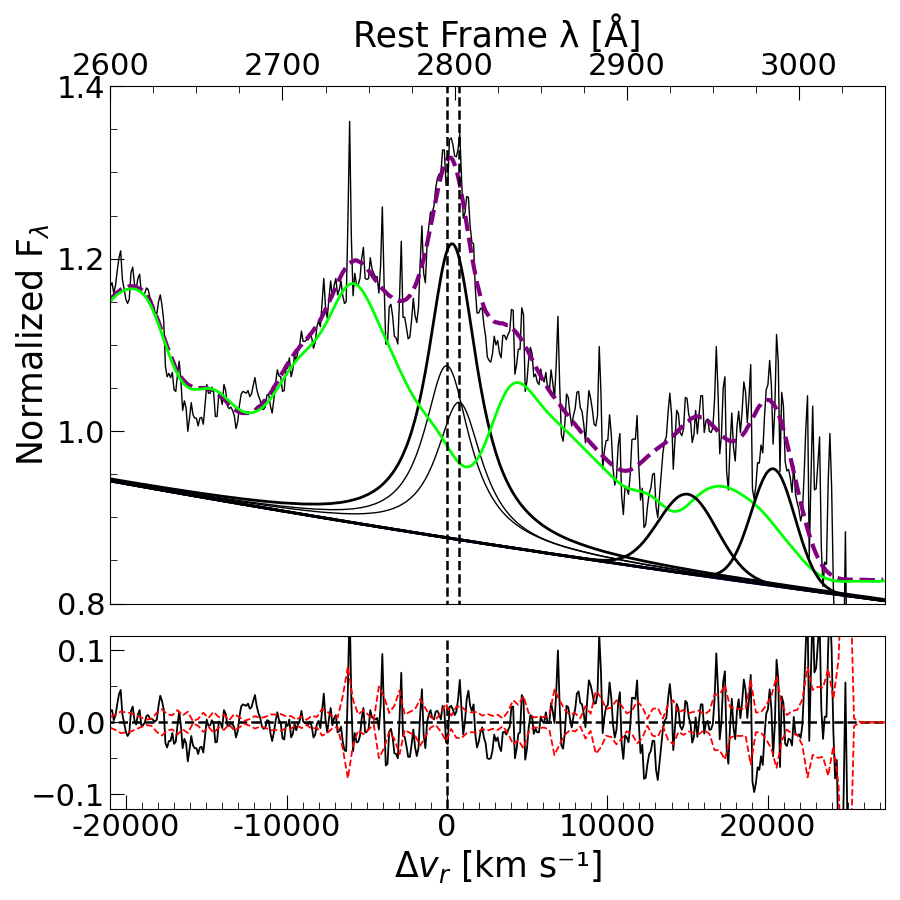}
    \includegraphics[scale=0.2]{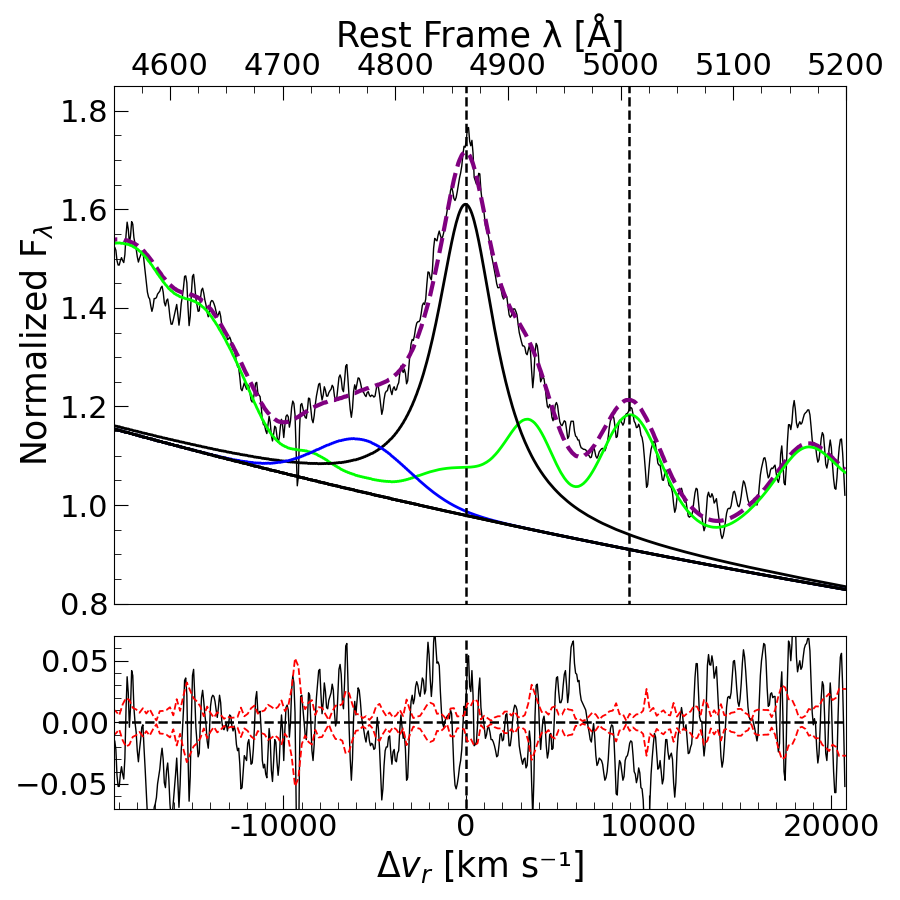} \\
    \includegraphics[scale=0.145]{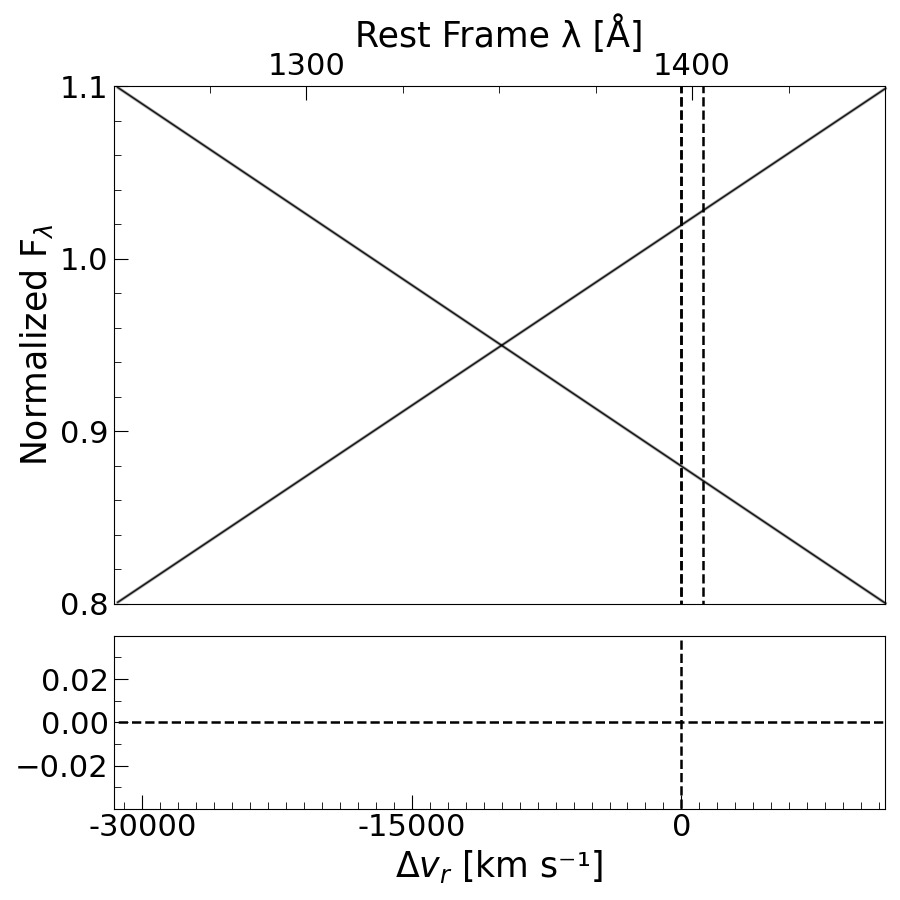}
    \includegraphics[scale=0.175]{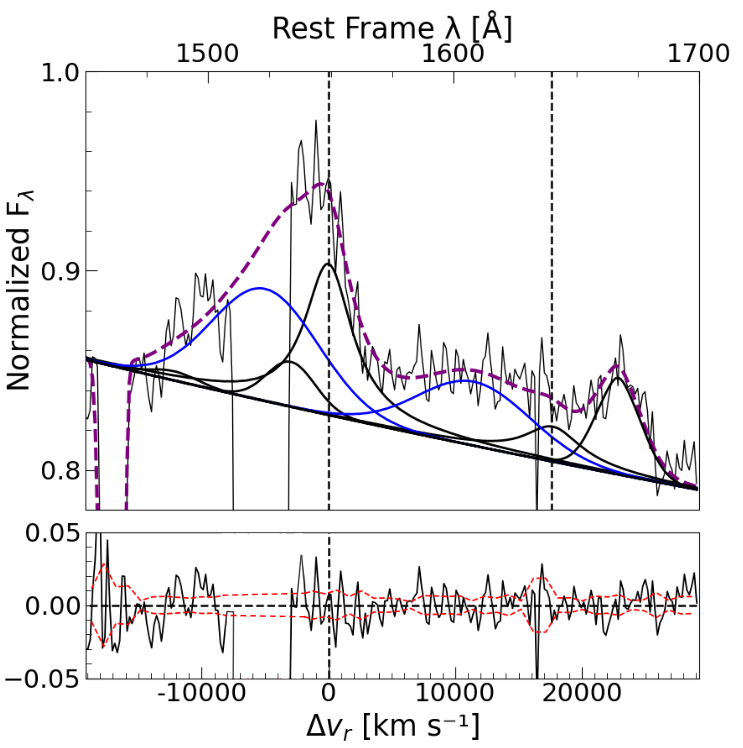} 
    \includegraphics[scale=0.2]{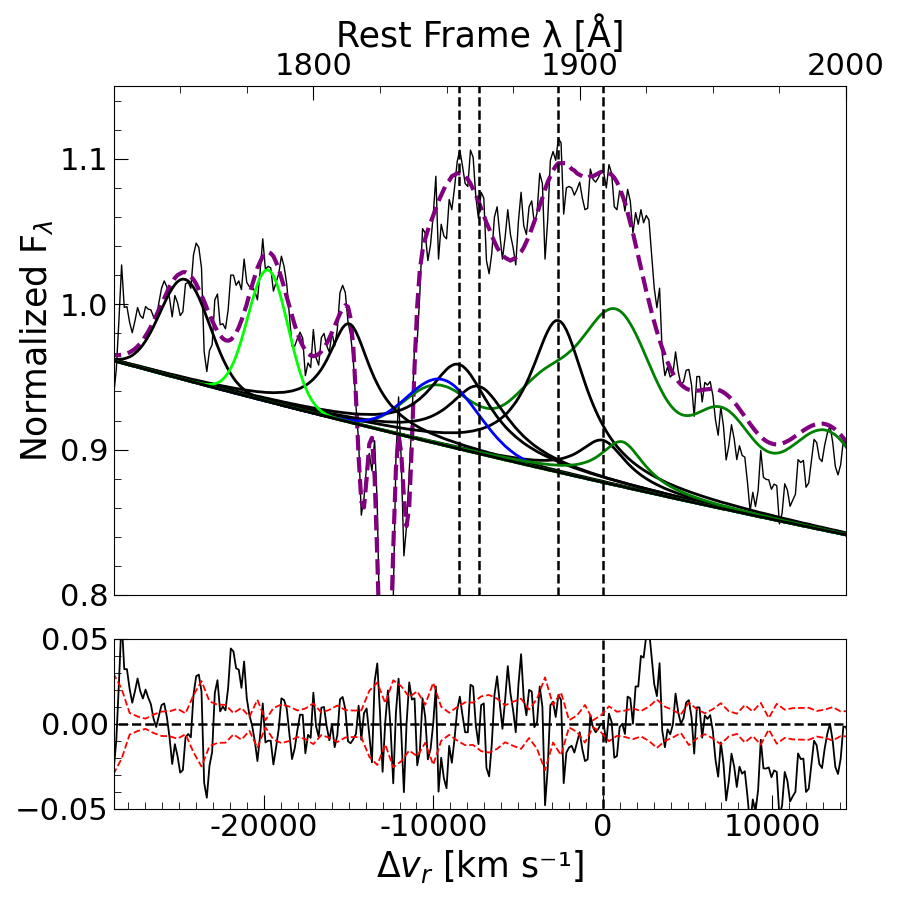}
    \includegraphics[scale=0.2]{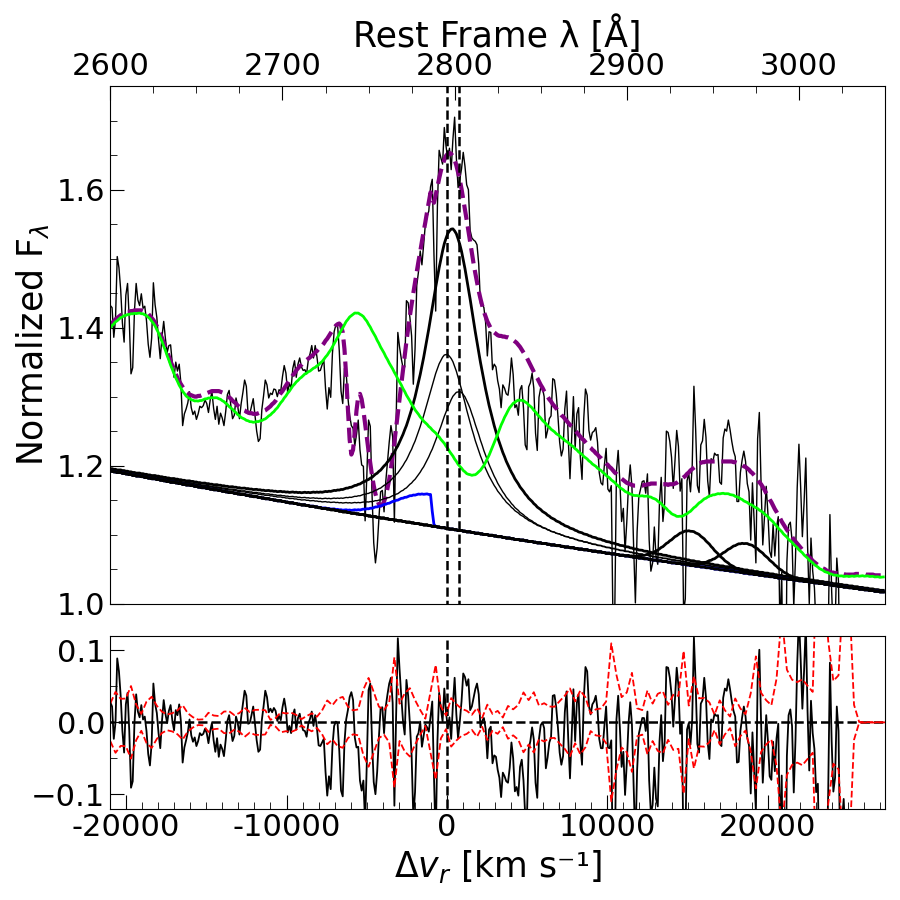}
    \includegraphics[scale=0.2]{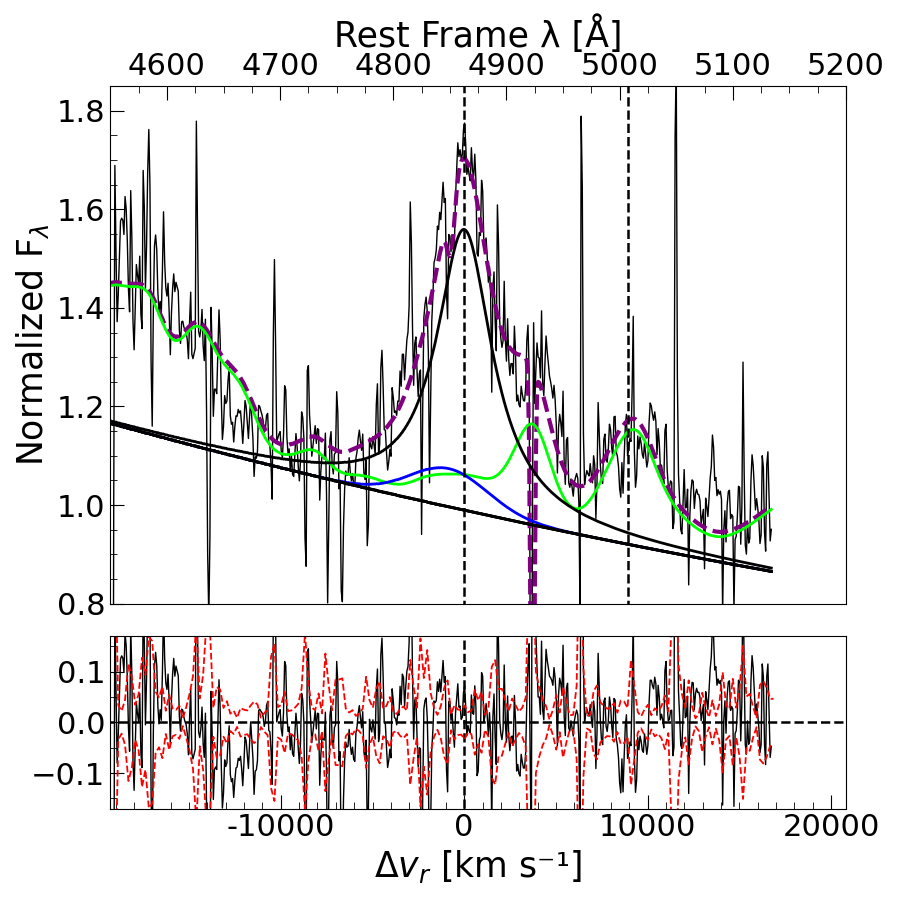}
  \end{tabular}}
    \rotatebox{90}{\begin{minipage}[c][\textwidth][c]{\wd0}
    \usebox0
    \caption{Multi-component fits, from left to right, of the 1400\AA\ blend, \civ, 1900\AA\ blend, \mgii, and \hb\ spectral regions as described in Sect. \ref{sec:multifitting}. From top to bottom SDSS quasars: J084502.73+081214.3, J093403.96+315331.3, and J105427.17+253600.8. The top abscissa scale is rest-frame wavelength in \AA. The ordinate scale is the normalized flux.
    In all the panels, a continuous black line marks the broad component at rest-frame, while the blue one corresponds to the blueshifted component associated with each emission. Dot-dashed vertical lines identify the position at rest-frame of the strongest emission lines. Dashed purple line marks the fit model obtained by {\tt specfit}. The \feiii\ and \feii\ contributions are traced by dark and pale green lines respectively. { The lower panels show  the residuals between spectrum and {\tt specfit} model (black line) and the noise at $1 \sigma$-level (see text for more details). The abscissa} is in radial velocity units (\kms). \label{fig:spec_fitsp1}
}
  \end{minipage}}
\end{figure*}   

\begin{figure*}
  \sbox0{\begin{tabular}{@{}cc@{}}
    \includegraphics[scale=0.2]{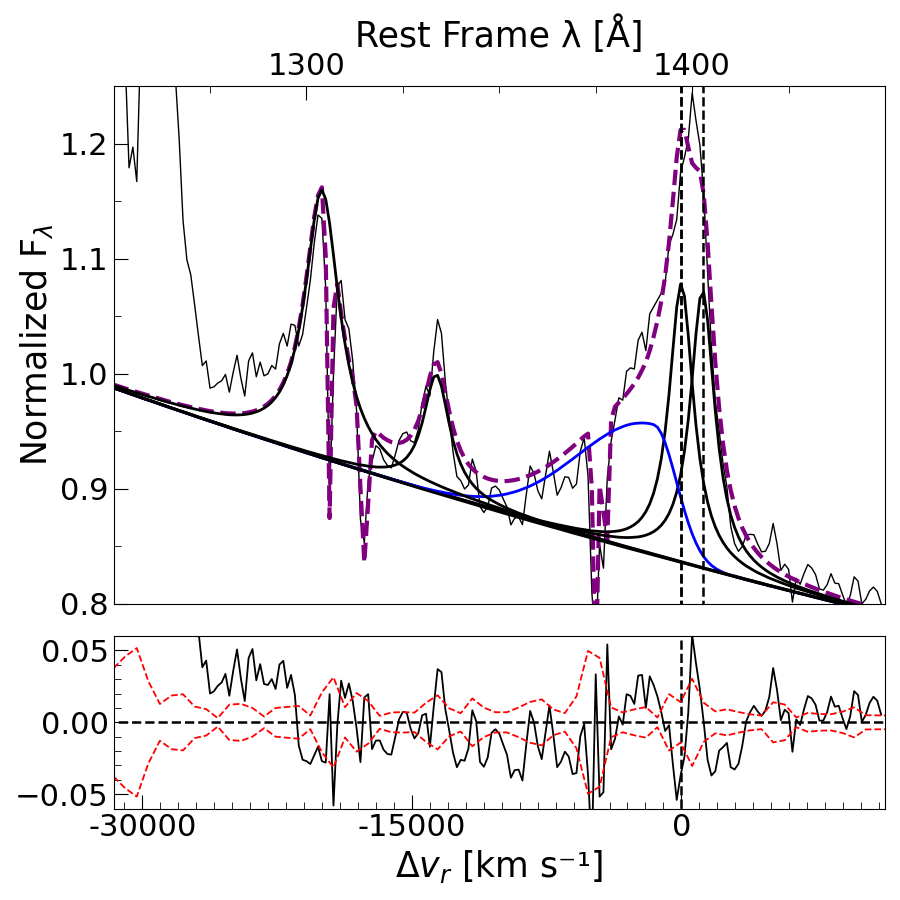}
    \includegraphics[scale=0.2]{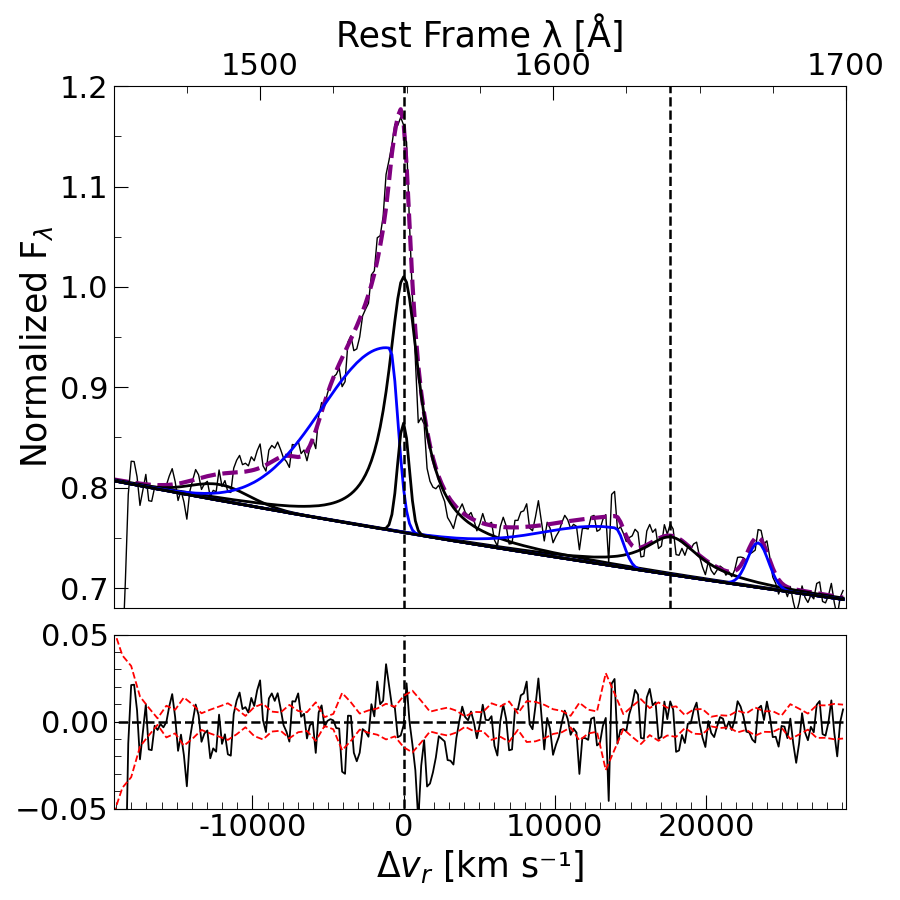}
    \includegraphics[scale=0.2]{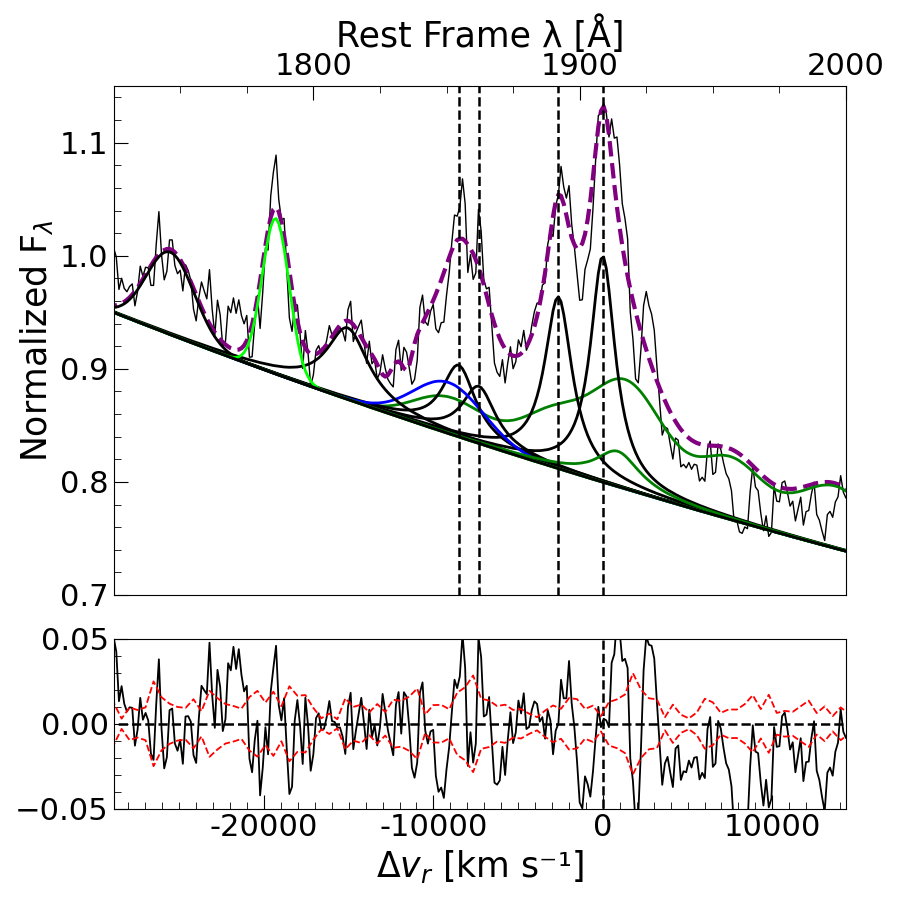}
    \includegraphics[scale=0.2]{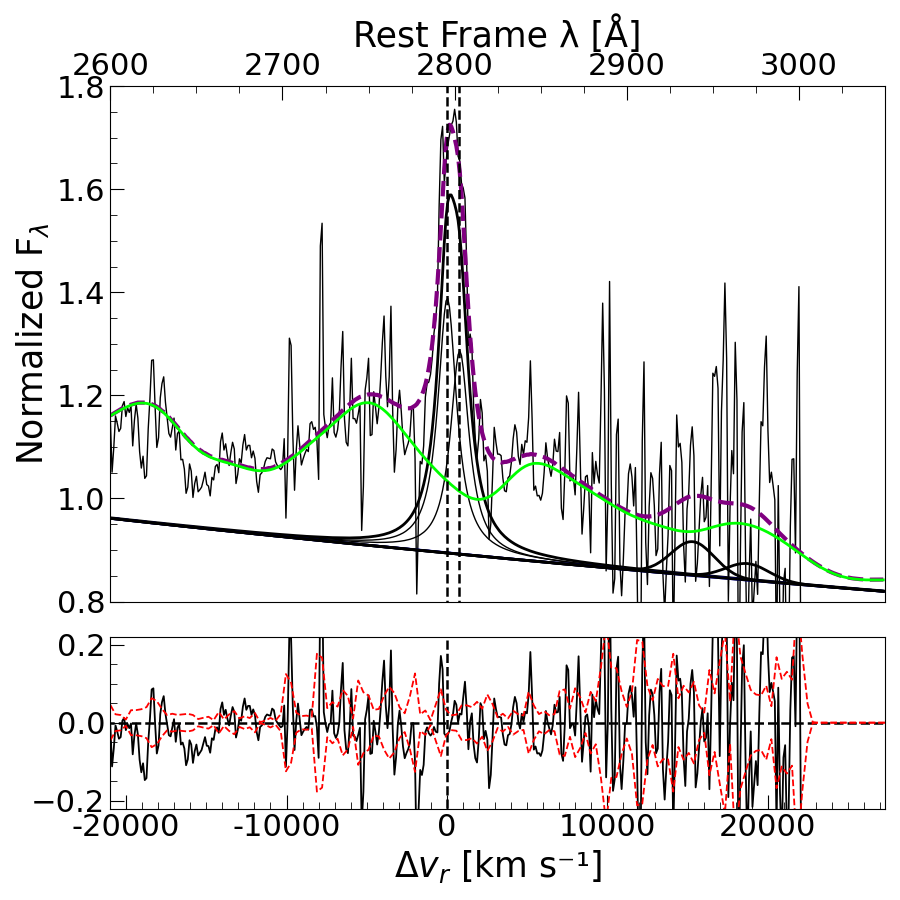} 
    \includegraphics[scale=0.2]{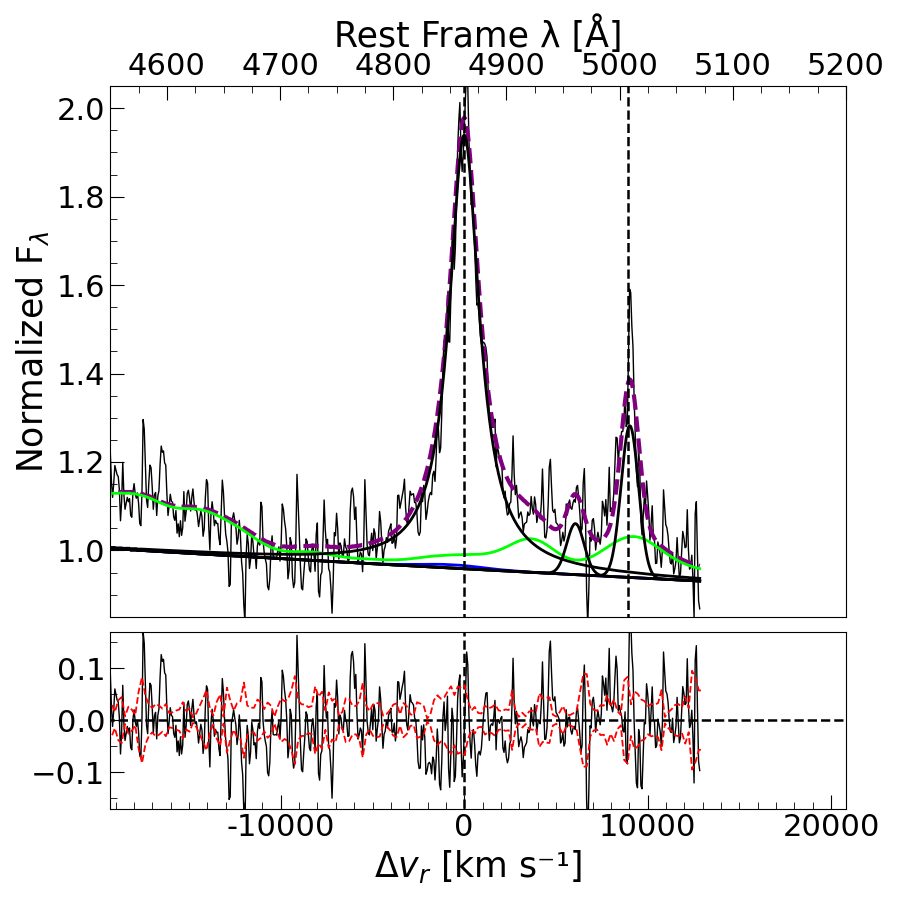} \\
    \includegraphics[scale=0.2]{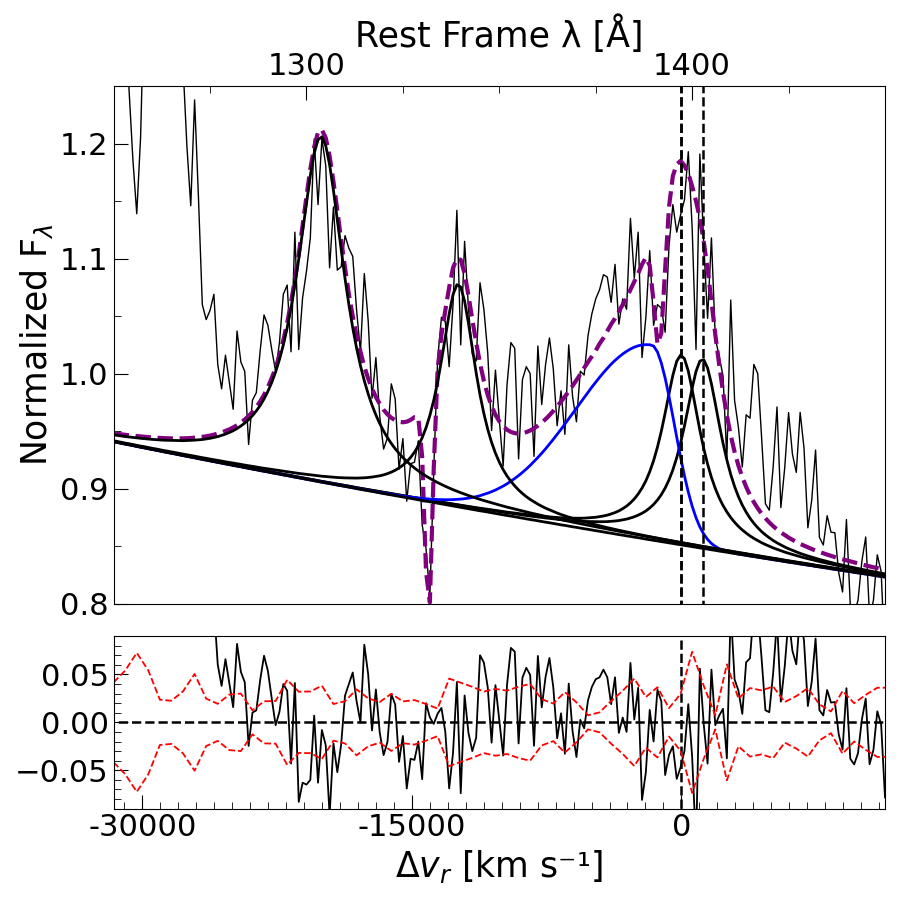}
    \includegraphics[scale=0.2]{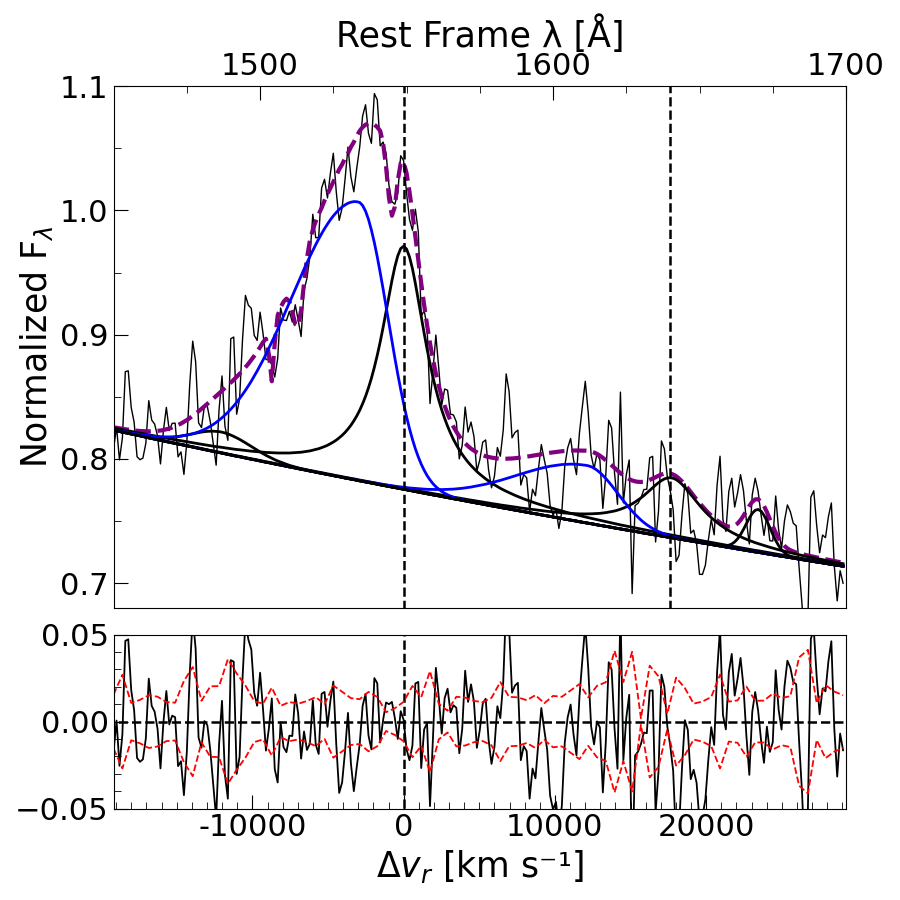}
    \includegraphics[scale=0.2]{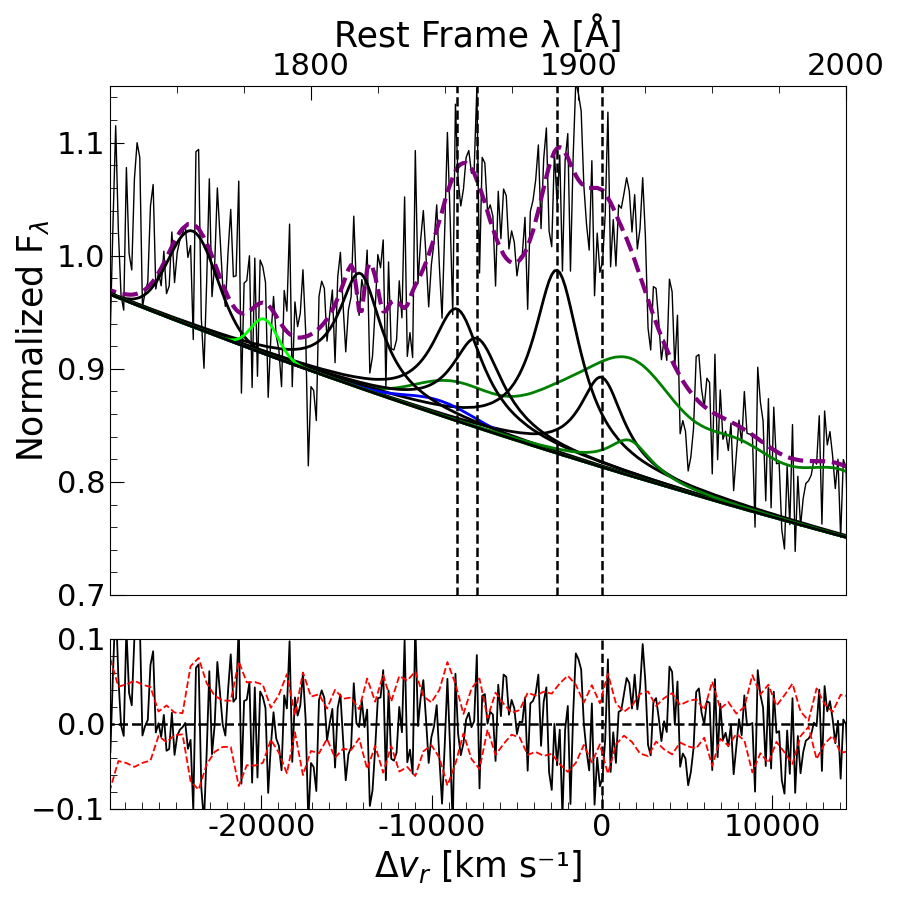}
    \includegraphics[scale=0.2]{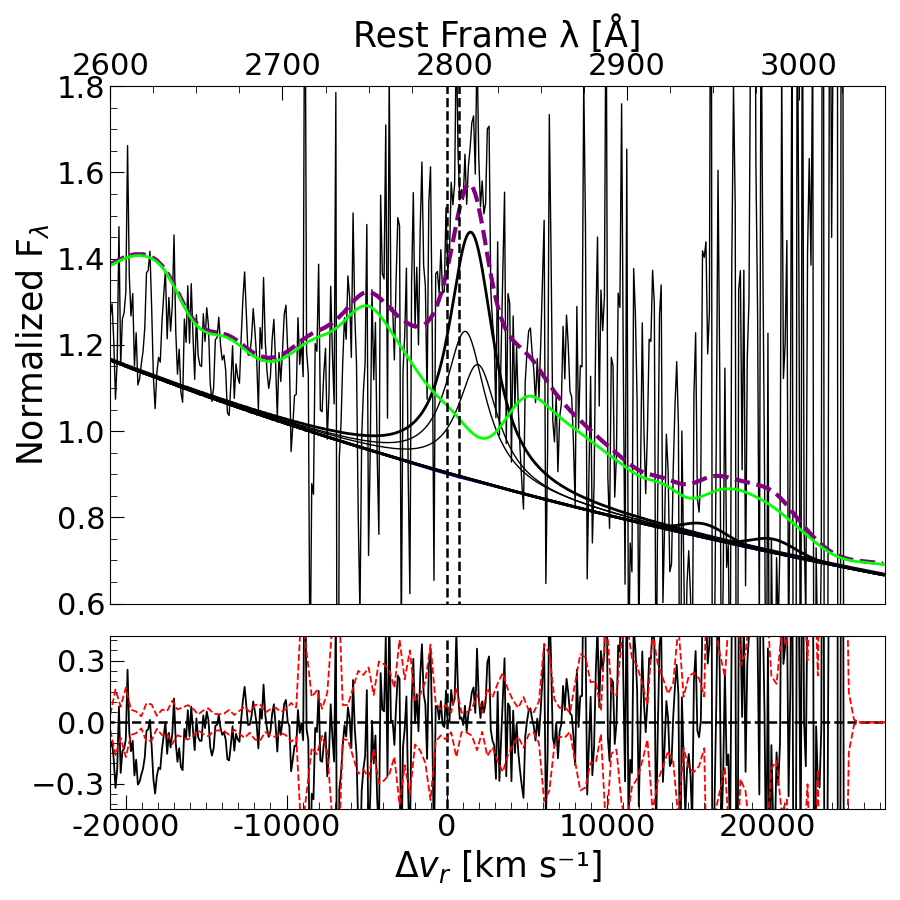}
    \includegraphics[scale=0.2]{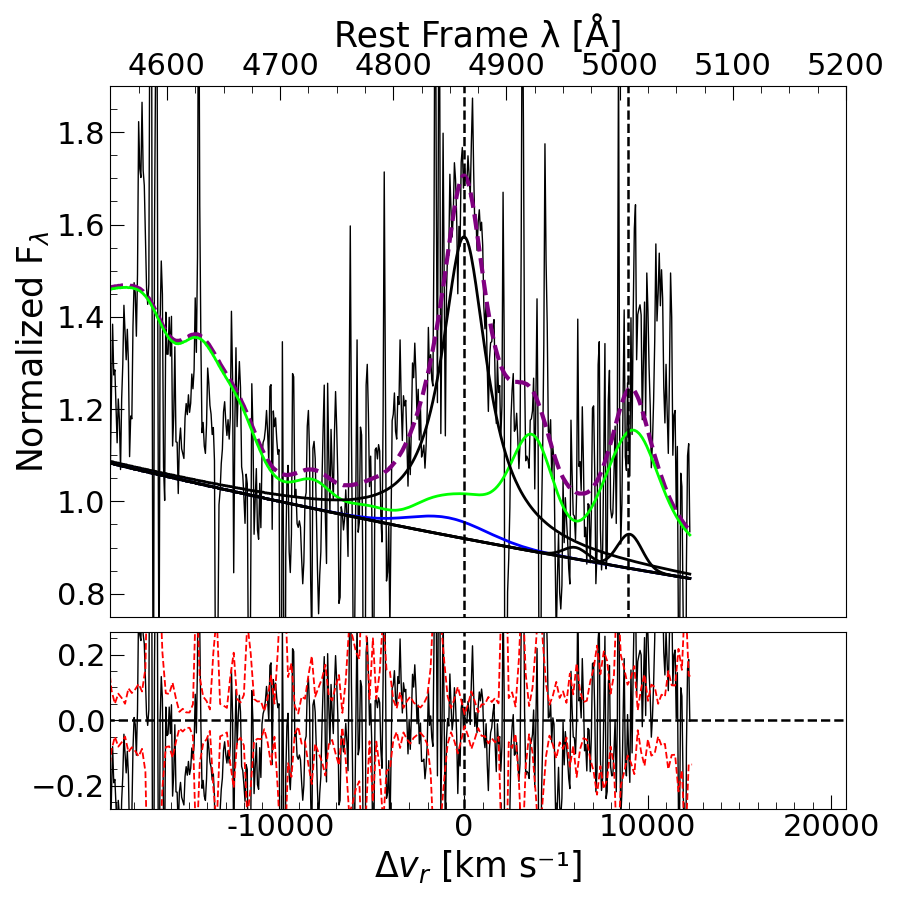} \\
    \includegraphics[scale=0.2]{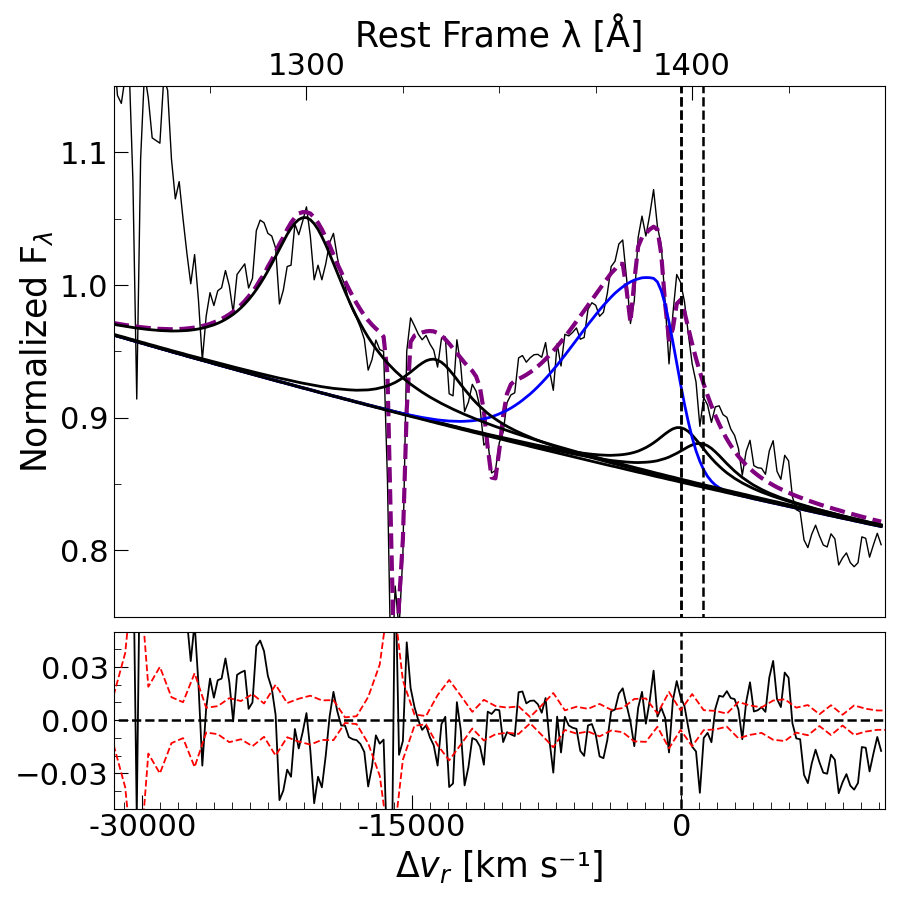} 
    \includegraphics[scale=0.2]{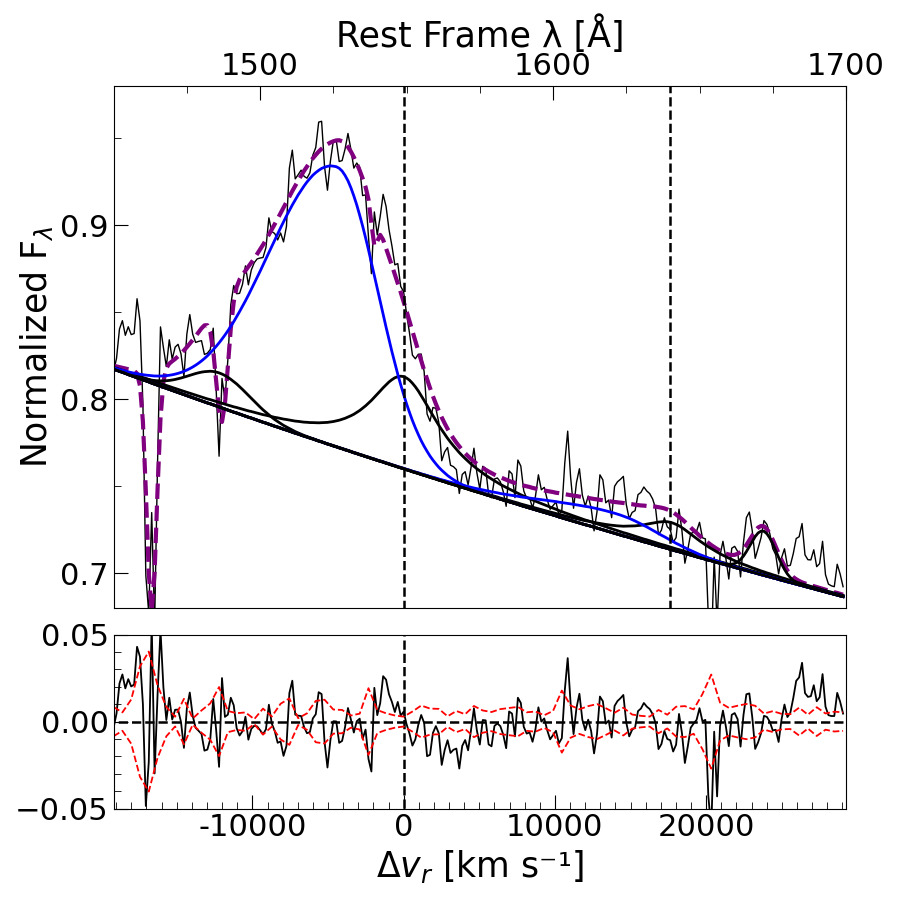} 
    \includegraphics[scale=0.2]{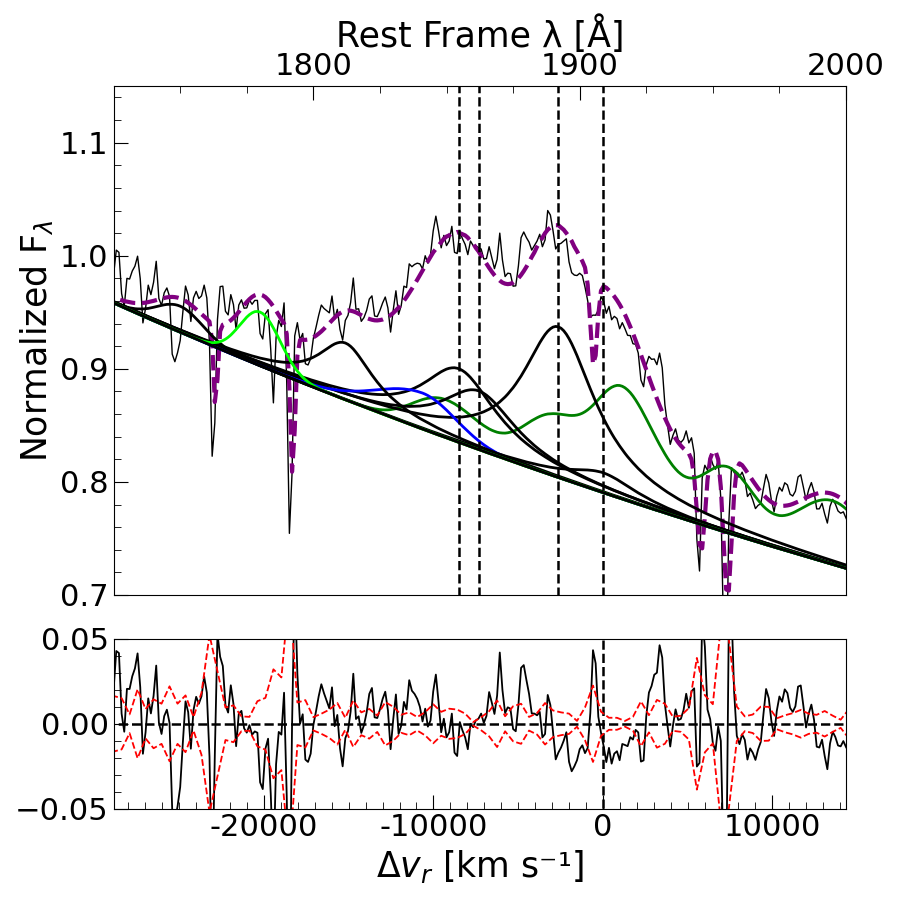}
    \includegraphics[scale=0.145]{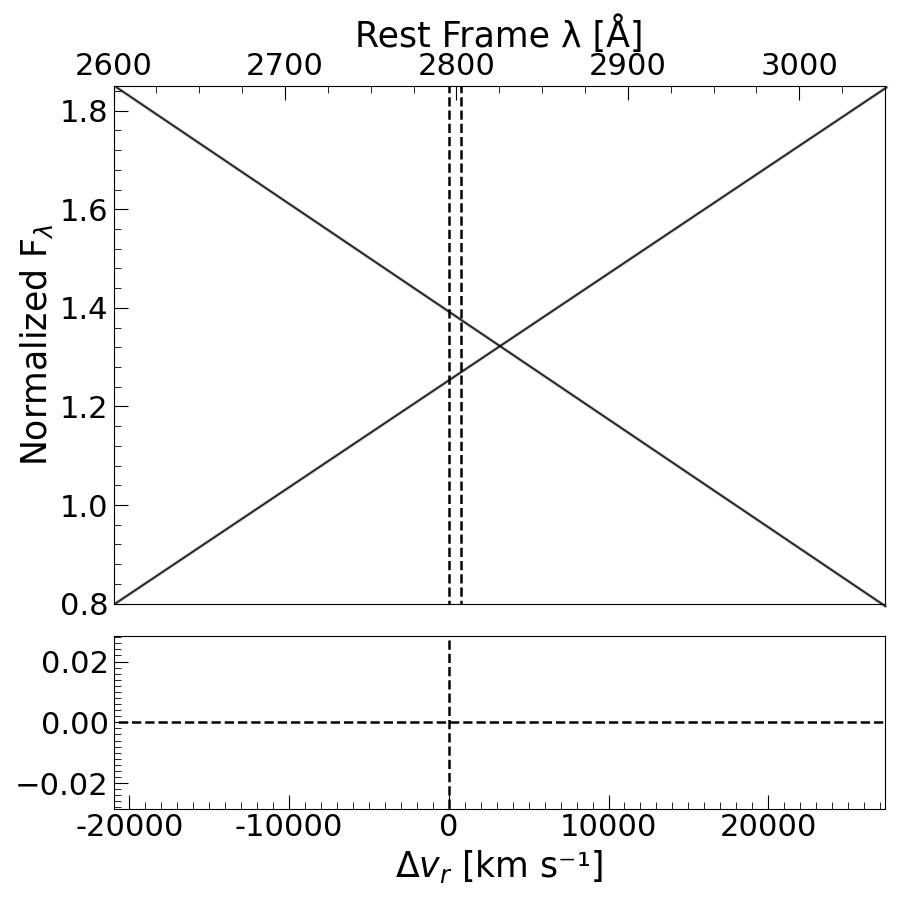}
    \includegraphics[scale=0.2]{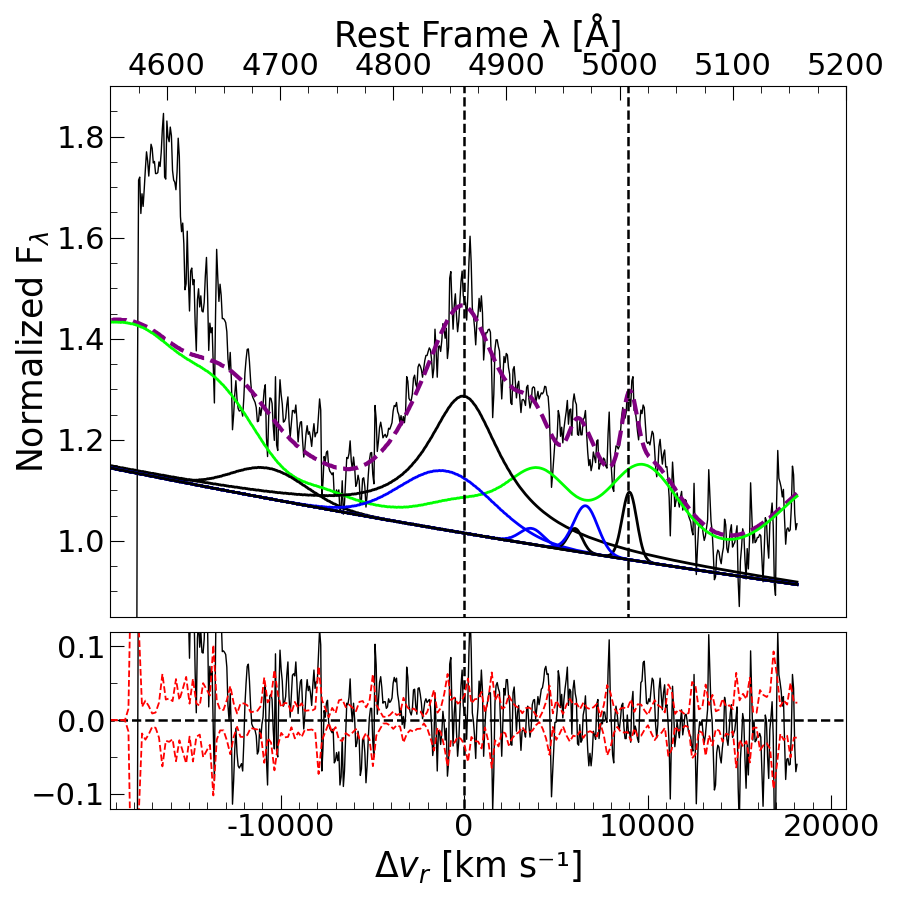} 
  \end{tabular}}
    \rotatebox{90}{\begin{minipage}[c][\textwidth][c]{\wd0}
    \usebox0
    \caption{continued. From top to bottom SDSS quasars: J125914.83+672011.8, J144218.09+484101.8, and J210831.56-063022.5.\label{fig:spec_fitsp2}
} 
  \end{minipage}}
\end{figure*}   

Figures \ref{fig:spec_fitsp1} and \ref{fig:spec_fitsp2} show the resultant fits to the line profiles. From left to right \siiv, \civ, \aliii, \mgii, and \hb, respectively, for each source. In two cases, we could not make a good fit due to absorption contamination in the 1400\AA\ blend: J093403.96+315331.3 and J105427.17+253600.8. As for the case of J210831.56-063022.5 \mgii, the data were not available in the SDSS.  
Table \ref{tab:quan_Hb} lists, in the following order: identification by the SDSS name, rest-frame ﬂux, equivalent width ($W$), FWHM, and \rfe\ of the \hb\ BC line. The following columns report the \hb\ Blue parameters: rest-frame ﬂux, $W$, FWHM, and shift. Here for shift we mean the radial velocity of the line peak with respect to the RF. {The lower panels in Figs. \ref{fig:spec_fitsp1} and \ref{fig:spec_fitsp2} show the residual of the subtraction of the model spectrum from the observed one (solid black line). This residual was processed using a boxcar filter with a width of 10 pixels. The filtered residual was then subtracted from the original residual, and the resulting difference was defined as noise. The noise was quantified as the RMS computed over a 10-pixel windows, yielding the noise vector at the continuum level (red dashed line). The application of the boxcar filter offers the advantage of removing features that do not correspond to random noise but instead arise from systematic discrepancies between the model and the observed spectrum.}
Tables \ref{tab:quan_Oiii} and \ref{tab:quan_UV} also describe the SDSS name rest-frame ﬂux, $W$, and FWHM of each emission line, including for the Blue parameters (if present). All uncertainties are computed using the quality parameter $\mathcal{Q}$ \citep{marziani22}. {Uncertainties of the continuum windows fluxes were estimated with $\pm 1\sigma$ along the local continuum.}

\subsubsection{1400 \AA\ blend}

As previously mentioned, for J093403 and J105427 it was impossible to save the \siivonly\ spectra due to significant absorptions contaminating the line profiles. Even in cases where the regions were available, absorptions were still present, making it somewhat challenging to accurately fit the emission of \siivonly +\oivonly. In almost all cases, we had to account for these absorptions in the {\tt specfit} multi-component model and statistically choose the best fits. In all cases, the dominant line profile (or at least similar to the BC) is the Blue component (FWHM up to 10,000 km/s). Only in the case of J125914 { the converse is true, still with an asymmetrical BLUE line profile but with a dominant core component. } %

\begin{table*}
\centering
\caption{Spectrophotometric Quantities of the H$\beta$ components of this work.}
\label{tab:quan_Hb}
\begin{tabular}{@{}ccccccccccc@{}}
\toprule
Jcode &  &  & \hb\ BC&  & \multicolumn{2}{c}{R$_{FeII}$} &  &  & \hb\ Blue &  \\ \cmidrule(l){3-11}
    & F$_{\lambda}(5100$ \AA)& F & W & FWHM & BC & BC+Blue & F & W & FWHM & shift \\
(1) & (2) & (3) & (4) & (5) & (6) & (7) & (8) & (9) & (10) & (11) \\ \midrule
J084502 & 207$\pm$12 & 14543$\pm$871 & 66$\pm$4 & 4193$\pm$503 & 0.784$\pm$0.06 & 0.717$\pm$0.07 & 2274$\pm$46 & 11$\pm$2 & 3422$\pm$274 & -3151$\pm$36 \\
J093403 & 772$\pm$5 & 49672$\pm$3463 & 61$\pm$4 & 4000$\pm$960 & 1.086$\pm$0.09 & 0.976$\pm$0.06 & 8332$\pm$49 & 11$\pm$1 & 5995$\pm$659 & -5524$\pm$82 \\
J105427 & 172$\pm$14 & 9203$\pm$736 & 50$\pm$4 & 3680$\pm$405 & 0.973$\pm$0.06 & 0.915$\pm$0.05 & 1144$\pm$18 & 7$\pm$0.2 & 4969$\pm$348 & -732$\pm$12 \\
J125914 & 172$\pm$10 & 8590$\pm$859 & 48$\pm$5 & 2000$\pm$300 & 0.464$\pm$0.05 & 0.483$\pm$0.08 & - & - & - & - \\
J144218 & 53$\pm$12 & 2623$\pm$157 & 47$\pm$3 & 3000$\pm$270 & 1.407$\pm$0.07 & 1.398$\pm$0.06 & 174$\pm$7 & 3$\pm$0.1 & 5005$\pm$250 & -779$\pm$23 \\
J210831 & 226$\pm$11 & 7774$\pm$544 & 32$\pm$2 & 4982$\pm$498 & 1.588$\pm$0.18 & 1.124$\pm$0.12 & 2714$\pm$25 & 12$\pm$1 & 5975$\pm$478 & -874$\pm$36 \\ \bottomrule
\end{tabular}
\\
\textbf{Notes}: All measurements are in the quasar rest-frame, and line ﬂuxes are all in units of 10$^{-17}$ erg s$^{-1}$ cm$^{-2}$. Uncertainties are computed using the quality parameter $\mathcal{Q}$ defined by \cite{marziani22}. Column (1): SDSS identiﬁcation of the object. Column (2): Continuum Flux at 5100\AA. Column (3): total ﬂux of \hb\ BC. Column (4): equivalent width of \hb\ BC, in \AA. Column (5): FWHM \hb\ BC, in \kms. Columns (6, 7): \rfe\ parameter with \hb\ BC only and with full profile (i.e. BC+Blue). Columns (8) to (10): same as \hb\ BC but for the blue component. Column (11): radial velocity shift of the line peak in \kms.
\end{table*} 

\begin{table*}
\centering
\caption{Spectrophotometric Quantities of the \oiii\ components of this work.}
\label{tab:quan_Oiii}
\begin{tabular}{@{}cccccccc@{}}
\toprule
Jcode &  & \oiii\ NC &  &  & \multicolumn{2}{c}{\oiii\ Blue} &  \\ \midrule
 & F & W & FWHM & F & W & FWHM & shift \\
(1) & (2) & (3) & (4) & (5) & (6) & (7) & (8) \\ \midrule
J084502 & 521$\pm$20 & 3$\pm$0.1 & 500$\pm$43 & 2581$\pm$998 & 12$\pm$1 & 3012$\pm$263 & -2257$\pm$379 \\
J093403 & - & - & - & - & - & - & - \\
J105427 & - & - & - & - & - & - & - \\
J125914 & 1517$\pm$64 & 9$\pm$0.4 & 1084$\pm$102 & - & - & - & - \\
J144218 & 140$\pm$11 & 3$\pm$0.2 & 1515$\pm$50 & - & - & - & - \\
J210831 & 650$\pm$35 & 3$\pm$0.2 & 905$\pm$104 & 750$\pm$42 & 3$\pm$0.2 & 1513$\pm$179 & -2215$\pm$433 \\ \bottomrule
\end{tabular}
\\
\textbf{Notes}: All measurements are in the quasar rest-frame, and line ﬂuxes are all in units of 10$^{-17}$ erg s$^{-1}$ cm$^{-2}$. Uncertainties are computed using the quality parameter $\mathcal{Q}$ defined by \cite{marziani22}. Column (1): SDSS identification of the object. Column (2): total ﬂux of \oiii\ NC. Column (4): equivalent width of \oiii\ NC, in \AA. Column (5): FWHM \oiii\ NC, in \kms.  Columns (8) to (10): same as \oiii\ NC but for the blue component (if detected). Column (8): radial velocity shift of the line peak in \kms.
\end{table*}

\subsubsection{\civonly}

The scenario observed for \civonly\ is similar to the blend 1400\AA\ for the sources J0845, J0934, and J1054. Sources J0845  and J1054 have been classified as   BAL quasars. For J0845 the BAL  trough may affect more  \siiv+\oiv, while for \civ\ the BAL  trough extends beyond the limit of the blueshifted emission component, making possible a reliable estimate.   The absorption trough of J1054 is more complex. Only the trough beyond the blue end of the emission component was included in the fit. However, the sharp profile of the absorption between 1505 and 1540, and the lack of other absorptions affecting the \civonly\ emission profile, made it possible to interpolate across the  range affected by the absorption (Fig. \ref{fig:spec_fitsp1}), and retrieve a reliable \civonly\ blueshifted component.  

Among the quasars displaying an asymmetrical profile in the blue component are J0845, J1442, and J1259. The last one is exhibiting a particularly pronounced asymmetry attributed to the narrow profile in the BCs observed in all prominent emission lines. The FWHM values of the blue component are large, $\sim$ 10,000 \kms, except in the case of J0934 where FWHM $\approx$ 5030 \kms, still a large value among quasar outflows \citep{bisognietal17,sulentic17,vietrietal18,vietrietal20}.  

\subsubsection{1900 \AA\ blend}

In general, the S/N ratio was good  (40-50) in the range of the 1900 \AA\ blend, and the only source that  exhibits significant absorptions is J105427. In this case, the absorption components were incorporated into the {\tt specfit} model. The mean FWHM values are: \cnl $\approx$ 3420 \kms, \si $\approx$3730 \kms, and \al $\approx$ 4153 \kms. All sources save two exhibit a strong blue component: weaker  in the case of J1442, and absent for  J1259. The FWHM of the \al\ Blue component is $\sim$5000\kms, except for J210831, which has a higher value ($\sim$ 6600 \kms). For this region, it is   possible to observe the significant contribution of \feiii\ emission, which extends across the entire pseudo-continuum due to \feii\ and \feiii. Additionally, it is apparent that at around 2050\AA, there is some emission not being accounted for by resonance lines, and instead attributed  to \feii\ and \feiii\ \citep{vestergaard01,MAetal18,mediavillaetal18,mediavillaetal19,templeetal20}. 

\subsubsection{\mgiionly}

The spectra of J084502 and J144218 exhibited a low S/N ($<$3), which complicated the fitting process for the redward wavelengths. However, for the remaining objects, the fitting quality was satisfactory. As mentioned earlier, the \mgiionly\ spectrum for J2108  was not available in the SDSS. Only in the case of J1054  were absorption components found that contaminated the  blue wing of the doublet profile, and they were incorporated into the {\tt specfit} model. The sources did not exhibit a blue component except for J0845 and J1054 (FWHM $\approx$ 5300\kms), which had low intensity and were asymmetrical (especially in the latter case), probably due to the detect absorption. 

\subsubsection{\hb}

This was the LBT sample consisting of good S/N spectra ($\sim$10-20, except for the case of J1442 which had a value of $\approx$ 4).  A precise separation of the BC and Blue components could be achieved, not only for \hb\ but also for \oiii\ (which, as previously mentioned, is a HIL that also shows outflows), as in  the J2108 case. On the other hand, due to the high contribution of \feii, it was impossible to detect the \oiii\ emission, for instance in J0934 and J1054. The case of J1259 is particularly noteworthy, as it has FWHM $\approx$ 2000\kms, which is almost at the limit of an NLSy1-type quasars \citep{osterbrockpogge85}. The narrowness of the lines is  peculiar due to the source high luminosity \citep{marzianietal18a}.  

\subsubsection{\oiiiopt}

\oiii\ emission is expected to be weak in this type of object, where \feii\ emission is very prominent and accretion is high \citep{BG92,sulentic00a,marzianietal16a,vietrietal18,templeetal24,deconto-machadoetal24}.  In J0934 and J1054 in fact, no \oiii\ emission is detected at all. In J1442 the detection is weak, being lost in the noise. On the other hand, in J0845, J1259 and J2108 a narrow \oiii\ component is observed which, although not very prominent, confirms  the rest-frame of the object based on the peak of \hb\ and \aliii. Only in J0845 a semi broad component of \oiii\ shifted to blue is detected. In J2108, we have also included a blueshifted \oiii\ component, although in this object it is weak and with a similar width to the \oiii\ component in the rest-frame. The fluxes of the \oiii\ components are reported in Table \ref{tab:quan_Oiii}.

\subsection{Emission line profiles in the xA quasars}
 \label{profiles}

In Table \ref{tab:line_prof} we report, for each object, the parameters associated with the full profile measurements of \civonly, \mgiionly, \siivonly, \al, and \hb. These parameters are the FWHM, the asymmetry index (AI) and the centroids of each emission line. The centroid at fraction $x$\ of the peak intensity is given by: 

\begin{equation}
    c(x) = \frac{\lambda_\mathrm{R(x)}+\lambda_\mathrm{B(x)}-2\lambda_0}{2\lambda_0} c,
\end{equation}

{where $\lambda_\mathrm{R(x)}$ and $\lambda_\mathrm{B(x)}$ are the wavelengths in the red (R) and blue (B) wings, respectively}. Values reported are for $x =$0.25, 0.5, 0.75, and 0.9.

\begin{landscape}
\begin{table}
\caption{Spectrophotometric Quantities of the UV regions of this work.}
\label{tab:quan_UV}
\begin{tabular}{@{}cccccccccccccc@{}}
\toprule
Jcode &  &  & \siivonly &  & \oivonly &  &  &  & \siivonly\ Blue &  &  &  &  \\ \cmidrule(r){2-11}
& $F_{\lambda}$(1350\AA) & F & W & FWHM & F & W & F & W & FWHM & shift &  &  &  \\
(1) & (2) & (3) & (4) & (5) & (6) & (7) & (8) & (9) & (10) & (11) &  &  & \\ \cmidrule(r){1-11}
J084502 & 631$\pm$16 & 1514$\pm$106 & 2$\pm$0.1 & 1587$\pm$159 & 1892$\pm$113 & 3$\pm$0.1 & 6368$\pm$382 & 10$\pm$0.6 & 10036$\pm$803 & -1777$\pm$404 &  &  &  \\
J093403 & 1390$\pm$29 & - & - & - & - & - & - & - & - & - &  &  &  \\
J105427 & 881$\pm$23 & - & - & - & - & - & - & - & - & - &  &  &  \\
J125914 & 1365$\pm$15 & 4003$\pm$200 & 3$\pm$0.1 & 1650$\pm$116 & 4003$\pm$280 & 3$\pm$0.1 & 4980$\pm$249 & 4$\pm$0.2 & 10008$\pm$801 & -1489$\pm$255 &  &  &  \\
J144218 & 322$\pm$11 & 1024$\pm$61 & 3$\pm$0.1 & 2626$\pm$236 & 1024$\pm$61 & 3$\pm$0.1 & 1754$\pm$88 & 5$\pm$0.2 & 10003$\pm$700 & -1604$\pm$795 &  &  &  \\
J210831 & 2211$\pm$50 & 2038$\pm$143 & 1$\pm$0.1 & 3830$\pm$421 & 2547$\pm$127 & 1$\pm$0.1 & 10546$\pm$527 & 5$\pm$0.2 & 9983$\pm$699 & -1498$\pm$330 &  &  &  \\ \cmidrule(r){1-11}
&  &  &  &  &  &  &  &  &  &  &  &  &  \\ \midrule
Jcode &  & \civonly &  &  & \heiionly &  &  & \civonly Blue &  &  &  & \heiionly Blue &  \\ \cmidrule(l){2-14} 
& F & W & FWHM & F & W & FWHM & F & W & FWHM & shift & F & W & shift \\
(1) & (2) & (3) & (4) & (5) & (6) & (7) & (8) & (9) & (10) & (11) & (12) & (13) & (14) \\ \midrule
J084502 & 7023$\pm$421 & 10$\pm$0.4 & 5664$\pm$510 & 1558$\pm$78 & 2$\pm$0.1 & 5234$\pm$366 & 7008$\pm$420 & 13$\pm$0.5 & 10388$\pm$935 & -2233$\pm$373 & 1447$\pm$72 & 3$\pm$0.1 & -3530$\pm$373 \\
J093403 & 9382$\pm$563 & 6$\pm$0.2 & 4352$\pm$392 & 5904$\pm$295 & 5$\pm$0.1 & 6037$\pm$483 & 695$\pm$28 & 0.5$\pm$0.05 & 5034$\pm$252 & -4527$\pm$103 & 1814$\pm$91 & 1$\pm$0.1 & -4524$\pm$103 \\
J105427 & 3084$\pm$111 & 2$\pm$0.1 & 4613$\pm$318 & 705$\pm$28 & 0.7$\pm$0.05 & 4299$\pm$258 & 3308$\pm$165 & 4$\pm$0.2 & 10639$\pm$745 & -3214$\pm$450 & 1533$\pm$61 & 2$\pm$0.1 & -4386$\pm$450 \\
J125914 & 7154$\pm$572 & 5$\pm$0.2 & 2534$\pm$304 & 1849$\pm$92 & 1$\pm$0.1 & 4195$\pm$336 & 8035$\pm$482 & 6$\pm$0.2 & 10561$\pm$950 & -918$\pm$846 & 1775$\pm$89 & 1$\pm$0.1 & -3436$\pm$846 \\
J144218 & 1907$\pm$114 & 6$\pm$0.2 & 3728$\pm$298 & 596$\pm$30 & 2$\pm$0.1 & 4468$\pm$268 & 2985$\pm$149 & 9$\pm$0.2 & 10550$\pm$844 & -2980$\pm$763 & 647$\pm$26 & 2$\pm$0.1 & -5190$\pm$763 \\
J210831 & 5313$\pm$266 & 2$\pm$0.1 & 5590$\pm$391 & 1327$\pm$53 & 0.5$\pm$0.05 & 4489$\pm$269 & 16690$\pm$835 & 7$\pm$0.2 & 10812$\pm$865 & -4424$\pm$509 & 1270$\pm$51 & 0.6$\pm$0.05 & -3395$\pm$509 \\ \midrule
&  &  &  &  &  &  &  &  &  &  &  &  &  \\ \cmidrule(r){1-14}
Jcode & & & \al &  & \si &  & \cnl &  & \feiii &  & \al\ Blue &  &  \\ \cmidrule(l){2-14}
& $F_{\lambda}$(1700\AA) & F & W & FWHM & F & FWHM & F & FWHM & F & F & W & FWHM & shift \\ 
(1) & (2) & (3) & (4) & (5) & (6) & (7) & (8) & (9) & (10) & (11) & (12) & (13) & (14) \\ \cmidrule(r){1-14}
J084502 & 516$\pm$12 & 3573$\pm$965 & 7$\pm$2 & 4536$\pm$253 & 3940$\pm$1379 & 4069$\pm$651 & 1988$\pm$119 & 3962$\pm$634 & 19$\pm$1 & 773$\pm$62 & 2$\pm$0.5 & 4893$\pm$489 & -1726$\pm$201  \\
J093403 & 1178$\pm$36 & 4452$\pm$1024 & 4$\pm$1 & 4073$\pm$484 & 5597$\pm$336 & 3501$\pm$560 & 5287$\pm$317 & 3272$\pm$523 & 45$\pm$2 & 1768$\pm$248 & 2$\pm$0.5 & 5087$\pm$458 & -1648$\pm$254   \\
J105427 & 710$\pm$14 & 3287$\pm$788 & 4$\pm$1 & 4552$\pm$360 & 3019$\pm$181 & 4110$\pm$658 & 718$\pm$7 & 3498$\pm$525 & 28$\pm$1 & 1065$\pm$107 & 2$\pm$0.5 & 5049$\pm$505 & -1855$\pm$217 \\  
J125914 & 941$\pm$18 & 2589$\pm$595 & 4$\pm$1 & 2478$\pm$167 & 2826$\pm$113 & 1994$\pm$319 & 3341$\pm$100 & 1776$\pm$284 & 30$\pm$2 & - & - & - &  \\
J144218 & 234$\pm$17 & 1518$\pm$228 & 6$\pm$2 & 3743$\pm$345 & 1274$\pm$115 & 3395$\pm$509 & 587$\pm$65 & 3156$\pm$473 & 9$\pm$1 & 100$\pm$1 & 0.4$\pm$0.05 & 4980$\pm$598 & -1881$\pm$129 \\
J210831 & 1568$\pm$27 & 9878$\pm$2074 & 6$\pm$2 & 5534$\pm$805 & 10976$\pm$440 & 5300$\pm$848 & 1333$\pm$147 & 4850$\pm$728 & 47$\pm$2 & 1720$\pm$206 & 1$\pm$0.1 & 6633$\pm$597 & -3032$\pm$237  \\  \cmidrule(r){1-14}
&  &  &  &  &  &  &  &  &  &  &  &  &  \\ \cmidrule(r){1-10}
Jcode &  & & \mgiionly &  & \feii &  & \mgiionly\ Blue &  &  &  &  &  &  \\ \cmidrule(l){2-10}
 & $F_{\lambda}$ (3000\AA) & F & W & FWHM & F & F & W & FWHM & shift &  &  &  &  \\
(1) & (2) & (3) & (4) & (5) & (6) & (7) & (8) & (9) & (10) &  &  &  &  \\ \cmidrule(r){1-10}
J084502 & 270$\pm$81 & 8592$\pm$430 & 30$\pm$2 & 3849$\pm$308 & 107$\pm$9 & 669$\pm$24 & 2$\pm$0.1 & 5384$\pm$185 & -1474$\pm$81 &  &  &  &  \\
J093403 & 949$\pm$49 & 17290$\pm$1141 & 17$\pm$1 & 3463$\pm$346 & 443$\pm$44 & - & - & - & - &  &  &  &  \\
J105427 & 352$\pm$78 & 8813$\pm$441 & 24$\pm$1 & 3769$\pm$302 & 169$\pm$14 & 434$\pm$14 & 1$\pm$0.05 & 5313$\pm$213 & -1499$\pm$76 &  &  &  &  \\
J125914 & 460$\pm$106 & 8487$\pm$509 & 18$\pm$1 & 1416$\pm$127 & 230$\pm$18 & - & - & - & - &  &  &   \\
J144218 & 101$\pm$57 & 1508$\pm$75 & 26$\pm$1 & 3000$\pm$210 & 56$\pm$4 & - & - & - & - &  &  &  &  \\
J210831 & - & - & - & - & - & - & - & - & - &  &  & &   \\ \bottomrule
\end{tabular}%
\\ \textbf{Notes:} All measurements are in the quasar rest-frame, and line ﬂuxes are all in units of 10$^{-17}$ erg s$^{-1}$ cm$^{-1}$. \textbf{Top:} Column (1): SDSS identification of the object. Column (2): continuum flux at 1350\AA. Columns (3,6,8): total ﬂux. Column (4,7,9): equivalent width in \AA. Column (5,10): FWHM in \kms. Column (11): radial velocity shift of the line peak. 
\textbf{Middle top}: Column (1): SDSS identification of the object. Columns (2,5,8,12): total ﬂux. Column (3,6,9,13): equivalent width in \AA. Column (4,7,10): FWHM in \kms. Column (11,14): radial velocity shift of the line peak. 
\textbf{Middle bottom}: Column (1): SDSS identification of the object. Columns (2,6,10,11): total ﬂux. Column (4,12): equivalent width in \AA. Column (5,7,9,13): FWHM in \kms. Column (14): radial velocity shift of the line peak. 
\textbf{Bottom}: Column (1): SDSS identification of the object. Columns (2,3,6,7): total ﬂux. Column (4,8): equivalent width in \AA. Column (5,9): FWHM in \kms. Column (10): radial velocity shift of the line peak in \kms. 
\end{table} 
\end{landscape}

The asymmetry index is different from zero (negative) only for objects with a blueshifted component. The asymmetry index at one quarter is defined as line displacement with respect to the  peak wavelength (in practice the $c(0.9)$ is used as a proxy for $\lambda_\mathrm{P}$) normalized by line width at $\frac{1}{4}$\ intensity,

\begin{equation}
    \mathrm{AI} =  \frac{\lambda_\mathrm{R(\frac{1}{4})}+\lambda_\mathrm{B(\frac{1}{4})}-2\lambda_\mathrm{P}}{\lambda_\mathrm{R(\frac{1}{4})}-\lambda_\mathrm{B(\frac{1}{4})}} 
\end{equation}

These parameters are measured on the full profiles, and therefore avoid the uncertainties associated with the decomposition of the line profiles. 
On the full profiles of \hb, \al, \mgiionly, it is necessary to compute the centroid at a low fractional intensity (i.e., $\frac{1}{4}$) to detect the effect of the BLUE component (Table \ref{tab:line_prof}). 

The full profile analysis confirms the result of the multi-component decomposition: the observed \civ\ and \siiv\ blueshifts spreads up to -4000, with value increasing toward low fractional intensity.   As for the ratio \rfe,  sources with a higher ratio, \civ\ and \siiv\  centroids tends to be highly blueshifted.  Whereas in \hb, \al\ and \mgiionly\ there is no strong dependence on line luminosity, \rfe, and FWHM;  data points are around zero radial velocity i.e., consistent with the rest frame of the quasar. 

 Overall, it seems that there is no systematic blueshift towards highest Eddington ratios, nor with luminosity. This is hardly surprising because the sources are only 5, at similar luminosity. The spread in Eddington ratio is also modest with all sources having \lledd (\hb) $\gtrsim 0.6$. J1259 is an outlier with \lledd   $\gtrsim 2$ in all estimates, due to its very narrow lines.

\begin{figure*}
    \centering
    \includegraphics[width=0.49\textwidth]{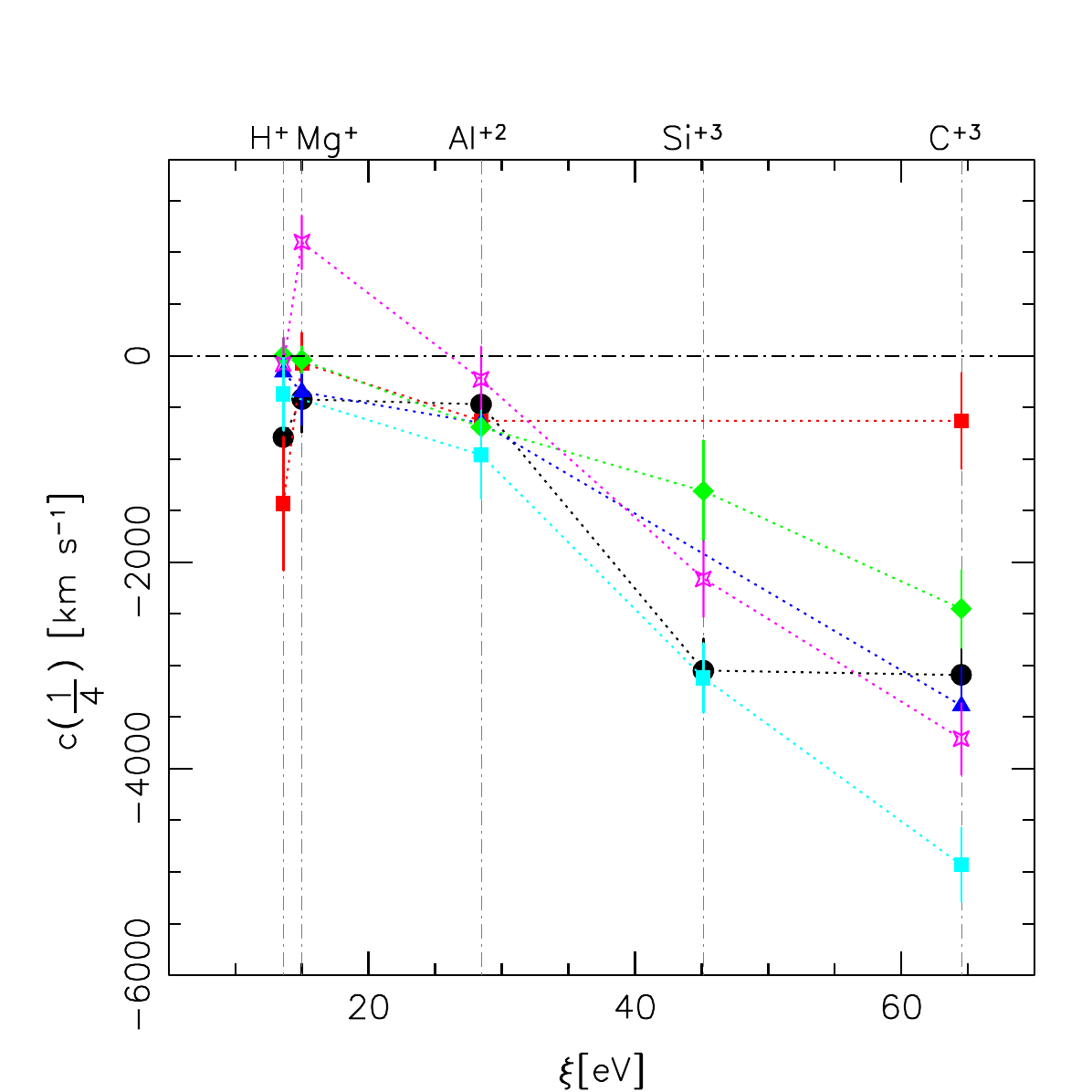}
    \includegraphics[width=0.49\textwidth]{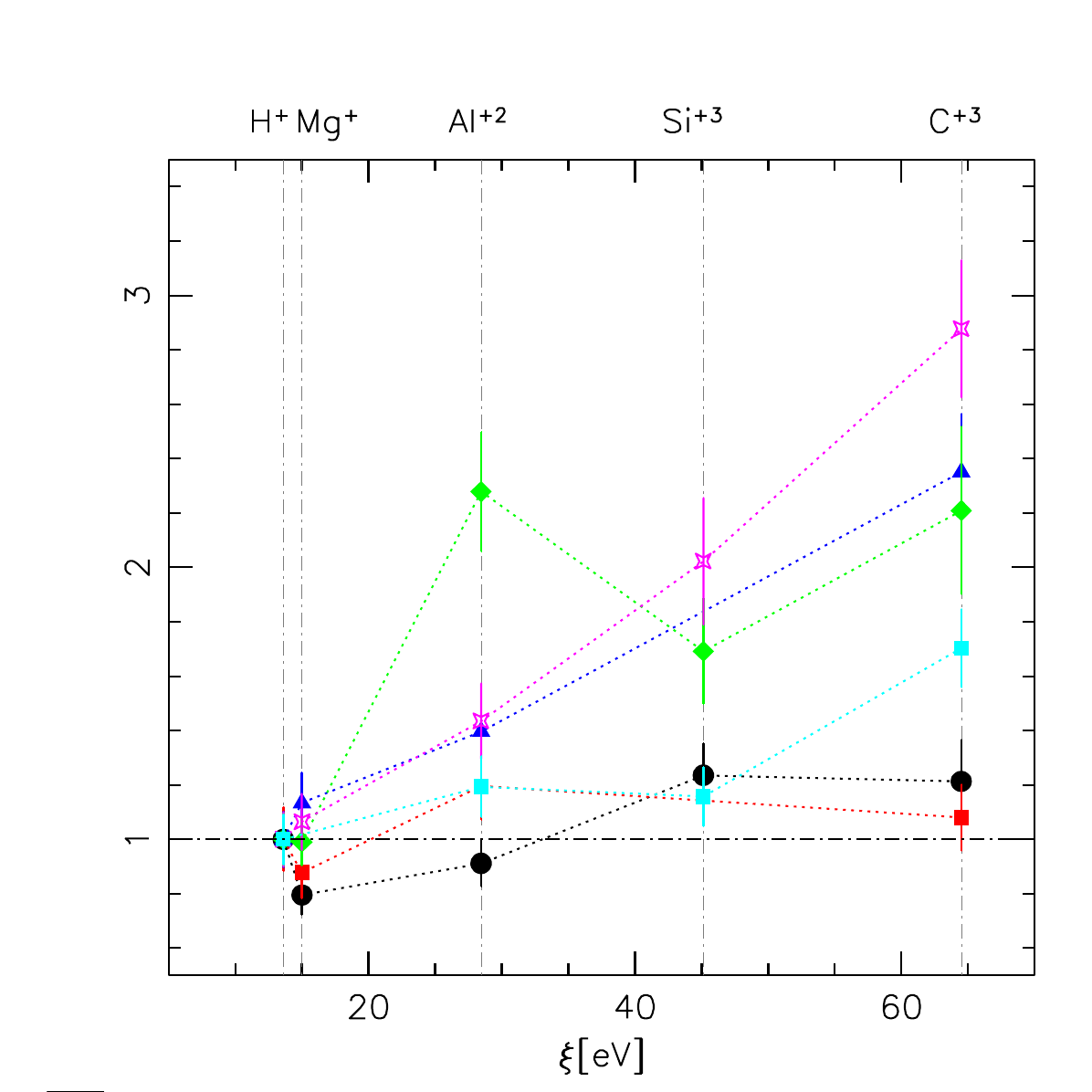}
    \caption{Shift amplitude of the centroid at $\frac{1}{4}$\ peak intensity (left) and FWHM / FWHM (\hb) ratios vs. ionization potential $\chi$  (right) for the ionic species associated with the emission lines considered in this study. Different colors and symbols identify different object in the sample. Black circles: J0845; red squares: J0934; blue triangles: J1054; green rhombuses: J1259; magenta stars: J1442; cyan squares: J2108.  }
    \label{fig:xishif}
\end{figure*}

It is interesting to note, however, that the shift amplitude tends to increase with ionization potential of the parent ionic species: from \hb\ and \mgii\ with $\chi \sim 10 $ eV showing modest shifts even toward the line base, we reach a maximum shift $\sim -5000$ \kms\ for \civ, with $\chi \sim 60$ eV. Fig. \ref{fig:xishif}   confirms that this trend is present in most of the sample sources. Additionally, we observe a systematic increase in the line width when comparing \hb\ and \civ: the FWHM ratio with respect to \hb\ is approximately 2. The high shift amplitude and larger FWHM indicate a spectroscopically resolved component affecting the high-ionization line profiles. In other words, the excess broadening in \civ\ is due to the prominence of its blueshifted component, which remains fainter in \hb\ (implying \civ/\hb $\gg 1$). In other words, the increase in FWHM from low to high ionization lines is not mainly due to a decrease in emissivity-weighted distance (which would be expected in case of a Keplerian velocity field), but to the onset of an addition component displaced in radial velocity \citep{sulentic07,marzianietal16a}.  The only exception in this respect appears J0934: it is  a BAL, and we cannot exclude that part or all of the blueshifted emission is removed by the broad absorption. 

\begin{table*}							
\centering								
\caption{Line profile parameters}		
\label{tab:line_prof}								
\begin{tabular}{@{}cccccccccc@{}}							
\toprule								
Jcode	&	Line 	&	FWHM			&	AI        			&	\coneq\			&	\ctwoq			&	\cthreeq			&	c(0.9)				\\	\midrule
J084502	&	\civonly\ 	&	7000	$\pm$	750	&	-0.12	$\pm$	0.07	&	-3090	$\pm$	420	&	-3050	$\pm$	380	&	-2790	$\pm$	230	&	-2340	$\pm$	120	&&	\\	
J084502	&	\mgiionly\	&	4590	$\pm$	300	&	-0.03	$\pm$	0.09	&	-420	$\pm$	330	&	-340	$\pm$	150	&	-290	$\pm$	110	&	-290	$\pm$	80	&&	\\	
J084502	&	\siivonly\	&	7130	$\pm$	480	&	-0.53	$\pm$	0.07	&	-3050	$\pm$	310	&	-2110	$\pm$	240	&	-1130	$\pm$	340	&	-340	$\pm$	160	&&	\\	
J084502	&	\aliiionly\	&	5260	$\pm$	330	&	-0.03	$\pm$	0.08	&	-470	$\pm$	320	&	-450	$\pm$	160	&	-370	$\pm$	140	&	-340	$\pm$	90	&&	\\	
J084502	&	\hb\	&	5770	$\pm$	390	&	-0.16	$\pm$	0.07	&	-790	$\pm$	300	&	-820	$\pm$	190	&	-240	$\pm$	180	&	-110	$\pm$	80	&&	\\	
	\\								
J093403	&	\civonly\ &	 	4580	$\pm$	360	&	-0.14	$\pm$	0.11	&	-630	$\pm$	470	&	-130	$\pm$	180	&	-50	$\pm$	120	&	-30	$\pm$	80	&&	\\	
J093403	&	\mgiionly\	&	3720	$\pm$	250	&	0.00	$\pm$	0.09	&	-70	$\pm$	300	&	-70	$\pm$	130	&	-80	$\pm$	100	&	-80	$\pm$	60	&&	\\	
J093403	&	\aliiionly\	&	5070	$\pm$	310	&	-0.06	$\pm$	0.07	&	-630	$\pm$	280	&	-550	$\pm$	160	&	-440	$\pm$	140	&	-400	$\pm$	90	&&	\\	
J093403	&	\hb\	&	4240	$\pm$	350	&	-0.29	$\pm$	0.14	&	-1430	$\pm$	650	&	-130	$\pm$	170	&	-40	$\pm$	120	&	-20	$\pm$	70	&&	\\	
		\\								
J105427	&	\civonly\  &	9120	$\pm$	510	&	-0.30	$\pm$	0.06	&	-3390	$\pm$	380	&	-2780	$\pm$	260	&	-2000	$\pm$	290	&	-1410	$\pm$	190	&&	\\	
J105427	&	\mgiionly\	&	4400	$\pm$	290	&	-0.07	$\pm$	0.09	&	-350	$\pm$	330	&	-270	$\pm$	150	&	-230	$\pm$	180	&	-80	$\pm$	70	&&	\\	
J105427	&	\aliiionly\	&	5420	$\pm$	330	&	-0.04	$\pm$	0.07	&	-650	$\pm$	300	&	-610	$\pm$	160	&	-510	$\pm$	150	&	-460	$\pm$	100	&&	\\	
J105427	&	\hb\	&	3880	$\pm$	280	&	-0.04	$\pm$	0.09	&	-160	$\pm$	290	&	-90	$\pm$	140	&	-50	$\pm$	110	&	-30	$\pm$	70	&&	\\	
			\\								
J125914	&	\civonly\ & 	4440	$\pm$	540	&	-0.43	$\pm$	0.10	&	-2450	$\pm$	380	&	-1390	$\pm$	270	&	-750	$\pm$	140	&	-670	$\pm$	60	&&	\\	
J125914	&	\mgiionly\	&	1990	$\pm$	110	&	0.03	$\pm$	0.08	&	-40	$\pm$	130	&	-40	$\pm$	50	&	-60	$\pm$	50	&	-80	$\pm$	40	&&	\\	
J125914	&	\siivonly\	&	3400	$\pm$	310	&	-0.43	$\pm$	0.14	&	-1310	$\pm$	490	&	50	$\pm$	150	&	300	$\pm$	70	&	280	$\pm$	110	&&	\\	
J125914	&	\aliiionly\	&	4580	$\pm$	310	&	-0.30	$\pm$	0.06	&	-690	$\pm$	210	&	-420	$\pm$	160	&	90	$\pm$	180	&	350	$\pm$	70	&&	\\	
J125914	&	\hb\	&	2010	$\pm$	140	&	0.00	$\pm$	0.10	&	0	$\pm$	170	&	10	$\pm$	70	&	10	$\pm$	60	&	10	$\pm$	30	&&	\\	
	\\								
J144218	&	\civonly\ 	&	8920	$\pm$	460	&	-0.30	$\pm$	0.06	&	-3710	$\pm$	350	&	-3080	$\pm$	230	&	-2370	$\pm$	250	&	-1850	$\pm$	160	&&	\\	
J144218	&	\mgiionly\	&	3300	$\pm$	220	&	0.01	$\pm$	0.09	&	1100	$\pm$	260	&	1100	$\pm$	110	&	1090	$\pm$	90	&	1090	$\pm$	60	&&	\\	
J144218	&	\siivonly\	&	6270	$\pm$	570	&	-0.39	$\pm$	0.08	&	-2160	$\pm$	370	&	-1190	$\pm$	280	&	-370	$\pm$	170	&	-200	$\pm$	110	&&	\\	
J144218	&	\aliiionly\	&	4450	$\pm$	290	&	-0.03	$\pm$	0.09	&	-230	$\pm$	320	&	-160	$\pm$	140	&	-130	$\pm$	120	&	-140	$\pm$	80	&&	\\	
J144218	&	\hb\	&	3100	$\pm$	220	&	-0.03	$\pm$	0.10	&	-80	$\pm$	260	&	-30	$\pm$	110	&	-10	$\pm$	80	&	0	$\pm$	50	&&	\\	
	\\							
J210831	&	\civonly\ 	&	9860	$\pm$	520	&	-0.07	$\pm$	0.05	&	-4930	$\pm$	360	&	-4660	$\pm$	260	&	-4580	$\pm$	280	&	-4430	$\pm$	190	&&	\\	
J210831	&	\siivonly\	&	6700	$\pm$	450	&	-0.29	$\pm$	0.07	&	-3120	$\pm$	340	&	-2660	$\pm$	230	&	-2080	$\pm$	230	&	-1640	$\pm$	140	&&	\\	
J210831	&	\aliiionly\	&	6910	$\pm$	460	&	-0.08	$\pm$	0.08	&	-960	$\pm$	430	&	-850	$\pm$	230	&	-640	$\pm$	190	&	-540	$\pm$	120	&&	\\	
J210831	&	\hb\	&	5790	$\pm$	380	&	-0.04	$\pm$	0.08	&	-370	$\pm$	360	&	-330	$\pm$	190	&	-240	$\pm$	160	&	-190	$\pm$	100	&&	\\	
\bottomrule\end{tabular}							
\\					
\textbf{Notes}:All measurements are in the quasar rest frame; centroids and FWHM are in \kms. 				
\end{table*}

\medskip
\subsection{Consistency of criteria to identify super-Eddington candidates}\label{consis_optUV}

\begin{figure*}
    \centering
    \includegraphics[width=0.49\textwidth]{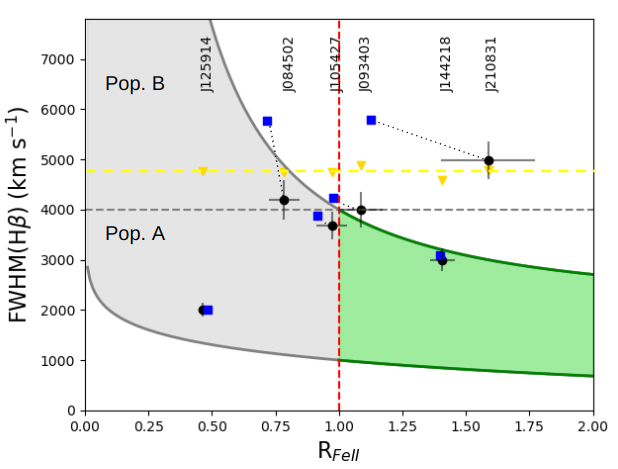}
    \includegraphics[width=0.47\textwidth]{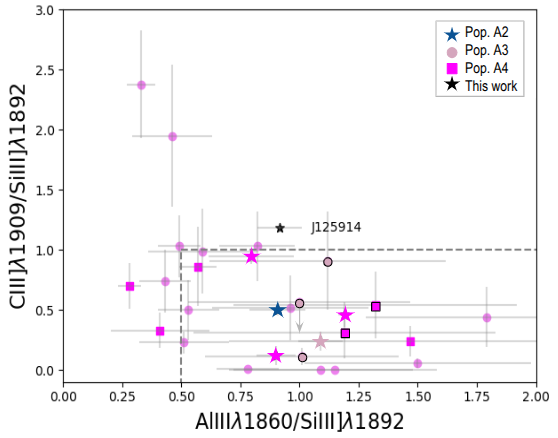}
    \caption{Distribution of the LBT sample in the optical (left) and UV (right) plane. $Left$: The luminosity-dependent limit between Pop. A and Pop. B. of \protect\citet{sulentic17} is shown in gold. The blue squares show the position within the sequence using only the  BC for the \hb\ FWHM, while  black circles make use of the full profile (i.e. BC$+$Blue) are in blue. The shaded area is a schematic representation of the Main Sequence of quasars for low-$z$\ samples. $Right$: Pop. A3 (circle) and A4 (square) quasars shown in the UV plane are from \protect\citet{marziani22,deconto-machadoetal23,deconto-machadoetal24}. The magenta and purple colors identify spectral type A3 and A4, respectively. Data outlined with a black contour represent quasars from \protect\cite{MS14}. Their rest-frame optical spectra were obtained from \protect\citet{matthewsetal21} and \protect\citet{templeetal24}.}

    \label{fig:xA_criteria}
\end{figure*}

The location of the six sources, both in the optical and UV planes, defined by FWHM (\hb) vs. \rfe, and the intensity ratios \ciii/\siiii\ vs. \aliii/\siiii, respectively, is shown in Figure \ref{fig:xA_criteria}. 
According to \citet{sulentic00a},  the xA quasars that meet the criteria \rfe\ $>$ 1.0 (red dashed line) are a sub-population of Population A.  Following \citetalias{MS14}, the extreme quasars (green shaded area) also meet the criteria: \al/\si $>$ 0.5 $\&$ \si/\cnl\ $>$ 1 (enclosed in grey dashed lines). The yellow upside-down triangles are the luminosity-dependent limit between Pop. A and Pop. B. of \citet[][gold]{sulentic17}, which brings the limit at FWHM $\approx$ 4000 \kms\ to significantly higher values for sources of bolometric luminosity $\log L_\mathrm{bol} \gtrsim 46$ \ergss:  FWHM$_\mathrm{AB} \approx 3500 + 500 (L_\mathrm{bol}/3.69\times 10^{44})^{0.15}$ \kms (applied to \hb). Each source is uniquely identified, enabling visualization of the object-position in the optical plane. Note that, if the FWHM \hb\ is cleaned of the excess emission due to the BLUE nonvirial component, all sources meet the line width criterion for being of Population A as seen in Fig. \ref{fig:xA_criteria} (black dots, left). 

{Of the six sources of the present sample, five meet both the UV and optical criteria (J1259 is consistently not an xA, since \rfe $\ll 1$ and \ciii/\siiii\ vs. \aliii/\siiii\ are too high and too low, respectively). The discordant source is J0845 which falls in the xA UV rectangle, albeit of spectral type A2 (\rfe $\approx 0.7 - 0.8$). To improve the statistics, in addition to the six sources of the LBT sample (stars), we added in Fig. \ref{fig:xA_criteria} (right), sources identified as Pop. A3 (circles) and Pop. A4 (squares) by \citet{marziani22,deconto-machadoetal23,deconto-machadoetal24}. We also included five quasars from the \citetalias{MS14} UV sample (black contours), which were observed in \cite{matthewsetal21} and \cite{templeetal24}.  }
{Fig. \ref{fig:xA_criteria} (right) shows that out of 28 sources confirmed as xA quasars (Pop. A3-A4) using the optical selection criterion \rfe $\gtrsim 1$, between three (if we exclude borderline cases) and seven do not meet the UV line ratios selection criteria, implying that between  $\sim75\%$ and 90\%\ of optically-classified  A3-A4 quasars fall within the UV xA region. } 

{The lack of a one-to-one correspondence may  due to  inaccuracies in the fitting of the \ciii\ emission line, especially because of contamination from \feiii$\lambda$1914 emission in its red wing, which may not have been properly taken into account. The epoch difference between optical and UV observations could also introduce inconsistencies.  The spectra obtained with LUCI are separated by several years from those acquired by the SDSS,   leaving open the possibility of considerable variation \citep{punslyetal16}. The MJD of the SDSS spectra ranges from 52,000 to 58,000, whereas most LBT observations are around 57,000. Variability is unlikely to significantly affect the classification: although emission lines do respond to continuum changes, a variation in their ratios would require a change in the physical conditions of the emitting gas — a scenario that appears improbable. }

{The expectation is that both selection criteria will help us identify the highest accreting quasars (super-Eddington candidates) for cosmological purposes, which will be discussed in Sect. \ref{ssec:hubble}.  The statistics is still limited, but our preliminary analysis suggests that  the two criteria might be consistently met in about $80$\%\ of cases. }

\subsection{General considerations}

The \hb\ blueshifted emission is detectable on \hb\   in three sources (J084502, J093403, and  J210831).   The \hb\ BLUE component is most likely associated with an outflow which produces the prominent blueshifted emission more
clearly observed in \civ\ profiles \citep[e.g.,][]{coatman16,sulentic17}.  In contrast, the \civonly\ blueshifted emission is clearly detectable in all sources (except in J093403 where it seems that the emission is severely contaminated by an absorption), implying that the intensity ratio \civonly/\hb\ can be expected to be $\gg$1.

The picture emerging from the previous analysis emphasizes the difference between the xA candidates and the Population A and B AGN at both low and high luminosity. Five of the six sources of the present sample show properties that make them similar   to other xA sources observed at low- and high-luminosity: strong \feii, weak \oiii, blue-shifted \civ\ and almost symmetric \hb\ \citep{marzianietal14,MS14,negrete18,MAetal18,deconto-machadoetal23,deconto-machadoetal24}. The properties are consistent with the one found at the high \rfe\ tip of the quasar main sequence \citep{BG92,sulentic00a,marzianietal18}. { In this respect, it is remarkable that 3 sources out of 5 have been detected in the radio. At redshift $\lesssim 1$, quasars of spectral types A3 and A4 often show significant radio emission, placing them in the radio-detected and intermediate domain \citep[][$10 \lesssim R_\mathrm{K} \lesssim 80$\ according to \citealt{zamfiretal08}]{gancietal19}. This radio emission has been ascribed to star formation processes because of its correlation with FIR parameters 
\citep{sanietal10,bonzinietal15,caccianigaetal15,netzeretal16,marzianietal21}.}

Only, at high luminosity, lines are broader (because of the higher black hole masses, \citealt{marzianietal18a}) and blueshifts reach the highest amplitudes ($\sim -5000$ \kms) because of the stronger radiation force and of the high \lledd\ \citep{sulentic17}.  Blueshifts appear ubiquitous and dominating in the high-ionization line profiles (see Fig. \ref{fig:spec_fitsp1} and Fig. \ref{fig:spec_fitsp2}), suggesting {\em dominance of outflows}, {and likely radiative acceleration}. At the same time, the \hb\ profile remains fairly symmetric and unshifted, suggesting the presence of a virialized component even at the highest luminosity \citep{MS14,sulentic07,gillettehamann24}. These results are consistent with other recent analyses, also for \oiii\ \citep{fioreetal17,bischettietal17,vietrietal18,kakkadetal20}.




\section{Discussion}\label{sec:discussionb}

In this section, we will discuss the estimation of physical parameters such as black hole mass and Eddington ratio, along with considerations and estimations regarding the winds/outflows of \civ\ and \oiii. Additionally, we will build a Hubble diagram using the highly accreting quasar sample from the LBT within the cosmological context.

\subsection{\lbol, \mbh\ and Eddington ratio}

\begin{table}
\caption{Accretion parameters.}
\label{tab:acc_param}\tabcolsep=2pt
\begin{tabular}{ccccccccc}
\cline{1-6}
Jcode &  &  & \al &  &  &  &  &  \\ \cline{2-6}
 &  & \multicolumn{2}{c}{log \lbol} & log \mbh & \lledd &  &  &  \\
 & BC N19 & a & N19 & M22 & N19 &  &  &  \\
(1) & (2) & (3) & (4) & (5) & (6) &  &  &  \\ \cline{1-6}
J084502 & 3.11 & 47.25 & 47.11 & 9.34 & 0.63 &  &  &  \\
J093403 & 2.90 & 47.56 & 47.39 & 9.43 & 1.06 &  &  &  \\
J105427 & 3.04 & 47.34 & 47.19 & 9.40 & 0.69 &  &  &  \\
J125914 & 2.97 & 47.45 & 47.29 & 8.93 & 2.59 &  &  &  \\
J144218 & 3.38 & 46.88 & 46.78 & 8.96 & 0.65 &  &  &  \\
J210831 & 2.78 & 47.73 & 47.54 & 9.79 & 0.68 &  &  &  \\ \cline{1-6}
 &  &  &  &  &  &  &  &  \\ \cline{1-8}
\multicolumn{1}{l}{Jcode} &  &  &  & \mgiionly &  &  &  & \multicolumn{1}{l}{} \\ \cline{2-8}
\multicolumn{1}{l}{} &  & \multicolumn{2}{c}{log \lbol} & \multicolumn{2}{c}{log \mbh} & \multicolumn{2}{c}{\lledd} &  \\ 
\multicolumn{1}{l}{} & BC N19 & b & N19 & TN12 & SL12 & b & N19 & \multicolumn{1}{l}{} \\
\multicolumn{1}{l}{(1)} & (7) & (8) & (9) & (10) & (11) & (12) & (13) & \multicolumn{1}{l}{} \\ \cline{1-8}
J084502 & 3.03 & 47.29 & 47.06 & 9.52 & 9.30 & 0.77 & 0.46 & \multicolumn{1}{l}{} \\
J093403 & 2.41 & 47.79 & 47.46 & 9.73 & 9.45 & 1.70 & 0.80 & \multicolumn{1}{l}{} \\
J105427 & 2.94 & 47.36 & 47.12 & 9.54 & 9.30 & 0.9 & 0.52 & \multicolumn{1}{l}{} \\
J125914 & 2.80 & 47.47 & 47.20 & 8.76 & 8.70 & 4.57 & 2.48 & \multicolumn{1}{l}{} \\
J144218 & 3.73 & 46.84 & 46.70 & 9.02 & 8.63 & 1.28 & 0.93 & \multicolumn{1}{l}{} \\
J210831 & - & - & - & - & - & - & - & \multicolumn{1}{l}{} \\ \cline{1-8}
\multicolumn{1}{l}{} & \multicolumn{1}{l}{} & \multicolumn{1}{l}{} & \multicolumn{1}{l}{} & \multicolumn{1}{l}{} & \multicolumn{1}{l}{} & \multicolumn{1}{l}{} & \multicolumn{1}{l}{} & \multicolumn{1}{l}{} \\ \hline
Jcode &  &  &  & \hb &  &  &  &  \\ \cline{2-9} 
 &  & \multicolumn{2}{c}{log \lbol} & \multicolumn{3}{c}{log \mbh} & \multicolumn{2}{c}{\lledd} \\
 & BC N19 & c & N19 & VP06 & SL12 & F24 & c & N19 \\
(1) & (14) & (15) & (16) & (17) & (18) & (19) & (20) & (21) \\ \hline
J084502 & 4.79 & 47.57 & 47.19 & 9.40 & 9.36 & 9.44 & 1.28 & 0.59 \\
J093403 & 3.67 & 48.15 & 47.51 & 9.56 & 9.54 & 9.63 & 3.25 & 0.97 \\
J105427 & 4.95 & 47.50 & 47.23 & 9.31 & 9.17 & 9.33 & 1.67 & 0.9 \\
J125914 & 4.94 & 47.51 & 47.53 & 8.78 & 8.64 & 9.22 & 5.72 & 3.05 \\
J144218 & 6.27 & 46.99 & 46.82 & 8.88 & 8.78 & 8.96 & 1.27 & 0.86 \\
J210831 & 4.71 & 47.61 & 47.32 & 9.63 & 9.40 & 9.41 & 1.29 & 0.65 \\ \hline
\end{tabular}
\\ \textbf{Notes:} Columns: (1) SDSS identification. (2,7,14) Bolometric correction (BC) computations following \cite{netzer19}. (3,8,15) and (4,9,16), Log of Bolometric luminosity in units of erg s$^{-1}$ using \citet[a=4.3, b=5.15, and c=9.26,][]{richards06}, and BC from \cite{netzer19}, respectively. Uncertainty is 10$\%$ the luminosity. (5,10,11,17,18) Black hole mass in unit of \msun. Below each column is detailed which scalar relation is used: \cite{marziani22} is M22, \cite{trakhtenbrot12} is TN12, \cite{vp06} is VP06, \cite{shen&liu12} is SL12, {and \cite{florisetal24} is F24}. (6,12,13,20,21). Eddington ratio with different BC factors using \cite{netzer19} or \cite{richards06}.
\end{table}

The bolometric correction  (B.C.)  for 1700, 3000 and 5100\AA\ were computed using {B.C.} $ = c \times [L/(10^{42} \mathrm{ergs\  s^{-1}})]^d$ and the constants of Table 1 by \cite{netzer19}, and the values of \cite{richards06} modified by the redshift. \lledd\ and \lbol\ values for the three lines are reported in Table \ref{tab:acc_param}, using the specified B.C. and the appropriate scaling relations \citep{marziani22,trakhtenbrot12,vp06}. The Eddington luminosity was computed for the  masses obtained from the FWHM of \al, \mgiionly, and \hb\ emission lines following the relation: \ledd $\approx 1.5\times10^{38} (M_\mathrm{BH}/M_{\odot})$ [\ergss] \citep[e.g.][]{NM10,netzer15}.

Using the FWHM of the \mgii\ line, we computed the virial \mbh\ using Eq. \ref{eq:mbh}:

\begin{equation}
    \mathrm{log}\left ( \frac{M_\mathrm{BH,vir}}{M_{\odot}} \right ) = a + b\ \mathrm{log} \left ( \frac{L}{10^{44} {\rm erg/s}} \right ) + c\ \mathrm{log} \left ( \frac{\rm FWHM}{\rm km/s} \right ),
    \label{eq:mbh}
\end{equation}

and the $a,\ b,\ c$\ values from Table 5 of \cite{shen&liu12}\footnote{This was done in order to minimize the difference in virial mass calibrations, using the one by \cite{vp06} as standard.}  {Using the  FWHM of \aliii, we applied  the} scaling law of  \citetalias{marziani22}  to compute the \mbh(\al). {Using the FWHM of \hb, we considered  Eq. 7 of \cite{florisetal24}, which is specifically tailored for high-accretion rate sources. The \mbh(\hb)\ formula is based on the \cite{vp06} relation with two key modifications: (1) a correction to the FWHM as proposed by \cite{mejiarestrepo18}, and (2) the inclusion of the dependence of the BLR radius on \rfe\ as described by \cite{duwang19}.} Table \ref{tab:acc_param} reports the \mbh\ estimates for the three lines.

\begin{figure}
    \centering
    \includegraphics[width=0.4\textwidth]{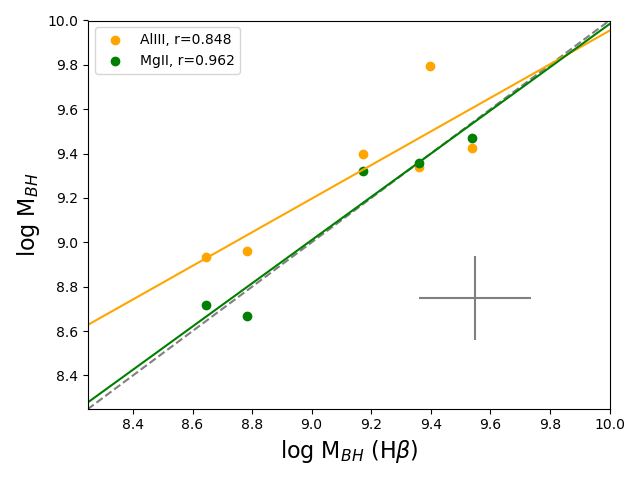}
    \caption{Distribution of log \mbh\ of \hb\ (abscissa) vs. log \mbh\ of \al\ and \mgiionly\ (ordinate), with the Pearson correlation of each computation. Dashed line is the equity line. 
    Error bar are the mean values of the \al\ and  \mgiionly\ uncertainties: 0.26 and 0.32, respectively.}
    \label{fig:Mbh_comp}
\end{figure}

The behavior of the \mbh\ values   is shown in Fig. \ref{fig:Mbh_comp}   with respect to \hb. In the abscissa scale is log \mbh\ (\hb) \citep[using ][]{shen&liu12} vs. log \mbh\ of \al\ and \mgiionly\ (using \citealt{marziani22} for the former, and \citealt{shen&liu12} for the latter). The Pearson correlation for each estimation is 0.89 for \al\ while for \mgiionly\ is 0.96. The values of these correlation coefficients show a very good agreement between the \mbh\ estimates employing different lines. The solid lines are the best fits obtained using the least-square method: $\log$\mbh(\mgiionly)$\approx 0.97 \log$ \mbh(\hb) + 0.25 and $\log$ \mbh(\al)$\approx$0.75 $\log$ \mbh(\hb) + 2.38. 
{As for the \hb\ \mbh\ estimations, \cite{florisetal24} and \cite{shen&liu12}, the values reported in Table \ref{tab:acc_param} (Cols. 17-19), the values from \citet{florisetal24} show a very good agreement with the estimates from \citet{vp06}, with average $\delta \log $ \mbh $\approx 0.07 \pm 0.21$, and a slight systematic difference with \citet{shen&liu12}: $\delta \log $ \mbh $\approx 0.18 \pm 0.20$. The scatter might be associated with the correction for orientation (included in F24, but not in SL12 and VP06): the largest $\delta \log $ \mbh $\approx 0.5$ occurs right for J125914, the object with the narrowest \hb\ line in the sample.  The small systematic effect between SL12 and \citet{florisetal24} is less clear, but the amplitude of the offset, about $50$\%, is well within the systematic uncertainty of the scaling laws based on \hb\ \citep{MS12,shen13}.} 

\subsection{Wind models}
\label{wm}

\begin{figure}
    \centering
    \includegraphics[width=0.5\textwidth]{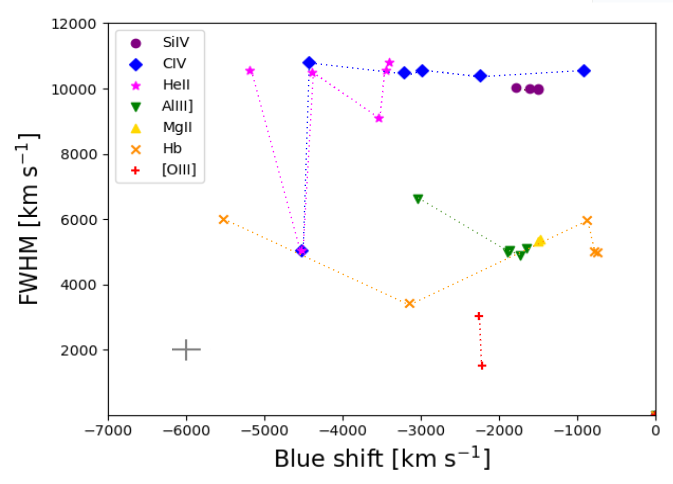}
    \caption{Distribution of shifts of the Blue component (if detected) on the profile lines of \civonly, \siivonly, \heiionly, \al, \mgiionly, \hb, and \oiii\ (in the case of \oiii\ is the semi-broad component blueshifted). Error bar are the mean values of the uncertainties of the FWHM and blueshifts: 310 and 250 \kms, respectively.}
    \label{fig:shift_Bl}
\end{figure}

Line profiles associated to xA quasars such as the ones of the LBT sample (as seen in Figs. \ref{fig:spec_fitsp1} and \ref{fig:spec_fitsp2}, 
with a presence of a BC and additional one)  can be tentatively modeled with a radial flow, possibly due to a radiation-driven wind \citep[e.g.][]{zamanov02,risaliti10}. Alternative models involve  magnetically driven centrifugal acceleration \citep{emmering92}. Both wind acceleration mechanisms are favored at high luminosity and  mass \citep{proga07}, and it is still not {fully} known which one is actually operating  \citep{elitzur14}, if they are concomitant, or applicable to different sources. The \civ\ line shift with respect to RF offers a fundamental test of disk wind models: winds driven by radiation pressure do not conserve the angular momentum of the gas flow and produce a mostly radial outflow \citep{elvis00} with very large blueshifts. Magnetically-driven acceleration  produces more symmetric, less shifted profiles. {The large shift amplitudes observed in the LBT sample ($\gtrsim 2000$ \kms) favor the radiation-driven hypothesis.}

The blueshift interpretation involves   emitting gas motion towards the observer causing   Doppler shift of   line radiation \citep{marzianietal17}. The distribution of the blueshifts (if detected) for the emission lines: \civonly, \siivonly, \heiionly, \al, \mgiionly, \hb, and \oiii\ is shown in Fig. \ref{fig:shift_Bl}. The UV lines (\civonly\ and \siivonly) shows larger shifts; on the converse,  \al, \hb, and \mgiionly\ present moderate shifts as indicated in \cite{MAetal18,buendia2023}. The semi-broad component of \oiii\ was only detected in two sources and showed a shift of $\sim$2000\kms. Error bar represents the mean values of the uncertainties of the FWHM and blueshifts: 310 and 250 \kms, respectively. Although the \oiii\ shift is large, we will restrict most of the following analysis to the line that is apparently most affected by the outflow, namely \civ. 

\subsubsection*{Outflow parameters determination}

Several caveats and assumptions accompany the computation of the kinetic power and thrust from single-epoch spectra.  In AGN outflows, the thrust is the force exerted by winds in the innermost regions. The outflows can be propelled by several mechanisms such as radiation pressure, magnetic fields, and accretion disk winds. They can exert significant thrust on the surrounding interstellar medium (ISM) or intracluster medium (ICM). The "thrust" of the wind generally refers to the force exerted by the wind itself.  The kinetic power refers to the rate at which kinetic energy is transported away and can be computed knowing the outflow mass flux and the outflow velocity. The kinetic power and thrust provides insights into the impact of outflows on their surrounding environments and in driving material away from the central black hole, regulating star formation and influence the gas content of the host galaxy \citep{liu13a,liu13b}.

Here, we follow the method for estimating the mass of ionized gas ($M^\mathrm{ion}$), the mass outflow rate ($\dot{M^\mathrm{ion}}$), the thrust ($\dot{M^\mathrm{ion}} k v_0$), and the kinetic power ($\dot{\epsilon}$) of the outflow for collisionally excited lines in photoionized gases \citep{canodiaz12} by \citet[][hereafter \citetalias{marzianietal17}]{marzianietal17}. \citetalias{marzianietal17} devised a framework that enables deriving outflow parameters independently from the covering and filling factors, assuming uniform gas density across all emitting regions. The following specific assumptions  are necessary to apply the following relation to measurements of \civonly\ obtained from single epoch spectra:
\begin{itemize}
\item constant electron density, $n$ [cm$^{-3}$];
\item all gas being in the same ionization stage;
\item well-defined chemical abundances, $Z$ [Z$_{\odot}$];
\item typical emitting radius, $r$ [$r_{\mathrm{g}}$];
\item outflow velocity, $v_o$ [\kms].
\end{itemize}

This simplified approach is suitable for elucidating the influence of  outflow parameters on thrust and kinetic power calculations with the equations of \citetalias{marzianietal17}. 

It should be noted that the outflow estimations derived from the Blue components of \civ\ using the equations proposed by \citetalias{marzianietal17}  assumed  a spectral energy distribution (SED) for  low-redshift and low-bolometric luminosity quasars. 
Given that the LBT sample are high-redshift  ($z\sim$2) and high-luminosity sources, the softening of the accretion disk emission due to larger \mbh\ makes the SED significantly different that the one of low-$L$ \citep{duras17,duras20}, reducing the amount of ionizing photons.
We therefore consider the recomputation of the conversion between line luminosity of \civ\ and amount of gas by \citet{deconto-machadoetal24}. 

\citet{deconto-machadoetal24}  considered observational constraints to estimate the outflow parameters from the \civonly\ emission line: (a)  the \civonly\ equivalent width of  is typically very low $W\sim$10\AA\ and in Pop. A $W\leq$30\AA\ \citep{MAetal18}, and (b) the \civ/\hb\ ratio can be assumed to be \civonly/\hb\ $\geq 5\sim 10$, due to the absence of a strong \hb\ blueshifted component.  These conditions are encountered for the LBT sample:  W(\civonly) $\approx$ 4 - 13.
Deconto-Machado et al. performed CLOUDY computations that uses a SED appropriate for high $L$,  from \cite{krawczyk13},  which yields estimates in agreement with those of \cite{vietrietal20}.  For the sake of clarity, we report below the expressions used for \civ.

$M^\mathrm{ion} \sim 
  5 \cdot 10^2 ~L_{45} \left(\frac{Z}{5Z_{\odot}}\right)^{-1} n_{9}^{-1}  
   M_{\odot}$ 

 $\dot{M}^\mathrm{ion} \sim  8  L_{45} v_\mathrm{o,5000} r^{-1}_{\rm 1pc} \left(\frac{Z}{5Z_{\odot}}\right)^{-1} n_{9}^{-1}   M_{\odot} \mathrm{yr}^{-1}$
 
$ \dot{M}^\mathrm{ion} k v_\mathrm{o} \sim 2.7 \cdot 10^{35}  L_{45} k v^{2}_\mathrm{o,5000} r^{-1}_{\rm 1pc}    \left(\frac{Z}{5Z_{\odot}}\right)^{-1} n_{9}^{-1}  $ g cm s$^{-2}$

 $ \dot{\epsilon}  \sim  6.6 \cdot 10^{43} L_{45} k^{2} v^{3}_\mathrm{o,5000}   r^{-1}_{\rm 1pc}   \left(\frac{Z}{5Z_{\odot}}\right)^{-1} n_{9}^{-1}$ erg s$^{-1}$    

where line luminosity, outflow velocity, radius, metallicity, and density have been normalized to 10$^{45}$ erg s$^{-1}$, 5000 \kms,  1 pc, five times solar, and $10^9$ cm$^{-3}$, in the same order.

Table \ref{tab:outflow_params} reports the values derived from the outflow parameters using the equations  above. The value  of $k$ = 5 for \civonly\ is the ratio of the $v$ measured on the profile  and the expected terminal velocity according to the simple model of \citet{NM10}. It is plausible that the emitting gas is still subject to acceleration by strong radiation forces within the BLR. In contrast, in the case of \oiii, assuming that the emission is due to gas expelled from the nucleus, both the kinetic power and thrust are dissipated as the material traverses greater distances towards the outer regions of the AGN, beyond the influence of the central black hole gravitational pull. The alternative is that \oiii\ is emitted by the host galaxy gas. In this case, the \oiii\ emitting gas in outflow would reflect the radiative feedback of the AGN. In either case, the observations imply a notable difference in the outflow conditions between \civonly\ and \oiii\ emission lines.

The estimations based on \civ\ suggest that in highly luminous quasars,  the mechanical feedback estimated from mildly ionized gas (which is part of the multi-frequency outflows observed in powerful quasars) may  be already close to achieve the required effect for evolutionary feedback on the host galaxy. The average ratio between the kinetic power of the \civ\ outflow is $|\dot{\epsilon}| \approx 0.04,$ close to the value expected from theoretical models, $5\%$ of the bolometric luminosity needed to account for the black hole mass - velocity dispersion correlation and  host-spheroid co-evolution \citep[e.g.,][]{dimatteoetal05}.   Effect on circum-nuclear and bulge star formation are expected to occur at a much lower threshold, $\dot{E}_{\textrm{kin}}/L_{\textrm{bol}}\sim 5\times 10^{-3}$ \citep{hopkinsetal10,hoplinselvis10}.  

\begin{figure*}
    \centering
    \includegraphics[width=0.3\textwidth]{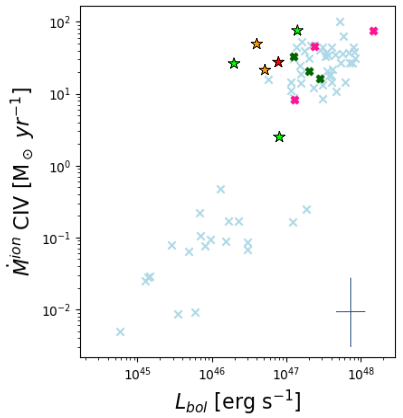}
    \includegraphics[width=0.3\textwidth]{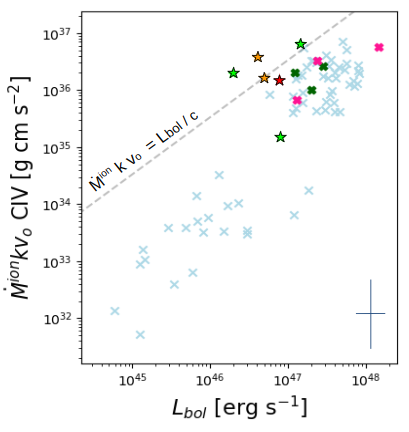}
    \includegraphics[width=0.31\textwidth]{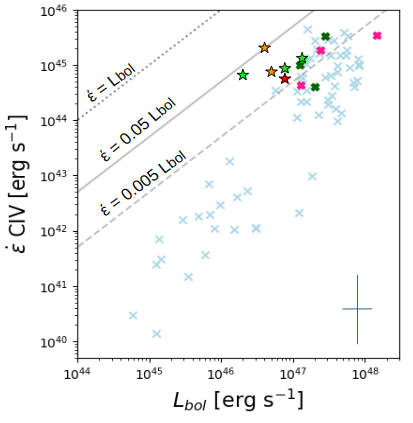}
    \caption{Distribution of the mass outflow rate (left), thrust (middle), and kinetic power (right) vs. bolometric luminosity for the \civ\ emission line. Black dotted, continuous and dashed lines show $\dot{\epsilon}$=\lbol,$\dot{\epsilon}$=0.05\lbol, and $\dot{\epsilon}$=0.005\lbol, respectively. The crosses correspond to data extracted from \citet{deconto-machadoetal24}, and stars for the sources of this work. Color scheme for \citet{deconto-machadoetal24} sources: light blue (Pop. A), green (bonafide Pop. xA), and pink (Pop. xA). Color scheme for the LBT sample: red (Pop. A), orange (bonafide xA), and lime (Pop. xA). Error bar are the typical uncertainties.}
    \label{fig:LbolvsKinpow}
\end{figure*}

\begin{table*}
\caption{Outflow physical parameters derived for \civonly: mass of ionized gas, mass outflow rate,
thrust and kinetic power}
\label{tab:outflow_params}
\begin{tabular}{cccccccc}
\hline
SDSS ID & \mbh & $v_o$ & L$_{\rm \civonly}$ & M$^{\rm ion}$ & $\dot{M}^{\rm ion}$ & $\dot{M}^{\rm ion} kv_o$ & $\dot{\epsilon}$ \\
 & $\pm$1.17 & $\pm$1712 & $\pm$0.21 & $\pm$66 & $\pm$17 & $\pm$2.26 & $\pm$0.332 \\
(1) & (2) & (3) & (4) & (5) & (6) & (7) & (8) \\ \hline
J084502 & 2.29 & -11293 & 44.52 & 165 & 50 & 3.81 & 2.106 \\
J093403$^\star$ & 3.44 & -8753 & 43.47 & 15 & 13 & 0.15 & 0.863 \\
J105427 & 1.49 & -11410 & 44.19 & 78 & 22 & 1.67 & 0.774 \\
J125914 & 0.44 & -8257 & 44.52 & 167 & 28 & 1.54 & 0.554 \\
J144218 & 0.60 & -10846 & 44.13 & 67 & 27 & 1.97 & 0.669 \\
J210831 & 2.49 & -12741 & 44.90 & 395 & 77 & 6.58 & 1.305 \\ \hline
\end{tabular}
\\ \footnotesize
\textbf{Notes}: All measurements are in the quasar rest-frame.  Columns: (1) SDSS identification. (2): black hole mass obtained with \hb, in units of $\times 10^{9}$\msun. (3): final outflow velocity \citep[following the definition by][]{deconto-machadoetal24}, in units of \kms. (4) Logarithm of the line luminosity, in units of \ergss.(5): mass of the ionized gas, in units of \msun. (6) mass outflow rate, in units of \msun\ yr$^{-1}$. (7): thrust, in units of $\times 10^{37}$ g cm s$^{-2}$ . (8): kinetic power, in units of $\times 10^{45}$ \ergss.  {$^\star$Source showing broad absorption lines in \civ\ and \siiv.
}
\end{table*}

\subsection{Metal enrichment}

The availability of the strongest UV lines, along with the emission features in the \hb\ spectral range make it possible to estimate the metal content of the line emitting gas \citep{florisetal24}. Here we follow the methodology already tested in several recent works \citep{sniegowskaetal20,garnicaetal22}.
An advantage with respect to them is the measurement of the \mgiionly, along with \hb\ and the \feiiopt. 

\begin{figure*}
    \includegraphics[width=0.45\textwidth]{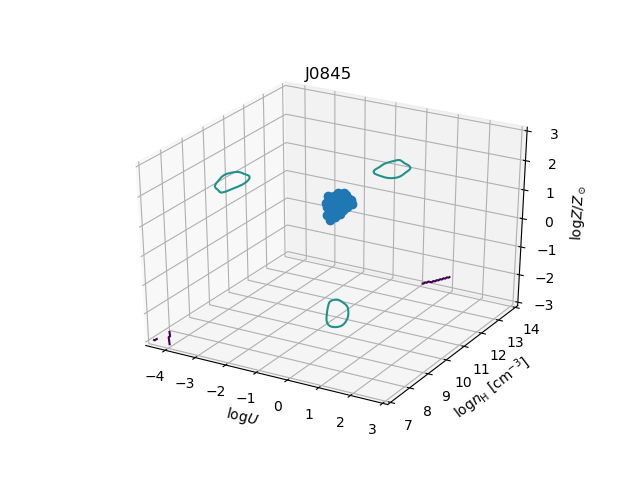}
     \includegraphics[width=0.25\textwidth]{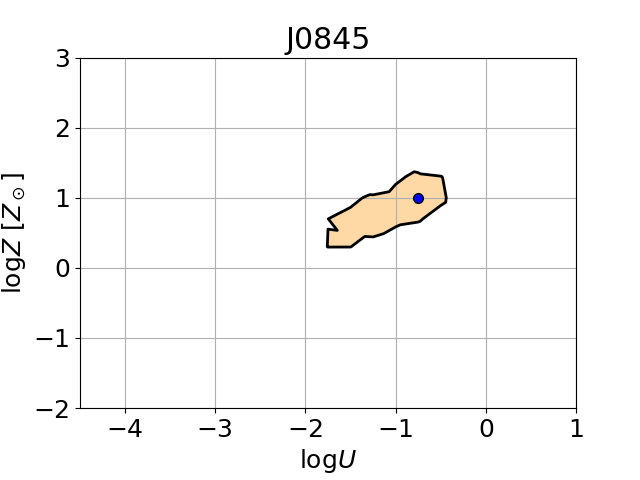}
    \includegraphics[width=0.25\textwidth]{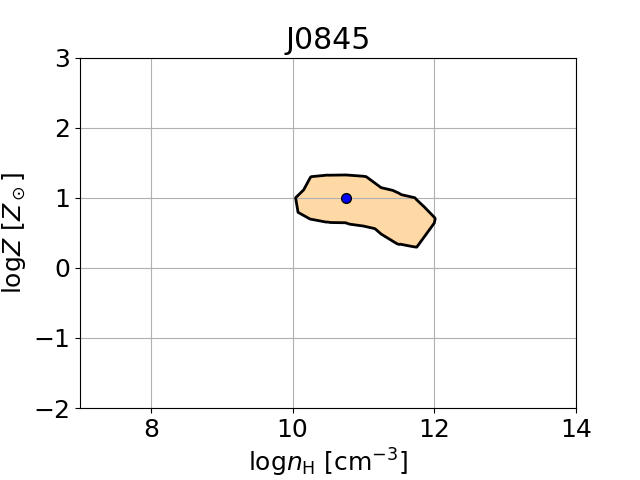}\\
    \caption{Parameter space $n_\mathrm{H}$, $U$, $Z$ for J084502. Left: data points in 3D space are elements in the grid of the parameter space selected for not being different from $\chi^2_\mathrm{min}$ by more than $1 \sigma $\ confidence level, { for the case turbulence $t = 0$ \kms}. Middle: projections on the plane ($\log U$, $\log Z$). Right: projections on the plane $\log n_\mathrm{H}$, $\log Z$.  The blue point marks the minimum $\chi^2$, and the contour delimits the $1\sigma$\ confidence range.}
    \label{fig:j0845z}
\end{figure*}

We utilize { 10}  diagnostic ratios, namely: \civonly/\heiiuv, \siivonly+\oivonly/\heiionly, \siivonly+\oivonly/\civonly, \ciii/\civonly, \aliiionly/\civonly, \aliiionly/\siiii, \siiii/\ciii, \mgiionly/\civonly, \feiiq/\hb, \al/\heiionly, { avoiding  \mgiionly/\hb\ and \civonly/\hb\ that might be affected by the non-simultaneity the of observations.} This set of  ratios is available for 3 objects. For J2018 the \mgiionly\ is not available. We exclude J0934 and J1054 for which the 1400 blend is not observable. We seek the model that yields the minimum $\chi^2$  between observed and predicted diagnostic ratios, in a grid of photoionization models covering the parameter space $n_\mathrm{H}$, $U$\ with a step of 0.25 dex, and for 12 $Z$ values between 0.01 and 1000 solar. We define the $\chi^2$\ as: 

   \begin{equation*}
\hspace{1cm}  
\chi^2(n_\mathrm{H}, U, Z) = \sum_\mathrm{k} w_\mathrm{k}\frac{(R_\mathrm{obs,k} - R_\mathrm{CLOUDY,k})^2}{R_\mathrm{obs,k}^2}
\end{equation*}

where $R_\mathrm{CLOUDY,k}$ identifies each diagnostic ratio computed with {\tt CLOUDY 23.01} \citep{gunasekeraetal23}, for a SED appropriate for xA sources. { The weight $w_\mathrm{k}$ has been assumed equal 1 save in one case (J2108), where a larger weight has been given to \feii\ emission.} The metallicity estimates are summarized in Table \ref{tab:z}. { Four cases with the same SED were considered, based on different turbulence values: $t = 0$ \kms, $t = 20$ \kms, $t = 300$ \kms. } Fig. \ref{fig:j0845z} shows the tight constraints in the parameter space $n_\mathrm{H}$, $U$, $Z$ { for J0845}. The values are consistent with previous estimates for high luminosity quasars at intermediate $z$, 10 -- 20 times the solar values \citep{hamannferland93,hamannferland99,dietrichetal03,xuetal18,sniegowskaetal21,garnicaetal22}, as well as for xA sources at low redshift and luminosity \citep{florisetal24}. { If we take a  prudential value, $Z \sim 10 Z_\odot$, as a reference, the mass outflow rate of metals is   very high $\dot{M}_\mathrm{metals} \sim  (50 \cdot 0.15) M_\odot $yr$^{-1}  \sim 10 M_\odot$ yr$^{-1}$.} 

The most intriguing case is perhaps the one of J210831. This source shows a spectrum where all the relative intensities of the broad components converge toward a scenario of high metallicity, along with low ionization and high density. The low ionization is suggested by the almost complete displacement toward the blue of the \siivonly\ and \civonly\ lines, implying  low values of all the ratios normalizing \civ.  The \rfe\ values is the highest in our sample, and \aliii\ is also very prominent. The O{\sc i} and Si{\sc ii} blend is almost as strong as the 1400 \AA\ blend. These indicators converge toward  extreme solutions \citep{negrete12,MAetal18}. However, the derived  metallicity value is extremely high, implying that both the relative abundances and mass fraction are dominated by metals and that the BLR are strongly polluted through selective enrichment that may yield a chemical composition different from the solar one \citep{wangetal23}.  The scaling of $\sim 500 $\ times the solar metallicity should therefore be viewed with { extreme } care. { The introduction of thermal broadening (i.e., a turbulent velocity dispersion of 20 \kms)  and 300 \kms\ \citep{templeetal20,templeetal21} in the {\tt CLOUDY} simulations is not changing the very high metallicity values which remain extreme.  The main consequence of the higher turbulence ($t = 300$ \kms) is a widening of the confidence range i.e., to somehow ``disperse" the result.  However, if we impose a higher weight on the \rfe\ ratio in the computation of the $\chi^2$, the metallicity of J2108 is reduced to $Z \sim 50 Z_\odot$, consistent with the values derived for the most extreme xA sources \citep{garnicaetal22}. This emphasizes the importance of involving high S/N data (higher S/N than the ones of the spectra in the present paper) that would permit a more precise decomposition of the blended line profiles), as well as   the largest possible number of ratios, possibly given different weight according to the accuracy of the measurements to compute of a minimum $\chi^2$\ that is representative of the best agreement between observed spectrum and model predictions.}   

{ In conclusion, the metallicity values are consistent with the values of the previous works which derived $\gtrsim 10 Z_\odot$ for super-Eddington quasars. This conclusion is reinforced by the stability of the estimates with respect to changes in line broadening. This is not to say that all quasars have BLR chemical composition with super-solar abundance. On the converse, type-1 AGN apparently show a well-defined trends along the quasar main-sequence, also involving slightly sub-solar metallicities ($Z \sim 0.1 - 0.5 Z_\odot$, \citealt{florisetal24}).} 

\begin{table*}
\begin{center}
\caption{Metal content estimates\label{tab:z}.  }
\begin{tabular}{lcccccccccccccccccccccl}\hline\tabcolsep=2pt
& \multicolumn{3}{c}{$t = 0$ \kms} &&  \multicolumn{3}{c}{$t = 20$ \kms} && 
\multicolumn{3}{c}{$t= 300$ \kms$^\mathrm{a}$} && \multicolumn{3}{c}{$t= 300$ \kms}  \\ \cline{2-4} \cline{6-8} \cline{10-12} \cline{14-16}
Ident. & $\log Z$ & 1$\sigma$ range  & $\chi^2_\mathrm{min}$  && $\log Z$ & 1$\sigma$ range  & $\chi^2_\mathrm{min}$  && $\log Z$ & 1$\sigma$ range  & $\chi^2_\mathrm{min}$  &&    $\log Z$ & 1$\sigma$ range  & $\chi^2_\mathrm{min}$  \\
&  [$Z_\odot$] & [$Z_\odot$]       &&&    [$Z_\odot$] & [$Z_\odot$] &&&  [$Z_\odot$] & [$Z_\odot$]    &&&  [$Z_\odot$] & [$Z_\odot$]     \\  \hline
J084502  & 1.0  & 0.3 -- 1.3  &   2.66  &             & 1.0  &  0.3 --  1.3& 2.445     &&  1.0 & 0.7 --  1.3 & 3.735  &&   0.7     & 0.0 -- 2.0& 2.840  &   \\
J125914  & 1.3  & 1.0 -- 1.3 & 2.815   &                & 1.3  & 1.0 --  2.0 & 1.992    &&  0.7 & 0.3 --  2.0    &3.214   &&   1.3   &    1.3 -- 2.0         & 2.027  &   \\
J144218  & 1.0:  & -1.0 -- 1.7 & 4.372 &       &  1.3  &  0.0 --  1.3& 3.416    && 1.3 & 0.0 --  3.0       &6.776   &  & 1.3   & 0.0 -- 2.0  &  3.864 &   \\
J210831$^\mathrm{b}$  & 2.7 & 2.0 -- 3.0 & 1.268 &  &   3.0&  3.0 --  3.0  & 0.939   &&  1.7 & 1.7 --  2.0     &6.565   &  & 3.0 & 2.7 -- 3.0 &     2.159 &    \\ \hline
\end{tabular}
\end{center}
\textbf{Notes}: Columns are  $\log Z$, 1$\sigma$ confidence range in $Z$, minimum $\chi^2$\ in this order. $^\mathrm{a}$: case with higher weight to the \feii/\hb\ ratio, otherwise identical to the other case with turbulence $t = 300$ \kms.   $^\mathrm{b}$ J2108: no \mgiionly. Double colon indicates poor constraints i.e., a wide 1$\sigma$\ range ($2 \times \sigma_{\log Z}\gtrsim 2$). 
\end{table*}

\subsection{Hubble diagram}\label{ssec:hubble}

These physical conditions for xA quasars: (1) Eddington ratio constant or close to the Eddington limit, (2) black hole mass  expressed in terms of virial motions and (3) spectral invariance, so that the product ionization parameter times density has to be approximately constant, make it possible  to write the following equation that yield a redshift-independent luminosity  \citep{MS14}: 

\begin{eqnarray*}\label{eq:vir}
L(\rm FWHM) & = & {\mathcal L_{\rm 0}} \cdot  (\rm FWHM)^{4}_{1000} \\ \nonumber 
 & = & 7.88 \cdot 10^{44} \left(\frac{L}{L_\mathrm{Edd}}\right)_{,1}^{2} \cdot  
  \frac{ \kappa_\mathrm{i,0.5}f_\mathrm{S,2}^2}{h \bar{\nu}_\mathrm{i,100 eV}} \\
  & \cdot &\frac{1}{(n_\mathrm{H}U)_{10^{9.6}}}(\rm FWHM)^{4}_{1000}\, {\rm erg  \, s}^{-1} 
\end{eqnarray*}

where the energy value has been normalized to 100 eV ($ \bar{\nu}_\mathrm{ i,100eV} \approx 2.42 \cdot 10^{16}$ Hz), $\kappa_\mathrm{i,0.5}$\ is the fraction of bolometric luminosity belonging to the ionizing continuum { scaled to 0.5},  the product ($n_{\rm H}U$) has been scaled to the typical value $10^{9.6} $cm$^{-3}$,  and the FWHM of the \hb\ BC  is expressed in units of 1000 \kms.  The FWHM of \hb\ broad component and of \aliii\ are hereafter adopted as a virial broadening estimator. Eq. \ref{eq:vir}  implies that, by a simple measurement of the FWHM of a LIL, one can derive a $z-$independent estimate of the accretion luminosity  \citep[][c.f. \citealt{teerikorpi05}]{MS14}. 

The distance modulus $\mu$ computed from the virial equation, $L(\delta v)$, can be written as:

\begin{equation}
\label{eq:mu0}
\mu  = 2.5 [\log  L(\delta v) - BC] - 2.5 \log  (f_{\lambda} \lambda)  -2.5 \log (4 \pi \delta_\mathrm{10pc}^{2}) + 5 \cdot \log (1+z)
\end{equation}

where the constant  $-2.5 \log (4 \pi \delta_\mathrm{10pc}^{2})$ =-100.19, with $\delta_\mathrm{10pc} \approx 3.08 \cdot 10^{19}$\, the distance of 10 pc expressed in cm. The  $f_{\lambda} \lambda$\ can be the flux at 5100 \AA\ for the \hb\ sample, or  the flux at 1700 \AA\ if the $\delta v$=FWHM comes from the \aliii\ and \siiii\ lines  \citep{dultzin20}. 

Two of the six targets observed with the LBT fully meet the criterion defining xA sources (\rfe\ $\gtrsim 1$); three of them can be considered {\em bona-fide} borderline xA, and one of them, J125914, a Pop. A but not meeting the xA selection criterion. 

\begin{figure*}
\centering
    \includegraphics[width=0.45\textwidth]{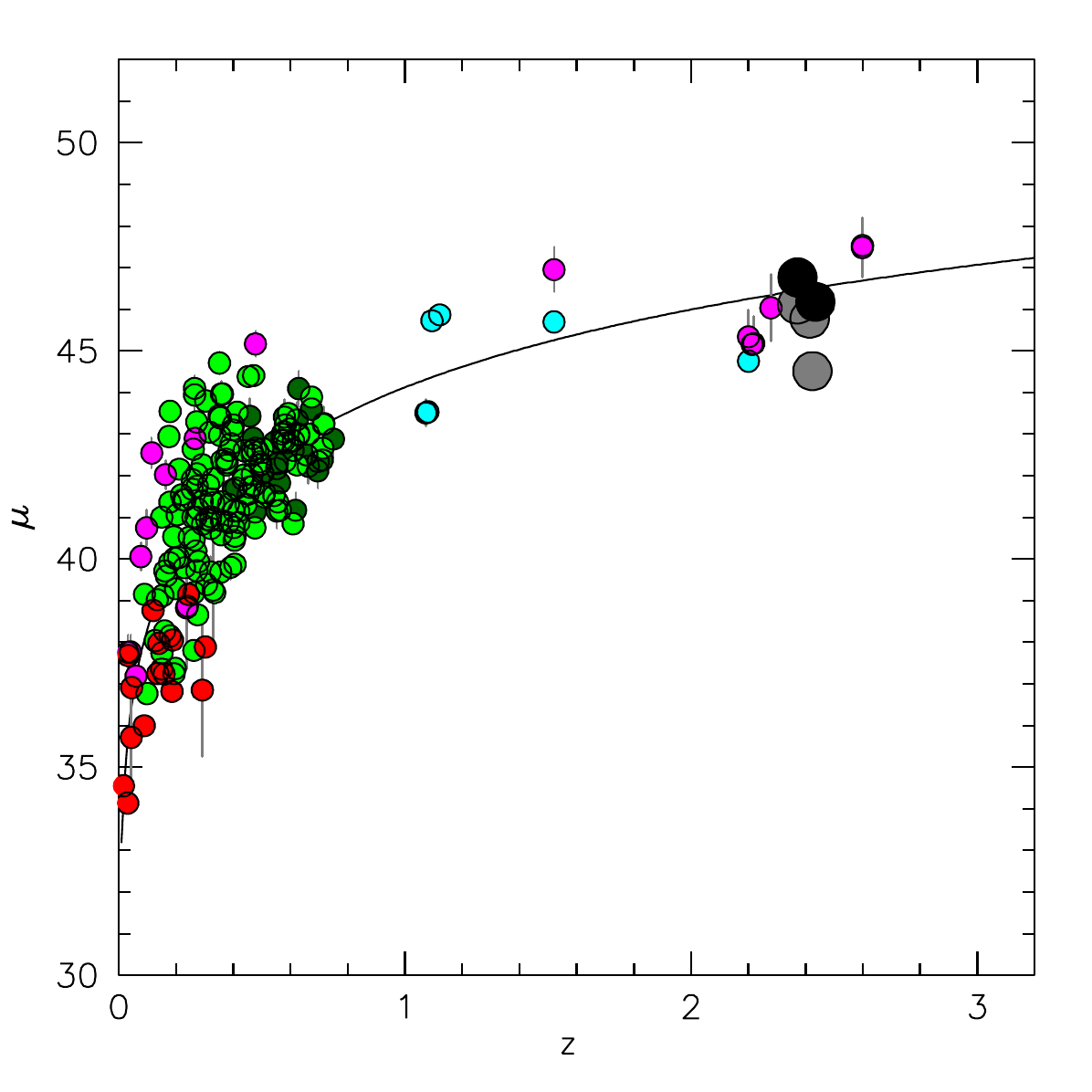}
    \includegraphics[width=0.45\textwidth]{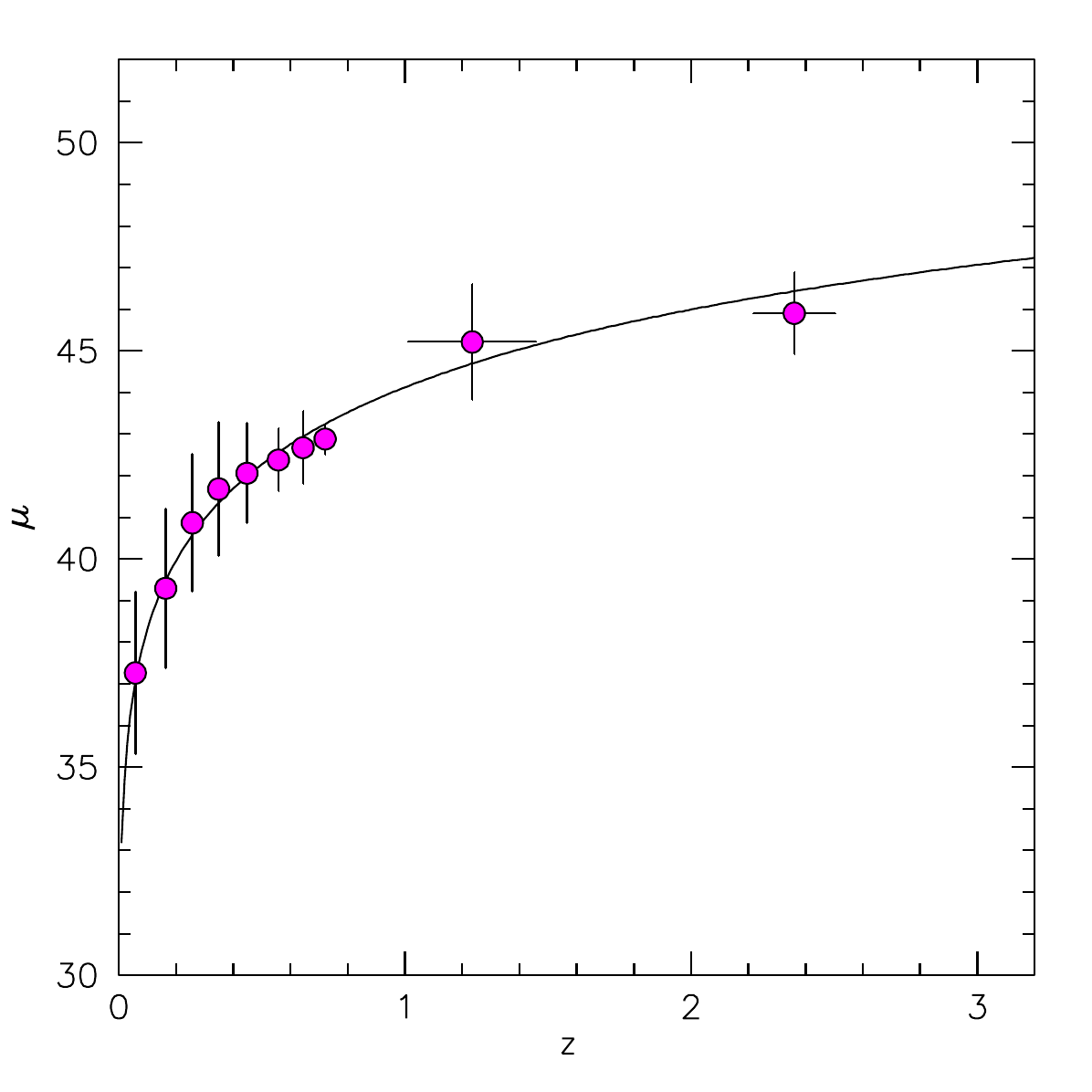}
    \vspace{-.5cm}\\
    \includegraphics[width=0.45\textwidth]{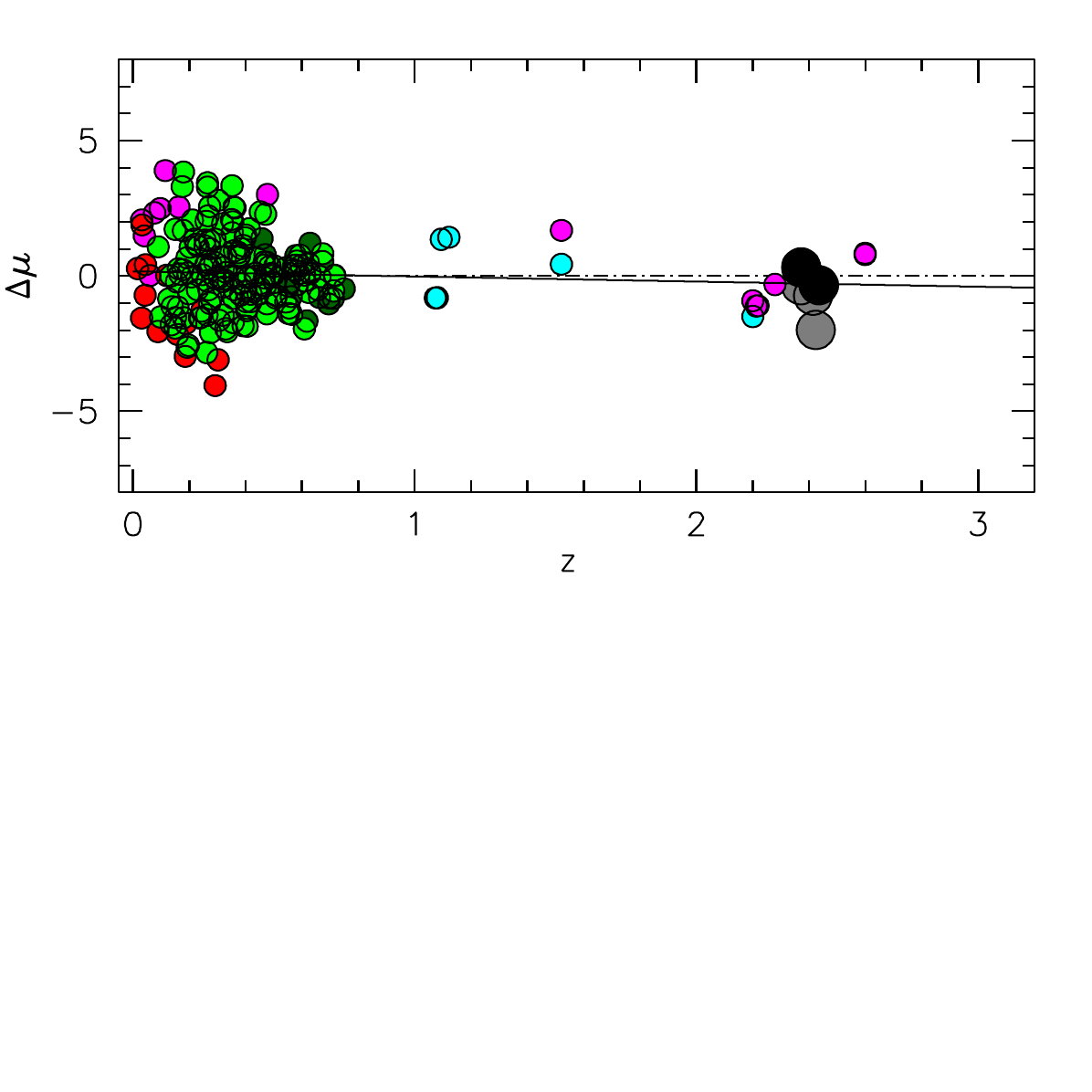}
    \includegraphics[width=0.45\textwidth]{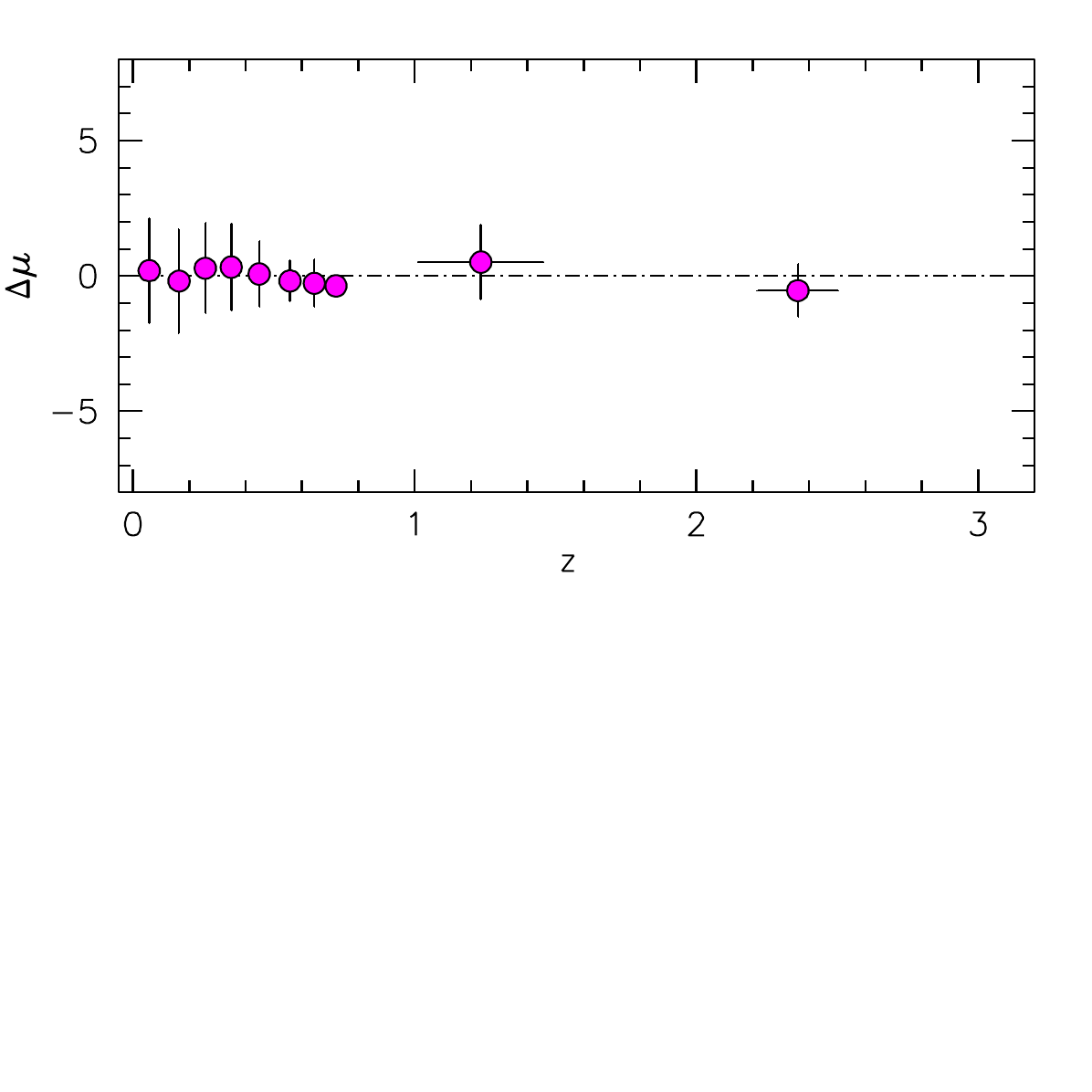}
    \vspace{-3.5cm}
    \caption{Left: {Hubble diagram distance modulus $\mu$\ versus redshift $z$ obtained from the virial luminosity equation. The five sources that are classified as xA (black) or borderline xA (gray) considered in this study are shown by larger symbols. Samples in \citet{dultzin20} are included: red \citet{duetal16}, green \citet{negrete18}, dark green \citet{marzianietal13a}. Cyan data points represents the xA quasars of the Hamburg-ESO Marziani-Sulentic (HEMS) survey \citep{marzianietal09}, magenta the  sample of \citet{marziani22}. Right: same as for the left panel, but with $\mu$ averaged over 10 redshift ranges. Lines: least-square on the residuals (gray) and $\Lambda$CDM cosmology model (black). }}
    \label{fig:hd}
\end{figure*}

The Figure \ref{fig:hd} shows the Hubble diagram $\mu$ versus $z$\ for several samples for which $\mu$\ has been derived from FWHM(\hb). The 5 objects analyzed in the present work are shown by larger symbols in Figure \ref{fig:hd} (black xA, gray borderline). It is notable that the two xA quasars closely agree with the prediction of standard cosmology and with the determinations from other samples. The averages of $\mu$ over redshift bins from 0 to $\approx 2.6$ (Figure \ref{fig:hd}, right panels)  for all determinations shown individually in the left panels) is following the trend expected for standard cosmology with some small deviations.  The agreement between the distance moduli estimated from the line width and the ones derived from standard cosmology is a restatement of a few basic aspects of the LIL emitting regions: a virial velocity field dominated by the central black hole mass, a scaling of the emitting region radius proportional to the square root of luminosity (to preserve spectral invariance), and of extreme Eddington ratios (Table \ref{tab:acc_param} reports \lledd $\gtrsim 1$).  These conditions imply an increase of FWHM with $L^\frac{1}{4}$ \citep{marziani22}, or $L \propto$FWHM$^4$ \citep{MS14}.     Clearly, more high quality, NIR observations -- such as the ones collected with LBT --  of the \hb\ spectral range for quasars at redshift $z \gtrsim 1$\ are needed to test the possibility of using the the distance moduli estimated from the line width as distance indicators.



\section{Summary and conclusions}\label{sec:conclusions}

The main results of the investigation can be summarized as follows:

\begin{itemize}
      \item We presented new LUCI  observations at the  LBT  covering a sample of 6 highly accreting quasars (super-Eddington candidates) at $z\sim2$. The NIR spectrographic camera LUCI allowed us to cover the \hb\ spectral region which yields a wealth of diagnostic information on the physical condition and dynamics of the line emitting region. SDSS spectra were used to analyze the spectral region of the UV emission lines from \siiv+\oiv\ to \mgii.  
     \item We carried out the data reduction of the original LUCI observations to ensure to maximize the S/N of the extracted spectra. 
    
     \item The bolometric luminosity of the sample source is very high, $\log L_\mathrm{bol} \gtrsim 47$ [erg s$^{-1}$], close to the domain of the WISE/SDSS selected hyper-luminous (WISSH)   \citep{bischettietal17} and HEMS surveys \citep{marzianietal09}. The Eddington ratio is close to one, with average $\log$\lledd $\sim  -0.1 \pm 0.1$\ ($\approx 0$ if the outlier J125914 is excluded), confirms the high radiative output of the sample sources.
    \item We considered the cross match between optical and UV selection criteria for the identification of xA sources.  Two sources are bonafide xA sources, three can be considered borderline xA, and one of them is a Pop. A quasar.  

    \item A multi-component fitting of the most prominent broad emission lines in the UV-optical ranges: 1400\AA\ blend, \civ, 1900\AA\ blend, \mgii, and \hb, permitted us to isolated a symmetric and unshifted component, strong in the low ionization lines,  along with a blueshifted component associated with a strong outflow dominating the emission of the high ionization lines. The  high amplitude \civ\ blueshifts $v \lesssim -2000$ \kms\ is detected in five sources { out of six, all the super-Eddington candidates of the LBT sample.}
     
    \item  Blueshifts are higher amplitudes in the lines (\civ, \heii, \siiv) whose parent species are of higher ionization potential.  
    \item At the same time, \hb,  and \mgii\ appear to be almost symmetric and unshifted or only marginally shifted. The intermediate ionization doublet \aliii\ shows modest shifts more consistent with the low-ionization lines. In larger samples, \aliii\ large shifts have been revealed although at a low prevalence \citep{marziani22,buendia2023}. Their absence might be a consequence of the small sample  size.
    
 \end{itemize}   

Several additional results are implied by the xA physical parameter estimates.
 
 \begin{itemize}   
    
    \item Estimates of the outflow dynamical parameters (mass of ionized gas, mass outflow rate, thrust and kinetic power) were made using the \civonly\ and \oiii\ emission lines. The wind parameter derived from the blueshifted \civ\ emission have kinetic power close to the one needed to account for host galaxy and black hole coevolution. Since AGN winds are multi-frequency \citep[e.g.,][]{lahaetal21,chartasetal21,pretisasikumar23,gianollietal24}, it appears likely that the ``bolometric" kinetic power may indeed be able to exert the feedback effect expected to drive the \mbh\ -- host galaxy velocity dispersion correlation. 
    \item The coverage of the UV and optical spectral ranges provided the measurements for $\sim 10 -  12 $\ diagnostic ratios suitable for the measurement of the metallicity of the broad line emitting gas. We confirm the high metallicity already found for xA systems at low- and high-luminosity \citep[][and references therein]{florisetal24}. 
    \item Finally, we applied the virial luminosity equation to built the Hubble diagram for a sample of xA sources. We added 5 genuine xA with the luminosity estimated from the line width of \hb. The results are consistent with the standard $\Lambda$CDM cosmology \citep{donofrioetal24}. More than of any cosmological implication, the Hubble diagram yields a representation of the virialized nature of the low-ionization emission preserved even  in extreme quasars. 
\end{itemize}

In conclusion, the LBT spectra of the \hb\ rest frame range allowed us to consider {several properties of super-Eddington candidates, already analyzed for sources selected from the \rfe$\gtrsim 1 $\ criterion}, as well as to achieve a proper estimation of the AGN outflow parameters by accurately measuring the redshift   quasar redshift, confirming the possibility of significant feedback. In addition, even if the sample included only 5  extreme quasars, the agreement of their \hb\ line widths with   cosmological expectations, and the absence of significant shifts signaling strong wind effects,  indicate that larger \hb\ samples may one day turn useful for an independent estimation of the cosmological parameters. 


\section*{Acknowledgements}

D. Dultzin and C. A. Negrete acknowledge the support form grant IN111422 PAPIIT UNAM. C. A. Negrete acknowledges the support form CONACyT project Paradigmas y Controversias de la Ciencia 2022-320020, CONAHCyT CBF2023-2024-1418, PAPIIT IA104325 and IN119123. The work of T. M. Buendia-Rios has been sponsored by CONACYT-Mexico through the Ph.D. scholarship No. 760641. T. M. B.-R. acknowledges the hospitality and support of INAF -- Astronomical Observatory of Padova where part of this work was done. This publication makes use of data products from the Two Micron All Sky Survey, which is a joint project of the University of Massachusetts and the Infrared Processing and Analysis Center/California Institute of Technology, funded by the National Aeronautics and Space Administration and the National Science Foundation. This publication used the facilities of the Italian Center for Astronomical Archive (IA2) operated by INAF at the Astronomical Observatory of Trieste.

The LBT is an international collaboration among institutions in the United States, Italy, and Germany. LBT Corporation partners are: The University of Arizona on behalf of the Arizona university system; Istituto Nazionale di Astrofisica, Italy; LBT Beteiligungsgesellschaft, Germany, representing the Max-Planck Society, the Astrophysical Institute Potsdam, and Heidelberg University; The Ohio State University, and The Research Corporation, on behalf of The University of Notre Dame, University of Minnesota and University of Virginia.
\section*{Data Availability}

IR spectra in ASCII format are provided as part of the paper.  For each spectrum, the rest frame wavelength and specific flux are reported in units of \AA\ and $10^{-17}$ erg s$^{-1}$ cm$^{-2}$ \AA$^{-1}$, respectively.   

Optical spectra are from the SDSS and can be downloaded from \href{https://skyserver.sdss.org/dr18/}{the data server of the SDSS Data Release 18}.



\bibliographystyle{mnras}
\bibliography{biblio} 











\bsp	
\label{lastpage}
\end{document}